%
%
%
\def\unredoffs{} \def\redoffs{\voffset=-.31truein\hoffset=-.48truein}
\def\speclscape{}
%
%
%
%
%
\newbox\leftpage \newdimen\fullhsize \newdimen\hstitle \newdimen\hsbody
\tolerance=1000\hfuzz=2pt
\catcode`\@=11 
\ifx\hyperdef\UNd@FiNeD\def\hyperdef#1#2#3#4{#4}\def\hyperref#1#2#3#4{#4}\fi
\def\bigans{b }
\def\answ{b }
%
\ifx\answ\bigans\message{(This will come out unreduced.}
\magnification=1200\unredoffs\baselineskip=16pt plus 2pt minus 1pt
\hsbody=\hsize \hstitle=\hsize 
\else\message{(This will be reduced.} \let\l@r=L
\magnification=1000\baselineskip=16pt plus 2pt minus 1pt \vsize=7truein
\redoffs \hstitle=8truein\hsbody=4.75truein\fullhsize=10truein\hsize=\hsbody
\output={\ifnum\pageno=0 
  \shipout\vbox{\speclscape{\hsize\fullhsize\makeheadline}
    \hbox to \fullhsize{\hfill\pagebody\hfill}}\advancepageno
  \else
  \almostshipout{\leftline{\vbox{\pagebody\makefootline}}}\advancepageno
  \fi}
\def\almostshipout#1{\if L\l@r \count1=1 \message{[\the\count0.\the\count1]}
      \global\setbox\leftpage=#1 \global\let\l@r=R
 \else \count1=2
  \shipout\vbox{\speclscape{\hsize\fullhsize\makeheadline}
      \hbox to\fullhsize{\box\leftpage\hfil#1}}  \global\let\l@r=L\fi}
\fi
%
\newcount\yearltd\yearltd=\year\advance\yearltd by -1900

\def\Title#1#2{\nopagenumbers\abstractfont\hsize=\hstitle\rightline{#1}%
\vskip 1in\centerline{\titlefont #2}\abstractfont\vskip .5in\pageno=0}
\def\Date#1{\vfill\leftline{#1}\tenpoint\supereject\global\hsize=\hsbody%
\footline={\hss\tenrm\hyperdef\hypernoname{page}\folio\folio\hss}}%
%

\def\draftmode{\message{ DRAFTMODE }\def\draftdate{{\rm preliminary draft:
\number\month/\number\day/\number\yearltd\ \ \hourmin}}%
\headline={\hfil\draftdate}\writelabels\baselineskip=20pt plus 2pt minus 2pt
 {\count255=\time\divide\count255 by 60 \xdef\hourmin{\number\count255}
  \multiply\count255 by-60\advance\count255 by\time
  \xdef\hourmin{\hourmin:\ifnum\count255<10 0\fi\the\count255}}}
\def\nolabels{\def\wrlabeL##1{}\def\eqlabeL##1{}\def\reflabeL##1{}}
\def\writelabels{\def\wrlabeL##1{\leavevmode\vadjust{\rlap{\smash%
{\line{{\escapechar=` \hfill\rlap{\sevenrm\hskip.03in\string##1}}}}}}}%
\def\eqlabeL##1{{\escapechar-1\rlap{\sevenrm\hskip.05in\string##1}}}%
\def\reflabeL##1{\noexpand\llap{\noexpand\sevenrm\string\string\string##1}}}
\nolabels
%
\global\newcount\secno \global\secno=0
\global\newcount\meqno \global\meqno=1
\def\s@csym{}
\def\newsec#1{\global\advance\secno by1%
{\toks0{#1}\message{(\the\secno. \the\toks0)}}%
\global\subsecno=0\eqnres@t\let\s@csym\secsym\xdef\secn@m{\the\secno}\noindent
{\bf\hyperdef\hypernoname{section}{\the\secno}{\the\secno.} #1}%
\writetoca{{\string\hyperref{}{section}{\the\secno}{\the\secno.}} {#1}}%
\par\nobreak\medskip\nobreak}
\def\eqnres@t{\xdef\secsym{\the\secno.}\global\meqno=1\bigbreak\bigskip}
\def\sequentialequations{\def\eqnres@t{\bigbreak}}\xdef\secsym{}
\global\newcount\subsecno \global\subsecno=0
\def\subsec#1{\global\advance\subsecno by1%
{\toks0{#1}\message{(\s@csym\the\subsecno. \the\toks0)}}%
\ifnum\lastpenalty>9000\else\bigbreak\fi
\noindent{\it\hyperdef\hypernoname{subsection}{\secn@m.\the\subsecno}%
{\secn@m.\the\subsecno.} #1}\writetoca{\string\quad
{\string\hyperref{}{subsection}{\secn@m.\the\subsecno}{\secn@m.\the\subsecno.}}
{#1}}\par\nobreak\medskip\nobreak}
\def\appendix#1#2{\global\meqno=1\global\subsecno=0\xdef\secsym{\hbox{#1.}}%
\bigbreak\bigskip\noindent{\bf Appendix \hyperdef\hypernoname{appendix}{#1}%
{#1.} #2}{\toks0{(#1. #2)}\message{\the\toks0}}%
\xdef\s@csym{#1.}\xdef\secn@m{#1}%
\writetoca{\string\hyperref{}{appendix}{#1}{Appendix {#1.}} {#2}}%
\par\nobreak\medskip\nobreak}
%
%
\def\checkm@de#1#2{\ifmmode{\def\f@rst##1{##1}\hyperdef\hypernoname{equation}%
{#1}{#2}}\else\hyperref{}{equation}{#1}{#2}\fi}
\def\eqnn#1{\DefWarn#1\xdef #1{(\noexpand\relax\noexpand\checkm@de%
{\s@csym\the\meqno}{\secsym\the\meqno})}%
\wrlabeL#1\writedef{#1\leftbracket#1}\global\advance\meqno by1}
\def\f@rst#1{\c@t#1a\em@ark}\def\c@t#1#2\em@ark{#1}
\def\eqna#1{\DefWarn#1\wrlabeL{#1$\{\}$}%
\xdef #1##1{(\noexpand\relax\noexpand\checkm@de%
{\s@csym\the\meqno\noexpand\f@rst{##1}}{\hbox{$\secsym\the\meqno##1$}})}
\writedef{#1\numbersign1\leftbracket#1{\numbersign1}}\global\advance\meqno by1}
\def\eqn#1#2{\DefWarn#1%
\xdef #1{(\noexpand\hyperref{}{equation}{\s@csym\the\meqno}%
{\secsym\the\meqno})}$$#2\eqno(\hyperdef\hypernoname{equation}%
{\s@csym\the\meqno}{\secsym\the\meqno})\eqlabeL#1$$%
\writedef{#1\leftbracket#1}\global\advance\meqno by1}
\def\xeqn{\expandafter\xe@n}\def\xe@n(#1){#1}
\def\xeqna#1{\expandafter\xe@n#1}
\def\eqns#1{(\e@ns #1{\hbox{}})}
\def\e@ns#1{\ifx\UNd@FiNeD#1\message{eqnlabel \string#1 is undefined.}%
\xdef#1{(?.?)}\fi{\let\hyperref=\relax\xdef\next{#1}}%
\ifx\next\em@rk\def\next{}\else%
\ifx\next#1\xeqn#1\else\def\n@xt{#1}\ifx\n@xt\next#1\else\xeqna#1\fi
\fi\let\next=\e@ns\fi\next}

\def\DefWarn#1{\ifx\UNd@FiNeD#1\else
\immediate\write16{*** WARNING: the label \string#1 is already defined ***}\fi}
%
\newskip\footskip\footskip14pt plus 1pt minus 1pt 
\def\footnotefont{\ninepoint}\def\f@t#1{\footnotefont #1\@foot}
\def\f@@t{\baselineskip\footskip\bgroup\footnotefont\aftergroup\@foot\let\next}
\setbox\strutbox=\hbox{\vrule height9.5pt depth4.5pt width0pt}
\global\newcount\ftno \global\ftno=0
\def\foot{\global\advance\ftno by1\def\foot@rg{\hyperref{}{footnote}%
{\the\ftno}{\the\ftno}\xdef\foot@rg{\noexpand\hyperdef\noexpand\hypernoname%
{footnote}{\the\ftno}{\the\ftno}}}\footnote{$^{\foot@rg}$}}
%
\newwrite\ftfile
\def\footend{\def\foot{\global\advance\ftno by1\chardef\wfile=\ftfile
\hyperref{}{footnote}{\the\ftno}{$^{\the\ftno}$}%
\ifnum\ftno=1\immediate\openout\ftfile=\jobname.fts\fi%
\immediate\write\ftfile{\noexpand\smallskip%
\noexpand\item{\noexpand\hyperdef\noexpand\hypernoname{footnote}
{\the\ftno}{f\the\ftno}:\ }\pctsign}\findarg}%
\def\footatend{\vfill\eject\immediate\closeout\ftfile{\parindent=20pt
\centerline{\bf Footnotes}\nobreak\bigskip\input \jobname.fts }}}
\def\footatend{}
%
%
\global\newcount\refno \global\refno=1
\newwrite\rfile
\def\ref{[\hyperref{}{reference}{\the\refno}{\the\refno}]\nref}
\def\nref#1{\DefWarn#1%
\xdef#1{[\noexpand\hyperref{}{reference}{\the\refno}{\the\refno}]}%
\writedef{#1\leftbracket#1}%
\ifnum\refno=1\immediate\openout\rfile=\jobname.refs\fi
\chardef\wfile=\rfile\immediate\write\rfile{\noexpand\item{[\noexpand\hyperdef%
\noexpand\hypernoname{reference}{\the\refno}{\the\refno}]\ }%
\reflabeL{#1\hskip.31in}\pctsign}\global\advance\refno by1\findarg}
\def\findarg#1#{\begingroup\obeylines\newlinechar=`\^^M\pass@rg}
{\obeylines\gdef\pass@rg#1{\writ@line\relax #1^^M\hbox{}^^M}%
\gdef\writ@line#1^^M{\expandafter\toks0\expandafter{\striprel@x #1}%
\edef\next{\the\toks0}\ifx\next\em@rk\let\next=\endgroup\else\ifx\next\empty%
\else\immediate\write\wfile{\the\toks0}\fi\let\next=\writ@line\fi\next\relax}}
\def\striprel@x#1{} \def\em@rk{\hbox{}}
\def\lref{\begingroup\obeylines\lr@f}
\def\lr@f#1#2{\DefWarn#1\gdef#1{\let#1=\UNd@FiNeD\ref#1{#2}}\endgroup\unskip}

\def\addref#1{\immediate\write\rfile{\noexpand\item{}#1}} 
\def\listrefs{\footatend\vfill\supereject\immediate\closeout\rfile\writestoppt
\baselineskip=\footskip\centerline{{\bf References}}\bigskip{\parindent=20pt%
\frenchspacing\escapechar=` \input \jobname.refs\vfill\eject}\nonfrenchspacing}
\def\startrefs#1{\immediate\openout\rfile=\jobname.refs\refno=#1}
\def\xref{\expandafter\xr@f}\def\xr@f[#1]{#1}
\def\refs#1{\count255=1[\r@fs #1{\hbox{}}]}
\def\r@fs#1{\ifx\UNd@FiNeD#1\message{reflabel \string#1 is undefined.}%
\nref#1{need to supply reference \string#1.}\fi%
\vphantom{\hphantom{#1}}{\let\hyperref=\relax\xdef\next{#1}}%
\ifx\next\em@rk\def\next{}%
\else\ifx\next#1\ifodd\count255\relax\xref#1\count255=0\fi%
\else#1\count255=1\fi\let\next=\r@fs\fi\next}
%

%
\newwrite\ffile\global\newcount\figno \global\figno=1
\def\fig{fig.~\hyperref{}{figure}{\the\figno}{\the\figno}\nfig}
\def\nfig#1{\DefWarn#1%
\xdef#1{fig.~\noexpand\hyperref{}{figure}{\the\figno}{\the\figno}}%
\writedef{#1\leftbracket fig.\noexpand~\xfig#1}%
\ifnum\figno=1\immediate\openout\ffile=\jobname.figs\fi\chardef\wfile=\ffile%
{\let\hyperref=\relax
\immediate\write\ffile{\noexpand\medskip\noexpand\item{Fig.\ %
\noexpand\hyperdef\noexpand\hypernoname{figure}{\the\figno}{\the\figno}. }
\reflabeL{#1\hskip.55in}\pctsign}}\global\advance\figno by1\findarg}
\def\listfigs{\vfill\eject\immediate\closeout\ffile{\parindent40pt
\baselineskip14pt\centerline{{\bf Figure Captions}}\nobreak\medskip
\escapechar=` \input \jobname.figs\vfill\eject}}
\def\xfig{\expandafter\xf@g}\def\xf@g fig.\penalty\@M\ {}
\def\figs#1{figs.~\f@gs #1{\hbox{}}}
\def\f@gs#1{{\let\hyperref=\relax\xdef\next{#1}}\ifx\next\em@rk\def\next{}\else
\ifx\next#1\xfig #1\else#1\fi\let\next=\f@gs\fi\next}
\def\figin{\epsfcheck\figin}\def\figins{\epsfcheck\figins}
\def\epsfcheck{\ifx\epsfbox\UNd@FiNeD
\message{(NO epsf.tex, FIGURES WILL BE IGNORED)}
\gdef\figin##1{\vskip2in}\gdef\figins##1{\hskip.5in}
\else\message{(FIGURES WILL BE INCLUDED)}%
\gdef\figin##1{##1}\gdef\figins##1{##1}\fi}
\def\DefWarn#1{}
\def\figinsert{\goodbreak\midinsert}
\def\ifig#1#2#3{\DefWarn#1\xdef#1{fig.~\noexpand\hyperref{}{figure}%
{\the\figno}{\the\figno}}\writedef{#1\leftbracket fig.\noexpand~\xfig#1}%
\figinsert\figin{\centerline{#3}}\medskip\centerline{\vbox{\baselineskip12pt
\advance\hsize by -1truein\noindent\wrlabeL{#1=#1}\footnotefont%
{\bf Fig.~\hyperdef\hypernoname{figure}{\the\figno}{\the\figno}:} #2}}
\bigskip\endinsert\global\advance\figno by1}
\newwrite\lfile
{\escapechar-1\xdef\pctsign{\string\%}\xdef\leftbracket{\string\{}
\xdef\rightbracket{\string\}}\xdef\numbersign{\string\#}}
\def\writedefs{\immediate\openout\lfile=\jobname.defs \def\writedef##1{%
{\let\hyperref=\relax\let\hyperdef=\relax\let\hypernoname=\relax
 \immediate\write\lfile{\string\def\string##1\rightbracket}}}}%
\def\writestop{\def\writestoppt{\immediate\write\lfile{\string\pageno
 \the\pageno\string\startrefs\leftbracket\the\refno\rightbracket
 \string\def\string\secsym\leftbracket\secsym\rightbracket
 \string\secno\the\secno\string\meqno\the\meqno}\immediate\closeout\lfile}}
\def\writestoppt{}\def\writedef#1{}
\def\seclab#1{\DefWarn#1%
\xdef #1{\noexpand\hyperref{}{section}{\the\secno}{\the\secno}}%
\writedef{#1\leftbracket#1}\wrlabeL{#1=#1}}
\def\subseclab#1{\DefWarn#1%
\xdef #1{\noexpand\hyperref{}{subsection}{\secn@m.\the\subsecno}%
{\secn@m.\the\subsecno}}\writedef{#1\leftbracket#1}\wrlabeL{#1=#1}}
\def\applab#1{\DefWarn#1%
\xdef #1{\noexpand\hyperref{}{appendix}{\secn@m}{\secn@m}}%
\writedef{#1\leftbracket#1}\wrlabeL{#1=#1}}
\newwrite\tfile \def\writetoca#1{}
\def\leaderfill{\leaders\hbox to 1em{\hss.\hss}\hfill}
\def\writetoc{\immediate\openout\tfile=\jobname.toc
   \def\writetoca##1{{\edef\next{\write\tfile{\noindent ##1
   \string\leaderfill {\string\hyperref{}{page}{\noexpand\number\pageno}%
                       {\noexpand\number\pageno}} \par}}\next}}}
\newread\ch@ckfile
\def\listtoc{\immediate\closeout\tfile\immediate\openin\ch@ckfile=\jobname.toc
\ifeof\ch@ckfile\message{no file \jobname.toc, no table of contents this pass}%
\else\closein\ch@ckfile\centerline{\bf Contents}\nobreak\medskip%
{\baselineskip=18pt
\parskip=2pt\catcode`\@=11\input\jobname.toc
\catcode`\@=12\bigbreak\bigskip}\fi}
\catcode`\@=12 
%
\edef\tfontsize{\ifx\answ\bigans scaled\magstep3\else scaled\magstep4\fi}
\font\titlerm=cmr10 \tfontsize \font\titlerms=cmr7 \tfontsize
\font\titlermss=cmr5 \tfontsize \font\titlei=cmmi10 \tfontsize
\font\titleis=cmmi7 \tfontsize \font\titleiss=cmmi5 \tfontsize
\font\titlesy=cmsy10 \tfontsize \font\titlesys=cmsy7 \tfontsize
\font\titlesyss=cmsy5 \tfontsize \font\titleit=cmti10 \tfontsize
\skewchar\titlei='177 \skewchar\titleis='177 \skewchar\titleiss='177
\skewchar\titlesy='60 \skewchar\titlesys='60 \skewchar\titlesyss='60
\def\titlefont{\def\rm{\fam0\titlerm}
\textfont0=\titlerm \scriptfont0=\titlerms \scriptscriptfont0=\titlermss
\textfont1=\titlei \scriptfont1=\titleis \scriptscriptfont1=\titleiss
\textfont2=\titlesy \scriptfont2=\titlesys \scriptscriptfont2=\titlesyss
\textfont\itfam=\titleit \def\it{\fam\itfam\titleit}\rm}
 \ifx\answ\bigans\else scaled\magstep1\fi
\ifx\answ\bigans\def\abstractfont{\tenpoint}\else
\font\absit=cmti10 scaled \magstep1
\font\abssl=cmsl10 scaled \magstep1
\font\absrm=cmr10 scaled\magstep1 \font\absrms=cmr7 scaled\magstep1
\font\absrmss=cmr5 scaled\magstep1 \font\absi=cmmi10 scaled\magstep1
\font\absis=cmmi7 scaled\magstep1 \font\absiss=cmmi5 scaled\magstep1
\font\abssy=cmsy10 scaled\magstep1 \font\abssys=cmsy7 scaled\magstep1
\font\abssyss=cmsy5 scaled\magstep1 \font\absbf=cmbx10 scaled\magstep1
\skewchar\absi='177 \skewchar\absis='177 \skewchar\absiss='177
\skewchar\abssy='60 \skewchar\abssys='60 \skewchar\abssyss='60
\def\abstractfont{\def\rm{\fam0\absrm}
\textfont0=\absrm \scriptfont0=\absrms \scriptscriptfont0=\absrmss
\textfont1=\absi \scriptfont1=\absis \scriptscriptfont1=\absiss
\textfont2=\abssy \scriptfont2=\abssys \scriptscriptfont2=\abssyss
\textfont\itfam=\absit \def\it{\fam\itfam\absit}\def\footnotefont{\tenpoint}%
\textfont\slfam=\abssl \def\sl{\fam\slfam\abssl}%
\textfont\bffam=\absbf \def\bf{\fam\bffam\absbf}\rm}\fi
\def\tenpoint{\def\rm{\fam0\tenrm}
\textfont0=\tenrm \scriptfont0=\sevenrm \scriptscriptfont0=\fiverm
\textfont1=\teni  \scriptfont1=\seveni  \scriptscriptfont1=\fivei
\textfont2=\tensy \scriptfont2=\sevensy \scriptscriptfont2=\fivesy
\textfont\itfam=\tenit \def\it{\fam\itfam\tenit}\def\footnotefont{\ninepoint}%
\textfont\bffam=\tenbf \def\bf{\fam\bffam\tenbf}\def\sl{\fam\slfam\tensl}\rm}
\font\ninerm=cmr9 \font\sixrm=cmr6 \font\ninei=cmmi9 \font\sixi=cmmi6
\font\ninesy=cmsy9 \font\sixsy=cmsy6 \font\ninebf=cmbx9
\font\nineit=cmti9 \font\ninesl=cmsl9 \skewchar\ninei='177
\skewchar\sixi='177 \skewchar\ninesy='60 \skewchar\sixsy='60
\def\ninepoint{\def\rm{\fam0\ninerm}
\textfont0=\ninerm \scriptfont0=\sixrm \scriptscriptfont0=\fiverm
\textfont1=\ninei \scriptfont1=\sixi \scriptscriptfont1=\fivei
\textfont2=\ninesy \scriptfont2=\sixsy \scriptscriptfont2=\fivesy
\textfont\itfam=\ninei \def\it{\fam\itfam\nineit}\def\sl{\fam\slfam\ninesl}%
\textfont\bffam=\ninebf \def\bf{\fam\bffam\ninebf}\rm}
%
%
\def\noblackbox{\overfullrule=0pt}
\hyphenation{anom-aly anom-alies coun-ter-term coun-ter-terms}
\def\inv{^{\raise.15ex\hbox{${\scriptscriptstyle -}$}\kern-.05em 1}}

\def\Dsl{\,\raise.15ex\hbox{/}\mkern-13.5mu D} 
\def\dsl{\raise.15ex\hbox{/}\kern-.57em\partial}

 \def\Tr{{\rm Tr}}
\def\lspace{\ifx\answ\bigans{}\else\qquad\fi}
\def\lbspace{\ifx\answ\bigans{}\else\hskip-.2in\fi} 
\def\boxeqn#1{\vcenter{\vbox{\hrule\hbox{\vrule\kern3pt\vbox{\kern3pt
	\hbox{${\displaystyle #1}$}\kern3pt}\kern3pt\vrule}\hrule}}}
\def\mbox#1#2{\vcenter{\hrule \hbox{\vrule height#2in
		\kern#1in \vrule} \hrule}}  
%
   
   \def\CU{{\cal U}}

\def\om#1#2{\omega^{#1}{}_{#2}}

\def\darr#1{\raise1.5ex\hbox{$\leftrightarrow$}\mkern-16.5mu #1}

\def\half{{\textstyle{1\over2}}} 
\def\roughly#1{\raise.3ex\hbox{$#1$\kern-.75em\lower1ex\hbox{$\sim$}}}
  
\input epsf
\noblackbox

\ifx\answ\bigans
\magnification=1200\baselineskip=14pt plus 2pt minus 1pt
\else\baselineskip=16pt 
\fi

\def\ie{{\it i.e.\ }}
\def\cf{{\it c.f.\ }}
\def\frac#1#2{{#1 \over #2}}
\def\al{{\alpha}}

\def\cf{{\it cf.\ }}
\def\ie{{\it i.e.\ }}
\def\eg{{\it e.g.\ }}
\def\eqq{{\it Eq.\ }}
\def\eqqs{{\it Eqs.\ }}

\def\ap{\alpha'}

\def\tb{type $IIB$\ }\def\ta{type $IIA$\ }
\def\ap{\alpha'}

\def\cf{{\it cf.\ }}
\def\ie{{\it i.e.\ }}
\def\eg{{\it e.g.\ }}
\def\eqq{{\it Eq.\ }}
\def\eqqs{{\it Eqs.\ }}
\def\th{\theta}

\def\al{\alpha}

\def\Om{\Omega}
\def\om{\omega}

\def\Om{\Omega}

\def\K{{\cal K}}
\def\CS{{\cal C}{\cal S}}

\newcount\figno
\figno=0
\def\fig#1#2#3{
\par\begingroup\parindent=0pt\leftskip=1cm\rightskip=1cm\parindent=0pt
\baselineskip=11pt \global\advance\figno by 1 \midinsert
\epsfxsize=#3 \centerline{\epsfbox{#2}} \vskip 12pt
\centerline{{\bf Figure \the\figno :}{\it ~~ #1}}\par
\endinsert\endgroup\par}
\def\figlabel#1{\xdef#1{\the\figno}}

\input epsf
\input psfig
\def\figin{\epsfcheck\figin}\def\figins{\epsfcheck\figins}
\def\epsfcheck{\ifx\epsfbox\UnDeFiNeD
\message{(NO epsf.tex, FIGURES WILL BE IGNORED)}
\gdef\figin##1{\vskip2in}\gdef\figins##1{\hskip.5in}
\else\message{(FIGURES WILL BE INCLUDED)}%
\gdef\figin##1{##1}\gdef\figins##1{##1}\fi}
\def\DefWarn#1{}
\def\figinsert{\goodbreak\midinsert}
\def\ifig#1#2#3{\DefWarn#1\xdef#1{fig.~\the\figno}
\writedef{#1\leftbracket fig.\noexpand~\the\figno}%
\figinsert\figin{\centerline{#3}}\medskip\centerline{\vbox{\baselineskip12pt
\advance\hsize by -1truein\noindent\footnotefont{\bf Fig.~\the\figno } #2}}
\bigskip\endinsert\global\advance\figno by1}


\def\appA{A}
\def\appB{B}\def\appBvii{B.7.}
\def\appC{C}
\def\appD{D}
\def\appE{E}

\def\tilde{\widetilde}

\def\h {{1\over 2}}

\def\ov {\overline}
\def\o {\over}
\def\fc#1#2{{#1 \o #2}}

\def\IZ{ {\bf Z}}
\def\IC{{\bf C}}
\def\IR{ {\bf R}}
\def\hat{\widehat}

\def\br{\hfill\break}

\def\det {{\rm det}}

\def\lf {\left}
\def\ri {\right}
\def\ra {\rightarrow}
\def\lra {\longrightarrow}
\def\re {{\rm Re}}
\def\im {{\rm Im}}
\def\p {\partial}

 \def\Sc {{\cal S}}
\def\Mc {{\cal M}}

\def\Kc {{\cal K}}
\def\Tc{{\cal T}}
\def\Uc{{\cal U}}
\newif\ifnref
\def\rrr#1#2{\relax\ifnref\nref#1{#2}\else\ref#1{#2}\fi}
\def\ldf#1#2{\begingroup\obeylines
\gdef#1{\rrr{#1}{#2}}\endgroup\unskip}
\def\nrf#1{\nreftrue{#1}\nreffalse}

\def\multrefvi#1#2#3#4#5#6{\nrf{#1#2#3#4#5#6}\refs{#1{--}#6}}

\def\multrefiv#1#2#3#4{\nrf{#1#2#3#4}\refs{#1{--}#4}}
\def\multrefvii#1#2#3#4#5#6#7{\nrf{#1#2#3#4#5#6}\refs{#1{--}#7}}
\def\doubref#1#2{\refs{{#1},{#2} }}
\def\threeref#1#2#3{\refs{{#1},{#2},{#3} }}
\def\fourref#1#2#3#4{\refs{{#1}{--}{#4} }}

\def\sevenref#1#2#3#4#5#6#7{\refs{{#1},{#2},{#3},{#4},{#5},{#6},{#7} }}
\nreffalse

\def\lref{\ldf}

\lref\zwart{G.~Zwart,
``Four-dimensional N = 1 Z(N) x Z(M) orientifolds,''
  Nucl.\ Phys.\ B {\bf 526}, 378 (1998)
  [arXiv:hep-th/9708040].
}

\lref\seealso{
 I.~Antoniadis, C.~Bachas, C.~Fabre, H.~Partouche and T.R.~Taylor,
 ``Aspects of type I - type II - heterotic triality in four dimensions,''
  Nucl.\ Phys.\ B {\bf 489}, 160 (1997)
  [arXiv:hep-th/9608012];\br
T.W.~Grimm and J.~Louis,
  ``The effective action of N = 1 Calabi-Yau orientifolds,''
  Nucl.\ Phys.\ B {\bf 699}, 387 (2004)
  [arXiv:hep-th/0403067];\br
H.~Jockers and J.~Louis,
``The effective action of D7-branes in N = 1 Calabi-Yau orientifolds,''
  Nucl.\ Phys.\ B {\bf 705}, 167 (2005)
  [arXiv:hep-th/0409098];
``D-terms and F-terms from D7-brane fluxes,''
  Nucl.\ Phys.\ B {\bf 718}, 203 (2005)
  [arXiv:hep-th/0502059].
}

\lref\review{
E.~Kiritsis,
``D-branes in standard model building, gravity and cosmology,''
Fortsch.\ Phys.\  {\bf 52}, 200 (2004)
[arXiv:hep-th/0310001];\br
D.~L\"ust,
``Intersecting brane worlds: A path to the standard model?,''
Class.\ Quant.\ Grav.\  {\bf 21}, S1399 (2004)
[arXiv:hep-th/0401156];\br
R.~Blumenhagen, M.~Cvetic, P.~Langacker and G.~Shiu,
``Toward realistic intersecting D-brane models,''
arXiv:hep-th/0502005.
}

\lref\BerglundDM{
P.~Berglund and P.~Mayr,
``Non-Perturbative Superpotentials in F-theory and String Duality,''
arXiv:hep-th/0504058.
}

\lref\IMNQ{
  L.E.~Ibanez, J.~Mas, H.P.~Nilles and F.~Quevedo,
``Heterotic Strings In Symmetric And Asymmetric Orbifold Backgrounds,''
  Nucl.\ Phys.\ B {\bf 301}, 157 (1988).
}

\lref\IntriligatorAU{
K.A.~Intriligator and N.~Seiberg,
``Lectures on supersymmetric gauge theories and electric-magnetic  duality,''
Nucl.\ Phys.\ Proc.\ Suppl.\  {\bf 45BC}, 1 (1996)
[arXiv:hep-th/9509066].
}

\lref\DeWolfeUU{
O.~DeWolfe, A.~Giryavets, S.~Kachru and W.~Taylor,
``Type IIA moduli stabilization,''
arXiv:hep-th/0505160.
}

\lref\KachruJR{
S.~Kachru and A.K.~Kashani-Poor,
``Moduli potentials in type IIA compactifications with RR and NS flux,''
JHEP {\bf 0503}, 066 (2005)
[arXiv:hep-th/0411279].
}

\lref\DouglasUM{
  M.R.~Douglas,
``The statistics of string / M theory vacua,''
  JHEP {\bf 0305}, 046 (2003)
  [arXiv:hep-th/0303194].
}

\lref\CurioSC{
G.~Curio, A.~Klemm, D.~L\"ust and S.~Theisen,
``On the vacuum structure of type II string compactifications on  Calabi-Yau
spaces with H-fluxes,''
Nucl.\ Phys.\ B {\bf 609}, 3 (2001)
[arXiv:hep-th/0012213].
}

\lref\DerendingerPH{
J.P.~Derendinger, C.~Kounnas, P.M.~Petropoulos and F.~Zwirner,
``Fluxes and gaugings: N = 1 effective superpotentials,''
arXiv:hep-th/0503229.
}

\lref\VilladoroCU{
G.~Villadoro and F.~Zwirner,
``N = 1 effective potential from dual type-IIA D6/O6 orientifolds with general
fluxes,''
arXiv:hep-th/0503169.
}

\lref\CurioEW{
G.~Curio, A.~Krause and D.~L\"ust,
``Moduli stabilization in the heterotic / IIB discretuum,''
arXiv:hep-th/0502168.
}

\lref\DerendingerJN{
J.P.~Derendinger, C.~Kounnas, P.M.~Petropoulos and F.~Zwirner,
``Superpotentials in IIA compactifications with general fluxes,''
Nucl.\ Phys.\ B {\bf 715}, 211 (2005)
[arXiv:hep-th/0411276].
}

\lref\CamaraDC{
P.G.~Camara, A.~Font and L.E.~Ibanez,
``Fluxes, moduli fixing and MSSM-like vacua in a simple IIA orientifold,''
arXiv:hep-th/0506066.
}

\lref\CIU{
P.G.~Camara, L.E.~Ibanez and A.M.~Uranga,
``Flux-induced SUSY-breaking soft terms,''
Nucl.\ Phys.\ B {\bf 689}, 195 (2004)
[arXiv:hep-th/0311241].
}

\lref\GGJL{
M.~Grana, T.W.~Grimm, H.~Jockers and J.~Louis,
``Soft supersymmetry breaking in Calabi-Yau orientifolds with D-branes and
fluxes,''
Nucl.\ Phys.\ B {\bf 690}, 21 (2004)
[arXiv:hep-th/0312232].
}

\lref\DenefMM{
F.~Denef, M.R.~Douglas, B.~Florea, A.~Grassi and S.~Kachru,
``Fixing All Moduli in a Simple F-Theory Compactification,''
arXiv:hep-th/0503124.
}

\lref\KalloshGS{
R.~Kallosh, A.K.~Kashani-Poor and A.~Tomasiello,
``Counting fermionic zero modes on M5 with fluxes,''
arXiv:hep-th/0503138.
}

\lref\SaulinaVE{
N.~Saulina,
``Topological constraints on stabilized flux vacua,''
arXiv:hep-th/0503125.
}

\lref\BerglundDM{
P.~Berglund and P.~Mayr,
``Non-perturbative superpotentials in F-theory and string duality,''
arXiv:hep-th/0504058.
}

\lref\KachruAW{
S.~Kachru, R.~Kallosh, A.~Linde and S.P.~Trivedi,
``De Sitter vacua in string theory,''
Phys.\ Rev.\ D {\bf 68}, 046005 (2003)
[arXiv:hep-th/0301240].
}

\lref\KachruSX{
S.~Kachru, R.~Kallosh, A.~Linde, J.~Maldacena, L.~McAllister and S.P.~Trivedi,
``Towards inflation in string theory,''
JCAP {\bf 0310}, 013 (2003)
[arXiv:hep-th/0308055].
}

\lref\TripathyHV{
P.K.~Tripathy and S.P.~Trivedi,
``D3 Brane Action and Fermion Zero Modes in Presence of Background Flux,''
arXiv:hep-th/0503072.
}

\lref\WittenBN{
E.~Witten,
``Non-Perturbative Superpotentials In String Theory,''
Nucl.\ Phys.\ B {\bf 474}, 343 (1996)
[arXiv:hep-th/9604030].
}

\lref\FKP{
  S.~Ferrara, C.~Kounnas and M.~Porrati,
  ``General Dimensional Reduction Of Ten-Dimensional Supergravity And
  Superstring,''
  Phys.\ Lett.\ B {\bf 181}, 263 (1986).
}

\lref\FT{
  S.~Ferrara and S.~Theisen,
``Moduli Spaces, Effective Actions And Duality Symmetry In String
Compactifications,''
CERN-TH-5652-90
{\it Based on lectures given at 3rd Hellenic Summer School, Corfu, Greece, Sep 13-23, 1989}
}

\lref\Ruben{
  M.~Marino, R.~Minasian, G.W.~Moore and A.~Strominger,
``Nonlinear instantons from supersymmetric p-branes,''
  JHEP {\bf 0001}, 005 (2000)
  [arXiv:hep-th/9911206].
}

\lref\FontCX{
A.~Font and L.E.~Ibanez,
``SUSY-breaking soft terms in a MSSM magnetized D7-brane model,''
JHEP {\bf 0503}, 040 (2005)
[arXiv:hep-th/0412150].
}

\lref\Denef{F.~Denef and M.R.~Douglas,
 ``Distributions of flux vacua,''
  JHEP {\bf 0405}, 072 (2004)
  [arXiv:hep-th/0404116];\br
B.S.~Acharya, F.~Denef and R.~Valandro,
  ``Statistics of M theory vacua,''
  arXiv:hep-th/0502060.
}

\lref\DenefDM{
F.~Denef, M.R.~Douglas and B.~Florea,
``Building a better racetrack,''
JHEP {\bf 0406}, 034 (2004)
[arXiv:hep-th/0404257].
}

\lref\RobbinsHX{
D.~Robbins and S.~Sethi,
``A barren landscape,''
Phys.\ Rev.\ D {\bf 71}, 046008 (2005)
[arXiv:hep-th/0405011].
}

\lref\LustBD{D.~L\"ust, P.~Mayr, S.~Reffert and S.~Stieberger,
   ``F-theory flux, destabilization of orientifolds and soft terms on
D7-branes,''
  Nucl.\ Phys.\ B {\bf 732}, 243 (2006)
  [arXiv:hep-th/0501139].
}

\lref\LustFI{D.~L\"ust, S.~Reffert and S.~Stieberger,
  ``Flux-induced soft supersymmetry breaking in chiral type IIB orientifolds
  with D3/D7-branes,''
  Nucl.\ Phys.\ B {\bf 706}, 3 (2005)
  [arXiv:hep-th/0406092].
}

\lref\LustDN{D.~L\"ust, S.~Reffert and S.~Stieberger,
 ``MSSM with soft SUSY breaking terms from D7-branes with fluxes,''
  Nucl.\ Phys.\ B {\bf 727}, 264 (2005)
  [arXiv:hep-th/0410074].
}

\lref\ASP{P.S.~Aspinwall and R.~Kallosh,
  ``Fixing all moduli for M-theory on K3 x K3,''
  arXiv:hep-th/0506014.
}

\lref\ChoiSX{
K.~Choi, A.~Falkowski, H.P.~Nilles, M.~Olechowski and S.~Pokorski,
``Stability of flux compactifications and the pattern of supersymmetry breaking,''
  JHEP {\bf 0411}, 076 (2004)
  [arXiv:hep-th/0411066].
}

\lref\HPN{K.~Choi, A.~Falkowski, H.P.~Nilles and M.~Olechowski,
``Soft supersymmetry breaking in KKLT flux compactification,''
  Nucl.\ Phys.\ B {\bf 718}, 113 (2005)
  [arXiv:hep-th/0503216].
}

\lref\FontCY{
A.~Font,
``Z(N) orientifolds with flux,''
JHEP {\bf 0411}, 077 (2004)
arXiv:hep-th/0410206.
}

\lref\AFIV{G.~Aldazabal, A.~Font, L.E.~Ibanez and G.~Violero,
``D = 4, N = 1, type IIB orientifolds,''
Nucl.\ Phys.\ B {\bf 536}, 29 (1998)
[arXiv:hep-th/9804026].
}

\lref\FIQ{A.~Font, L.E.~Ibanez and F.~Quevedo,
  ``Z(N) X Z(M) Orbifolds And Discrete Torsion,''
  Phys.\ Lett.\ B {\bf 217}, 272 (1989).
}

\lref\LMRS{
D. L\"ust, P.~Mayr, R.~Richter and S.~Stieberger,
``Scattering of gauge, matter, and moduli fields from intersecting branes,''
Nucl.\ Phys.\ B {\bf 696}, 205 (2004)
[arXiv:hep-th/0404134].
}

\lref\BCS{R.~Blumenhagen, J.P.~Conlon and K.~Suruliz,
``Type IIA orientifolds on general supersymmetric Z(N) orbifolds,''
JHEP {\bf 0407}, 022 (2004)
[arXiv:hep-th/0404254].
}

\lref\GorlichQM{
L.~G\"orlich, S.~Kachru, P.K.~Tripathy and S.P.~Trivedi,
 ``Gaugino condensation and nonperturbative superpotentials in flux
arXiv:hep-th/0407130.
}

\lref\MS{F.~Marchesano and G.~Shiu,
  ``Building MSSM flux vacua,''
  JHEP {\bf 0411}, 041 (2004)
  [arXiv:hep-th/0409132].}

\lref\DHVW{L.J.~Dixon, J.A.~Harvey, C.~Vafa and E.~Witten,
``Strings On Orbifolds,''
Nucl.\ Phys.\ B {\bf 261}, 678 (1985);
``Strings On Orbifolds. 2,''
Nucl.\ Phys.\ B {\bf 274}, 285 (1986).
}

\lref\BlumenhagenXX{
  R.~Blumenhagen, F.~Gmeiner, G.~Honecker, D.~Lust and T.~Weigand,
 ``The statistics of supersymmetric D-brane models,''
  Nucl.\ Phys.\ B {\bf 713}, 83 (2005)
  [arXiv:hep-th/0411173].
}

\lref\structure{J.A.~Casas, F.~Gomez and C.~Munoz,
``Complete structure of Z(n) Yukawa couplings,''
Int.\ J.\ Mod.\ Phys.\ A {\bf 8}, 455 (1993)
[arXiv:hep-th/9110060].
}

\lref\BailinNK{
D.~Bailin and A.~Love,
``Orbifold compactifications of string theory,''
Phys.\ Rept.\  {\bf 315}, 285 (1999).
}

\lref\BreitenlohnerJF{
P.~Breitenlohner and D.Z.~Freedman,
``Stability In Gauged Extended Supergravity,''
Annals Phys.\  {\bf 144}, 249 (1982).
}

\lref\GRIMM{T.W.~Grimm and J.~Louis,
  ``The effective action of type IIA Calabi-Yau orientifolds,''
  Nucl.\ Phys.\ B {\bf 718}, 153 (2005)
  [arXiv:hep-th/0412277].
}

\lref\future{
  D.~L\"ust, S.~Reffert, E.~Scheidegger, W.~Schulgin and S.~Stieberger,
``Moduli stabilization in type IIB orientifolds. II,''
  arXiv:hep-th/0609013.
}

\lref\GKP{S.B.~Giddings, S.~Kachru and J.~Polchinski,
``Hierarchies from fluxes in string compactifications,''
Phys.\ Rev.\ D {\bf 66}, 106006 (2002)
[arXiv:hep-th/0105097].
}

\lref\CIUii{P.G.~Camara, L.E.~Ibanez and A.M.~Uranga,
  ``Flux-induced SUSY-breaking soft terms on D7-D3 brane systems,''
  Nucl.\ Phys.\ B {\bf 708}, 268 (2005)
  [arXiv:hep-th/0408036].
}

\lref\CU{J.F.G.~Cascales and A.M.~Uranga,
``Chiral 4d N = 1 string vacua with D-branes and NSNS and RR fluxes,''
JHEP {\bf 0305}, 011 (2003)
[arXiv:hep-th/0303024].
}

\lref\FP{A.R.~Frey and J.~Polchinski,
``N = 3 warped compactifications,''
Phys.\ Rev.\ D {\bf 65}, 126009 (2002)
[arXiv:hep-th/0201029].
}

\lref\KST{S.~Kachru, M.B.~Schulz and S.~Trivedi,
``Moduli stabilization from fluxes in a simple IIB orientifold,''
JHEP {\bf 0310}, 007 (2003)
[arXiv:hep-th/0201028].
}

\lref\BLT{R.~Blumenhagen, D.~L\"ust and T.R.~Taylor,
``Moduli stabilization in chiral type IIB orientifold models with fluxes,''
Nucl.\ Phys.\ B {\bf 663}, 319 (2003)
[arXiv:hep-th/0303016].
}

\lref\Klemm{L.E.~Ibanez, J.~Mas, H.P.~Nilles and F.~Quevedo,
  ``Heterotic Strings In Symmetric And Asymmetric Orbifold Backgrounds,''
  Nucl.\ Phys.\ B {\bf 301}, 157 (1988);\br
J.~Erler and A.~Klemm,
``Comment on the generation number in orbifold compactifications,''
Commun.\ Math.\ Phys.\  {\bf 153}, 579 (1993)
[arXiv:hep-th/9207111].
}

\lref\Slansky{
R.~Slansky,
``Group Theory For Unified Model Building,''
Phys.\ Rept.\  {\bf 79}, 1 (1981).
}

\lref\FontCY{
A.~Font,
``Z(N) orientifolds with flux,''
JHEP {\bf 0411}, 077 (2004)
[arXiv:hep-th/0410206].
}

\lref\mspal{M.~Spalinski,
 ``Duality transformations in twisted Narain compactifications,''
  Nucl.\ Phys.\ B {\bf 377}, 339 (1992).
``On the discrete symmetry group of Narian orbifolds,''
  Phys.\ Lett.\ B {\bf 275}, 47 (1992).
}

\lref\CveticYW{
  M.~Cvetic, J.~Louis and B.A.~Ovrut,
  ``A String Calculation Of The Kahler Potentials For Moduli Of Z(N)
  Orbifolds,''
  Phys.\ Lett.\ B {\bf 206}, 227 (1988).
}

\lref\IbanezHC{
  L.E.~Ibanez and D.~L\"ust,
  ``Duality anomaly cancellation, minimal string unification and the effective
  low-energy Lagrangian of 4-D strings,''
  Nucl.\ Phys.\ B {\bf 382}, 305 (1992)
  [arXiv:hep-th/9202046].
}

\lref\TV{S.~Gukov, C.~Vafa and E.~Witten,
  ``CFT's from Calabi-Yau four-folds,''
  Nucl.\ Phys.\ B {\bf 584}, 69 (2000)
  [Erratum-ibid.\ B {\bf 608}, 477 (2001)]
  [arXiv:hep-th/9906070];\br
T.R.~Taylor and C.~Vafa,
``RR flux on Calabi-Yau and partial supersymmetry breaking,''
Phys.\ Lett.\ B {\bf 474}, 130 (2000)
[arXiv:hep-th/9912152];\br
P.~Mayr,
``On supersymmetry breaking in string theory and its realization in brane
worlds,''
Nucl.\ Phys.\ B {\bf 593}, 99 (2001)
[arXiv:hep-th/0003198].
}

\lref\WessCP{
  J.~Wess and J.~Bagger,
  ``Supersymmetry and supergravity,''
Princeton, USA: Univ. Pr. (1992) 259 p.
}


\Title{\vbox{\rightline{LMU--ASC 45/05} \rightline{MPP--2005--59}
\rightline{\tt hep-th/0506090}}}
{\vbox{\centerline{Moduli Stabilization in Type $IIB$ Orientifolds (I)}
}}
\smallskip
\centerline{D. L\"ust$^{a,b}$,\ \ S. Reffert$^b$,\ \ W. Schulgin$^b$\ \ and\ \ 
S. Stieberger$^a$}
\bigskip
\centerline{\it $^a$ Arnold--Sommerfeld--Center for Theoretical Physics,}
\centerline{\it  Department f\"ur Physik, 
Ludwig--Maximilians--Universit\"at M\"unchen,}
\centerline{\it Theresienstra\ss e 37, 80333 M\"unchen, Germany}
\vskip7pt
\centerline{\it $^b$ Max--Planck--Institut f\"ur Physik,}
\centerline{\it F\"ohringer Ring 6, 80805 M\"unchen, Germany}

\bigskip\bigskip
\centerline{\bf Abstract}
\vskip .2in
\noindent
We discuss flux quantization and moduli stabilization in toroidal 
type $IIB$ $\IZ_N$-- or $\IZ_N\times \IZ_M$--orientifolds, 
focusing mainly on their orbifold limits.
After presenting a detailed discussion of their moduli spaces and effective actions,
we study the supersymmetric vacuum structure of these models and derive 
criteria for the existence of stable minima.
Furthermore, we briefly investigate the models away from their orbifold points and comment
on the microscopic origin of their non--perturbative superpotentials.

\Date{}
\noindent

\goodbreak

\listtoc 
\writetoc
\break

\newsec{Introduction}

Type $II$--orientifold models are phenomenologically 
interesting for a number of reasons. First, they provide 
a viable framework to derive models that exhibit many features
of the standard model of particle physics, such as non-Abelian
gauge bosons, chiral fermions and family repetition (for some reviews see \review).
The present  paper will deal with type $IIB$ orientifold 
compactifications\foot{For some recent work on type $IIA$
orientifolds with fluxes and fixed moduli see
\multrefvi\KachruJR\DerendingerJN\GRIMM\VilladoroCU\DeWolfeUU\CamaraDC.}. 
In this context, the standard model
fields are living on the world--volume of $D3$-branes
and/or $D7$--branes which may intersect
each other in the internal dimensions.
Second, by adding closed string 3--form fluxes via
an effective superpotential $W_{\rm flux}$ \TV, many or even all of the geometric complex
structure moduli plus the
dilaton can be frozen \threeref\CurioSC\GKP\KST.
At the same time,
space-time supersymmetry may be spontaneously broken. 
Since the $3$--form fluxes 
arise already at string tree-level, it is possible to compute
the pattern of the soft supersymmetry breaking parameters in the
matter sector on the $D$-branes \multrefvii\CIU\GGJL\LustFI\CIUii\LustDN\FontCX\LustBD.
Last but not least, the proposal of KKLT \KachruAW\
showed a promising possibility to freeze besides
the complex structure moduli and the dilaton also all K\"ahler
moduli through the effects of a non-perturbative superpotential $W_{\rm np}$.
This superpotential can be induced by Euclidean $D3$-instantons
and/or by gaugino condensation in some hidden gauge group sector.
The scenario of KKLT has also far-reaching consequences in cosmology,
since by lifting the potential by some other effects, \eg by adding 
anti-branes, a small positive cosmological constant can be generated.
Also string backgrounds with inflation in the early phase of the
universe may have a similar origin \KachruSX .
As pointed out in {\it Ref.} \DouglasUM, there exists a huge number of string ground states,
called the string landscape. An observation, which motivates a statistical survey of all
possible string vacua with fluxes \Denef\ and also with $D$--branes \BlumenhagenXX.

Certainly being a very attractive scheme, there
are still many open issues which should be addressed
in the context of the KKLT proposal.
The main question basically is, whether all moduli can be really stabilized
in concrete and perhaps even semi--realistic orientifold
models. Specifically, a semi--realistic model generically
contains the following  
ingredients: after choosing a geometrical background space, in type $IIB$ a
(warped) compactification on $M_4\times X_6$,
with $X_6$ being an orbifold or a Calabi-Yau space, an orientifold action
is defined which reduces space-time supersymmetry from
N=2 down to N=1 in four space-time dimensions.
Orientifolds imply orientifold planes ($O3/O7$--planes) with negative $D$--brane charges, which 
require the presence
of $D3/D7$-branes and/or $3$--form fluxes, such that all tadpole cancelation
conditions are satisfied, and N=1 supersymmetry is preserved also in the
open string sector of the theory. The next and also very important
question for KKLT is whether Euclidean $D3$-instantons and/or 
gaugino condensation within the effective Yang-Mills theory on the
$D7$-branes can generate a non-vanishing non-perturbative
superpotential $W_{\rm np}$. The answer 
to this question is contained in the topology of the 4-cycles 
(divisors) inside $X_6$,
around which the $D3/D7$-branes  are wrapped. More specifically, 
for the Euclidean $D3$-brane instantons, the number of fermionic zero modes
living on the world volume of a $D3$-brane
 must be equal to two. For the wrapped $D7$-branes,
the dynamics
of open string non-Abelian gauge theory on a stack of $D7$-branes
must allow for a non-vanishing gaugino condensate.
Whether all sufficient requirements for $W_{\rm np}$
are indeed satisfied  depends however not only on the topology and geometry of
the divisors inside $X_6$, but also on the chosen 3--form flux, which affects
the number of fermionic zero modes on the $D3$-branes as well as
the open string spectrum on the $D7$-branes 
\multrefvii\DenefDM\GorlichQM\TripathyHV\KalloshGS\SaulinaVE\BerglundDM\ASP.  

Therefore the tree--level 
superpotential $W_{\rm flux}$ and the non-perturbative superpotential
$W_{\rm np}$ are  not independent from each other but are rather
intimately related.
This observation really tells us that it is not useful
to disentangle the vacuum structure due to the 3--form fluxes from
the non-perturbative effects and vice versa.

In this paper, we will not discuss in very great detail the microscopic
origin of the non-perturbative superpotential  $W_{\rm np}$,
but leave this for a future publication \future.
(Recently a concrete type $IIB$ orientifold with all moduli fixed was constructed
in \DenefMM.) 
Only a few remarks at the end of the paper will address this question.
Our main emphasis in this work is the investigation of
the moduli-spaces and the vacuum structure of type $IIB$ orientifold
compactifications in their various toroidal orbifold limits. Hence we will simply assume
the existence of a non-perturbative superpotential $W_{\rm np}$,
which depends only on the untwisted K\"ahler moduli $T^i$.
The effects of blowing up the orbifold or the presence of blowing up
K\"ahler moduli will be neglected, respectively postponed for 
future work \future\ (except for
a short discussion at the end of the paper). So, $W_{\rm np}(T^i)$ 
can be viewed as being the truncation of a more complete superpotential
that contains all K\"ahler moduli.
Nevertheless, several interesting questions can be addressed within the
orbifold framework.

The first issue which we will discuss is the
careful parameterization of the moduli spaces of type $IIB$ orientifolds
in the various orbifold limits.
The proper N=1 supergravity definition of the closed
string K\"ahler moduli $T^i$ and complex structure  
moduli $U^j$ 
in terms of the background metrics $g_{ij}$ 
and Ramond $4$--form $C_4$ depends of course on the considered orbifold
twists, but also on the underlying 6-dimensional group lattices.
Moreover, the definition of the K\"ahler moduli 
of type $IIB$  orientifolds 
is different from the one in heterotic compactifications.
We also determine the associated tree-level effective action, and
the possible 3-form fluxes in type $IIB$ orbifolds.

Second, in KKLT one assumes that the complex structure 
moduli are fixed by $W_{\rm flux}$ alone and then are integrated
out assuming that they are heavy. In particular, the assumption
is made that the flux vacua are still given through 3--form fluxes which are
still imaginary self dual ($ISD$) and are of the Hodge types $G_{(2,1)}$ and
$G_{(0,3)}$.
We will see however
that 
the inclusion of the additional non-perturbative effects in the superpotential
besides the 3--form fluxes has the effect of generic
supersymmetric AdS ground-states being described by
fluxes which are not anymore $ISD$ with only
$G_{(2,1)}$ components, but will rather include 
also all $IASD$ (imaginary
anti self-dual) types as well (see also the discussion in \CurioEW).

The third problem is related to the stability of the obtained
supersymmetric vacua. Although {\sl stable}
 AdS vacua generically allow for scalar
fields with negative (mass)$^2$, provided the masses still fulfill
the Breitenlohner-Freedman bound \BreitenlohnerJF, 
the KKLT framework only works
if all (mass)$^2$ eigenvalues of the fixed scalar fields are already
positive in the AdS ground state. The reason for this stronger
requirement is that otherwise, the uplift to a dS vacuum by adding a positive
constant to the scalar potential would not work, i.e. would not lead
to a stable dS ground state. However in concrete
orientifold models, this stability criterion is
far from being automatically satisfied, as already observed in \ChoiSX.
We will discuss in which orbifold compactifications there is a chance
to obtain stable AdS ground states with positive scalar (mass)$^2$.

\vskip0.6cm

This paper is organized as follows. Section 2
deals with the complex structures and 
K\"ahler moduli of the different $\IZ_N$ and $\IZ_N\times\IZ_M$--orbifolds. 
Here, only orbifolds are treated 
which lead to models that allow tadpole cancellation. Without introducing
torsion or vector structure, this is possible for the following orbifolds:
$\IZ_3,\,\IZ_{6-I},\,\IZ_{6-II}$, $\IZ_{12-I}$, $\IZ_2\times \IZ_2$, $\IZ_3\times 
\IZ_3$, $\IZ_2\times \IZ_3$, $\IZ_2\times \IZ_6$, $\IZ_2\times \IZ_{6'}$, $\IZ_3\times \IZ_6$ 
and $\IZ_6\times \IZ_6$.
As for many of the orbifold twists, several 
torus lattices are possible and the 
complex structure depends on the lattice, the complex structures and K\"ahler 
moduli of different lattices are worked out.
We also discuss the closed string 
low energy effective action,
namely the K\"ahler potential of the moduli fields.
Then,
we determine 3-form fluxes 
and the 3-form flux induced tree level superpotential.
In particular, 
we give a general 
introduction to 3--form fluxes and a 
list of all invariant flux 
components under all possible orbifold groups without discrete torsion.
As a result, it turns out that only the 
$IASD$--flux $G_{(3,0)}$ and the $ISD$ flux $G_{(0,3)}$
are generic for all orbifolds. 

Section 3 will deal with the 
discussion of the (discrete) vacuum structure of the considered
orientifold compactifications. 
Due to the presence of the 3-form fluxes as well as the
non-perturbative superpotential, 
all complex structure and K\"ahler moduli will
be generically fixed at discrete values. 
We will see that due to the K\"ahler moduli
dependent non-perturbative superpotential, 
the supersymmetry conditions on 3-form fluxes are
changed compared to the pure flux case. 
Analyzing the stability properties of the associated scalar
potential we will see that stable AdS ground states are
not possible if the orbifold
does not contain complex structure
moduli regardless of the number of untwisted K\"ahler
moduli present in $W_{\rm np}$.
One needs at least one untwisted complex structure modulus $U$
for stability. We also comment on the integrating out procedure
of the complex structure moduli, which can lead to inconsistent
results, as it is the case for orbifolds with just one
complex structure modulus. Furthermore, we briefly discuss open string moduli
and soft--supersymmetry breaking terms in the presence of $W_{\rm np}$. We show that
\eg the ratio of gaugino masses after inclusion of $W_{\rm np}$ is given 
by the same quotient as without $W_{\rm np}$ in a certain region in the parameter/moduli 
space.

Section 4 contains a preliminary discussion on the
microscopic origin of $W_{\rm np}$.
We will discuss a set of conditions for a non-vanishing
$W_{\rm np}$, like that for the case of gaugino condensation
in the hidden gauge sectors on stacks of $D7$-branes, the associated $D7$-brane 
divisors inside $X_6$ must not intersect each other.
These conditions are most naturally met if the orbifold
singularities are resolved.
These resolved orbifolds, their vacuum structures and other related
issues will then be
the topic of the forthcoming publication \future.

Finally the two appendices A and B contain the technical details
about the moduli spaces of the orientifolds, the definitions
of the complex structure, the geometrical lattice structures which underlie
the various orbifolds, and the possible flux solutions for
the ${\bf Z}_N$ and ${\bf Z}_N\times {\bf Z}_M$-orbifolds.
Appendix C gives the Cartan matrices of the 
relevant lattice Lie groups, and for completeness appendix D shows 
the definitions of the K\"ahler moduli on the heterotic side.

\newsec{Moduli space, tree-level effective action and 3-form fluxes in \tb orientifolds}

\subsec{Type $IIB$ orientifolds of toroidal orbifolds}

We concentrate on orientifolds of \tb compactified on the toroidal orbifolds
\eqn\orbi{
X_6=\fc{T^6}{\IZ_N}\ \ \ ,\ \ \ X_6=\fc{T^6}{\IZ_N\times \IZ_M}\ ,}
with the orbifold groups $\Gamma=\IZ_N$ and $\Gamma=\IZ_N\times \IZ_M$. 
To define the orbifold compactification $X_6$, we must specify the six--torus $T^6$ and the
discrete point group $\Gamma$.  We will restrict ourselves to orbifolds with
Abelian point group without discrete torsion. The point group element $\theta$
can then be written as 
$\theta=\exp[2\pi i(v^1M^{12}+v^2M^{34}+v^3M^{56})],$ where the $M^{ij}$ are
the generators of the Cartan sub-algebra and $0\leq|v^i|<1,\ i=1,2,3$.
To obtain N=2 supersymmetry, the point group $\Gamma$
must be a subgroup of $SU(3)$. This gives us 
$\pm v^1\pm v^2\pm v^3=0.$
This condition together with the requirement that $\Gamma$ must act
crystallographically on the lattice specified by $T^6$ leads to $\Gamma$ being
either $\IZ_N$ with $N=3,4,6,7,8,12\,$ or $\IZ_M\times \IZ_N$ with $N$ a multiple of
$M$ and $N=2,3,4$. $\IZ_6,\ \IZ_8$ and $\IZ_{12}$ have two inequivalent embeddings
in $SO(6)$.
We will use the standard embeddings, as given \eg  in \DHVW.

To obtain an N=1 (closed) string spectrum, one introduces an 
orientifold projection $\Om I_n$, with $\Om$ describing a reversal of the 
orientation of the closed string world--sheet and $I_n$ a reflection of $n$ internal
coordinates. For $\Om I_n$ to represent a symmetry of the original theory, $n$
has to be an even integer in \tb.
Generically, this projection produces orientifold fixed planes [$O(9-n)$--planes],
placed at the orbifold fixed points of $T^6/I_n$. They have negative tension, which 
has to be balanced by introducing positive tension objects.
Candidates for the latter may be collections of $D(9-n)$--branes and/or non--vanishing
three--form fluxes $H_3$ and $C_3$.
The orbifold group $\Gamma$ mixes with the orientifold group $\Om I_n$.
As a result, if the group $\Gamma$ contains $\IZ_2$--elements $\th$, 
which leave one complex plane fixed, we obtain additional $O(9-|n-4|)$-- or 
$O(3+|n-2|)$--planes from the element $\Om I_n \th$. 

In the following, only the two cases of $n=6$ ($O3$--plane) and $n=2$
($O7$--planes) will be relevant
to us. Without introducing torsion or vector structure, tadpoles may be completely 
cancelled by adding $D3/D7$--branes, 
provided the orbifold twist $\Gamma$ is $\IZ_3,\,\IZ_{6-I},\,\IZ_{6-II},\,\IZ_7$ or $\IZ_{12-I}$ 
\AFIV.
Furthermore, from the  $\IZ_N\times \IZ_M$--orbifolds,  only the twists $\IZ_2\times\IZ_2,
\IZ_3\times\IZ_3, \IZ_6\times\IZ_6, \IZ_2\times\IZ_3,\IZ_2\times \IZ_6, \IZ_2\times \IZ_6'$ and $\IZ_3\times \IZ_6$ allow
for tadpole cancellation in the above setup \zwart. 
This is to be contrasted with \ta intersecting $D6$--brane
constructions, where it has been recently shown that essentially all orbifold
groups $\Gamma$ allow for tadpole cancellation due to the appearance of only
untwisted and $\IZ_2$--twisted sector tadpoles \BCS.

\subsec{Closed string moduli space}

The geometry of the orbifold $X_6$ is described by 
$h_{(1,1)}(X_6)$ K\"ahler moduli $\Tc^i$ and $h_{(2,1)}(X_6)$ complex
structure moduli $\Uc^i$, which split into twisted and untwisted moduli. In the following,
the dimension of the latter is denoted by $h^{\rm untw.}_{(1,1)}(X_6)$ and 
$h^{\rm untw.}_{(2,1)}(X_6)$, respectively.
In addition, there is the complex dilaton field $S$:
\eqn\Dilaton{
S=i\ C_0+e^{-\phi_{10}}\ ,}
with $\phi_{10}$ the dilaton field and $C_0$ the Ramond scalar in $D=10$.
The parameter space of $S$ is locally spanned by the coset
\eqn\Scoset{
\Mc_S=\fc{SU(1,1)}{U(1)}\ .}
Furthermore, we have: $e^{-\phi_{10}}=e^{-\phi_4}\ Vol(X_6)^{-1/2}$, with $Vol(X_6)$
the volume of the compactification manifold $X_6$.

Without $D$--brane moduli, locally the closed string moduli space $\Mc$ is a direct product of
the complex dilaton field $S$, the K\"ahler $\Mc_K$ and complex structure moduli 
$\Mc_{CS}$ \LMRS\ (see also {\it Refs.} \seealso):
\eqn\DIRECT{
\Mc=\Mc_S\otimes \Mc_K\otimes \Mc_{CS}\ .}
The K\"ahler metric for the moduli space $\Mc$ derives from the K\"ahler potential
\eqn\KAE{\eqalign{
K&=-\ln(S+\ov S)-2\ K_\Kc-K_{\cal CS}\cr
&=-\ln(S+\ov S)-2\ \ln Vol(X_6)-\ln\lf(-i \int_{X_6}\Omega_3\wedge \ov\Omega_3\ri)\ ,}}
with $\Omega_3$ the holomorphic $3$--form of the manifold $X_6$ and $Vol(X_6)$ its volume.
Above $K_\Kc$ represents the K\"ahler potential
for the K\"ahler moduli $\Tc^i$ and $K_{\cal CS}$ the corresponding K\"ahler potential
for the complex structure moduli $\Uc^j$.

Depending on the numbers $h_{(1,1)}^{\rm untw.}, h_{(2,1)}^{\rm untw.}$
of untwisted K\"ahler $\Tc^i$ and complex structure moduli $\Uc^j$, 
the generic (untwisted) moduli spaces $\Mc_\Kc,\ \Mc_{\cal CS}$ 
appearing in toroidal orbifold compactifications are described 
by the following six different cosets \multrefiv\FKP\CveticYW\IbanezHC\FT:
\eqn\cosets{\eqalign{
h_{(1,1)}^{\rm untw.}=3\ \ ,\ \ h_{(2,1)}^{\rm untw.}=0,1,3&:\ \   
\Mc_\K=\lf(\fc{SU(1,1)}{U(1)}\ri)^3\ \ ,\ \ 
\Mc_{\CS}=\lf(\fc{SU(1,1)}{U(1)}\ri)^{h_{(2,1)}^{\rm untw.}}\ ,\cr
h_{(1,1)}^{\rm untw.}=5\ \ ,\ \ h_{(2,1)}^{\rm untw.}=0,1&:\ \  
\Mc_\K=\fc{SU(2,2)}{SU(2)\times SU(2)\times U(1)}\times
\lf(\fc{SU(1,1)}{U(1)}\ri)\ ,\cr
&\hskip0.6cm\Mc_{\CS}=\lf(\fc{SU(1,1)}{U(1)}\ri)^{h_{(2,1)}^{\rm untw.}}\ ,\cr
h_{(1,1)}^{\rm untw.}=9\ \ ,\ \ h_{(2,1)}^{\rm untw.}=0&:\ \   
\Mc_\K=\fc{SU(3,3)}{SU(3)\times SU(3)\times U(1)}\ .\cr}}
The numbers $h_{(1,1)}^{\rm twist.}, h_{(2,1)}^{\rm twist.}$ depend both on the orbifold
group $\Gamma$ and the underlying torus lattice $T^6$ \Klemm. 
A list of representative orbifold examples, relevant to us in the following, is given in Table 1
and 2.
\vskip0.5cm
{\vbox{\ninepoint{$$
\vbox{\offinterlineskip\tabskip=0pt
\halign{\strut\vrule#
&~$#$~\hfil 
&\vrule$#$ 
&~$#$~\hfil 
&\vrule$#$ 
&~$#$~\hfil 
&\vrule$#$
&~$#$~\hfil 
&\vrule$#$ 
&~$#$~\hfil 
&\vrule$#$
&~$#$~\hfil 
&\vrule$#$
&~$#$~\hfil 
&\vrule$#$\cr
\noalign{\hrule}
&\ \IZ_N&&\ {\rm Twist}\ \Gamma&&\ {\rm Lattice}\ T^6&&h_{(1,1)}^{\rm untw.}&&h_{(2,1)}^{\rm untw.}&&
h_{(1,1)}^{\rm twist.}&&h_{(2,1)}^{\rm twist.}&\cr
\noalign{\hrule}\noalign{\hrule}
&\ \IZ_3      &&\   (Z_3^{(2)},Z_3^{(2)},Z_3^{(2)})&&\  SU(3)^3           &&9 && 0 && 27 && 0&\cr
&\ \IZ_{6-I}    &&\ (Z_3^{(2)},Z_6^{(2)},Z_6^{(2)})&&\  SU(3)\times G_2^2 &&5 && 0 && 24 && 5&\cr
&\ \IZ_{6-I}    &&\ (Z_6^{(2)},Z_3^{(4)})&&\ G_2\times SU(3)^{2}_{\flat}  &&5 && 0 && 20 && 1&\cr
&\ \IZ_{6-II}   &&\ (Z_2^{(1)},Z_2^{(1)},Z_3^{(2)},Z_6^{(2)})&&\ SU(2)^2\times SU(3)\times G_2
                                                                          &&3 && 1 && 32 && 10&\cr
&\ \IZ_{6-II}   &&\  (Z_2^{(1)},Z_3^{(2)},Z_6^{(3)})&&\ SU(3)\times SO(8) &&3 && 1 && 26 && 4&\cr
&\ \IZ_{6-II}   &&\  (Z_2^{(1)},Z_2^{(1)},Z_3^{(4)})&&\ SU(2)^2\times SU(3)
\times SU(3)_{\sharp}                                                     &&3 && 1 && 28 && 6&\cr
&\ \IZ_{6-II}   &&\         (Z_2^{(1)},Z_6^{(5)})&&\ SU(2)\times SU(6)    &&3 && 1 && 22 && 0&\cr
&\ \IZ_7      &&\       (Z_7^{(6)})&&\ SU(7)                              &&3 && 0 && 21 && 0&\cr
&\ \IZ_{12-I} &&\      (Z_3^{(2)},Z_{12}^{(4)})&&\ SU(3)\times F_4        &&3 && 0 && 26 && 5&\cr
&\ \IZ_{12-I} &&\      (Z_{12}^{(6)})&&\ E_6                              &&3 && 0 && 22 && 1&\cr
\noalign{\hrule}}}$$
\vskip-6pt
\centerline{\noindent{\bf Table 1:}
{\sl Twists, lattices and Hodge numbers for $\IZ_N$ orbifolds.}}
\vskip10pt}}}
\vskip-0.5cm\ \br
\vskip0.5cm
\br
The Hodge-numbers for the $\IZ_N\times\IZ_M$--orbifolds may be found in {\it Ref.} \FIQ.
\vskip0.5cm
{\vbox{\ninepoint{$$
\vbox{\offinterlineskip\tabskip=0pt
\halign{\strut\vrule#
&~$#$~\hfil 
&\vrule$#$ 
&~$#$~\hfil 
&\vrule$#$ 
&~$#$~\hfil 
&\vrule$#$
&~$#$~\hfil 
&\vrule$#$ 
&~$#$~\hfil 
&\vrule$#$
&~$#$~\hfil 
&\vrule$#$
&~$#$~\hfil 
&\vrule$#$\cr
\noalign{\hrule}
&\ \IZ_N\times \IZ_M&&\ \theta&&\ \omega&&
h_{(1,1)}^{\rm untw.}&&h_{(2,1)}^{\rm untw.}&&
h_{(1,1)}^{\rm twist.}&&h_{(2,1)}^{\rm twist.}&\cr
\noalign{\hrule}\noalign{\hrule}\noalign{\hrule}
&\ \IZ_2 \times\IZ_2     &&\   \fc{1}{2}(1,0,-1)&&\ \fc{1}{2}(0,1,-1)   &&3 && 3 && 48 && 0&\cr
&\ \IZ_3 \times\IZ_3     &&\   \fc{1}{3}(1,0,-1)&&\ \fc{1}{3}(0,1,-1)   &&3 && 0 && 81 && 0&\cr
&\ \IZ_6 \times\IZ_6     &&\   \fc{1}{6}(1,0,-1)&&\ \fc{1}{6}(0,1,-1)   &&3 && 0 && 81 && 0&\cr
&\ \IZ_3 \times\IZ_6     &&\   \fc{1}{3}(1,0,-1)&&\ \fc{1}{6}(0,1,-1)   &&3 && 0 && 70 && 1&\cr
&\ \IZ_2 \times\IZ_3     &&\   \fc{1}{2}(1,0,-1)&&\ \fc{1}{3}(0,1,-1)   &&3 && 1 && 32 && 10&\cr
&\ \IZ_2 \times\IZ_6     &&\   \fc{1}{2}(1,0,-1)&&\ \fc{1}{6}(0,1,-1)   &&3 && 1 && 48 && 2&\cr
&\ \IZ_2 \times\IZ_{6'}&&\   \fc{1}{2}(1,0,-1)&&\ \fc{1}{6}(1,1,-2)   &&3 && 0 && 33 && 0&\cr
\noalign{\hrule}}}$$
\vskip-6pt
\centerline{\noindent{\bf Table 2:}
{\sl Twists $\theta,\omega$ and Hodge numbers of  $\IZ_N\times \IZ_M$ orbifolds.}}
\vskip10pt}}}
\vskip-0.5cm
\br
The corresponding K\"ahler potentials for the spaces \cosets\ are known from heterotic string 
compactifications \CveticYW:
\eqn\CosetsKAE{\eqalign{
h_{(1,1)}^{\rm untw.}=3\ \ ,\ \ h_{(2,1)}^{\rm untw.}=0,1,3&:\ \   
K_\Kc=-\sum_{i=1}^3\ln(\Tc^i+\ov \Tc^i)\ \ ,\ \ K_{\CS}=
-\sum_{j=1}^{h_{(2,1)}^{\rm untw.}}\ln(\Uc^j+\ov\Uc^j)\ ,\cr
h_{(1,1)}^{\rm untw.}=5\ \ ,\ \ h_{(2,1)}^{\rm untw.}=0,1&:\ \  
K_\Kc=-\ln\det(\Tc^{ij}+\ov \Tc^{ij})-\ln(\Tc^5+\ov \Tc^5)\ ,\cr 
&\ \ \ \ \ K_{\CS}=-\sum_{j=1}^{h_{(2,1)}^{\rm untw.}}\ln(\Uc^j+\ov\Uc^j)\ ,\cr
h_{(1,1)}^{\rm untw.}=9\ \ ,\ \ h_{(2,1)}^{\rm untw.}=0&:\ \   
K_\Kc=-\ln\det(\Tc^{ij}+\ov\Tc^{ij})\ .\cr}}
What is less known is the parameterization of the moduli fields $\Tc^i,\Uc^i$ 
in terms of the data of the torus,
\ie the real metric $g$ and the discrete symmetries of the underlying effective
field theory. It is this aspect among others we shall elaborate in more detail in this section.

There is one important difference when compactifing the heterotic and \tb string 
on the same six--manifold $X_6$.
In the heterotic string, the complexification of the K\"ahler moduli $\Tc^i$ is achieved through 
the Neveu--Schwarz antisymmetric tensor $B_2$, while in
the orientfolds we discuss here, this is accomplished  with the Ramond $4$--form $C_4$.
Moreover, while the string--theoretical moduli fields $\Tc^i$ define proper
complex scalars of chiral N=1 multiplets in $D=4$ heterotic compactifications, 
they do not enjoy this property in \tb orientifolds.
More precisely, 
in \tb the axionic part of the complexified K\"ahler modulus $\Tc^i$ is given by some internal 
component of the Ramond $4$--form $C_4$, \ie the $4$--cycle integral $\int_{C_i} C_4$, 
while for the heterotic compactification on the same manifold $X_6$, the K\"ahler moduli are 
complexified
with some internal part of the $NS$ $2$--form $B_2$, \ie $\int_{C_j} B_2$, with some
$2$--cycle $C_j$.
Since $h_{(2,2)}(X_6)=h_{(1,1)}(X_6)$, from the cohomological point of view, there is not much
difference, as the $2$--form $\omega_i$, which appears in the expansion of $B_2$, 
is the Poincar\'e dual of the $4$--cycle $C_i$.
An other peculiarity in \tb orientifold compactifications with wrapped $D7$--branes is that
the K\"ahler moduli $\Tc^i$ following from the geometry of the manifold $X_6$ do not represent
scalars of chiral N=1 multiplets in $D=4$. One has to define new moduli $T^i$, which  refer to
the underlying effective field theory and lead to the correct effective field theory description.
In fact, a quite general formula may be given, which relates the $h_{(1,1)}$ string theoretical moduli
fields $\Tc^i$ to their field--theoretical analogs $T^i$ (for more details \cf subsection 2.5):
\eqn\NICEE{
T^i=\fc{\p}{\p \re(\Tc^i)}\ Vol(X_6(\Tc^j))+i\ \int_{C_i} C_4\ .}
Here, $Vol(X_6(\Tc^j))$ is the volume (in string units) 
of the internal manifold $X_6$ expressed in terms of the 
K\"ahler moduli $\Tc^j$, defined in \tb on $X_6$.

As we may see from the list \cosets, the complex structure moduli space is much simpler, as
this space only consists of factors of $\fc{SU(1,1)}{U(1)}$.
Furthermore, in many of the orbifold examples, the complex structure moduli $\Uc^i$ 
are fixed through the orbifold twist, \ie $h_{(2,1)}^{\rm untw.}=0$. 
Only in the case when the orbifold has $\IZ_2$--subelements, some $\Uc^i$ remain
unfixed. Except for the twist $\IZ_2\times \IZ_2$, there may only exist one
such $\IZ_2$--element in order to preserve N=1 supersymmetry in $D=4$.
Hence, for $\IZ_2\times \IZ_2$ we have $h_{(2,1)}^{\rm untw.}=3$, while {\it all} other orbifolds
with $\IZ_2$--elements have $h_{(2,1)}^{\rm untw.}=1$.
On the other hand, in \tb orientifolds the complex structure moduli 
$\Uc^i$ following from the string background $X_6$  
already describe scalars $U^i$ of N=1 chiral multiplets in $D=4$. Hence, we have:
\eqn\wehave{
U^i=\Uc^i\ \ \ ,\ \ \ i=1,\ldots,h_{(2,1)}^{\rm untw.}\ .}

\subsec{Background parameterization of geometric moduli in orientifolds of \tb}

For the above orbifolds (with $(-1)^{F_L}I_6\Omega$ orientifold projection) 
we now want to  find the parameterization of  the (untwisted)  
K\"ahler $\Tc^i$ and complex structure moduli $\Uc^i$ in terms of their torus metric 
$g$ and their $4$--form background $C_4$.
This classification is interesting on its own as it allows to study the modular symmetries 
of those compactifications.
As we have already reported, in the \tb orientifolds we are discussing, 
the K\"ahler moduli $\Tc^i$ are complexified with components of the Ramond $4$--form $C_4$, while
for heterotic compactification the Neveu--Schwarz $2$--form $B_2$ is relevant.
Since $h_{(2,2)}(X_6)=h_{(1,1)}(X_6)$, from the cohomological point of view, there is not much
difference, as the $2$--form $\omega_i$, which appears in the expansion of $B_2$, 
is the Poincar\'e dual of the $4$--cycle $C_i$.
Hence, in the following, in particular in appendix \appA\ we shall very often derive benefits from
this correspondence.

All we need to know about the lattices $T^6$ is contained in the Cartan matrix of the 
respective Lie algebra. The matrix elements of the Cartan matrix are defined as follows:
\eqn\cartan{
A_{ij}=2\ {\langle e_i, e_j\rangle\over \langle e_j, e_j\rangle}\ ,}
where the $e_i$ are the simple roots. All Cartan matrices that will be needed 
here can be found in appendix \appD. 
The above orbifolds are all Coxeter orbifolds, i.e. the twist $\Gamma$ corresponds to the 
Coxeter element of the Lie algebra. It is given by successive Weyl reflections with respect to all 
simple roots: $Q=S_1S_2...S_{rank}$, where the Weyl reflections are given by
\eqn\Weyl{
S_i({\bf x})= {\bf x}-2{\langle{\bf x}, e_i\rangle\over \langle e_i,e_i\rangle}e_i.}
Generalized Coxeter automorphisms are obtained by combining Weyl reflections with outer 
automorphisms, which are generated by transpositions of roots that are symmetries of the Dynkin 
diagram. $P_{ij}$ denotes the transposition of the $i$'th and $j$'th roots. 
The lattices marked with $\flat$ and $\sharp$ are realized as generalized Coxeter twists, the 
automorphism is in the first case $S_1S_2S_3S_4P_{36}P_{45}$ and in the second $S_1S_2S_3P_{34}S_5S_6$.

The orbifold twist $\Gamma$ may be represented by a matrix $Q_{ij}$, which rotates
the six lattice basis vectors: $e_i\ra Q_{ji}\ e_j$.
Once we have determined the Coxeter element $Q$ via \Weyl\ and the Cartan matrix of the lattice 
in question, its metric $g$ can be obtained through the requirement that the orbifold twist must 
leave the scalar product invariant \mspal, \ie:
\eqn\twi{Q^tg\,Q=g\ .}
We obtain the form of the metric in the lattice basis\foot{Speaking of the lattice basis is a bit 
misleading, as the orbifold twists generically allow more degrees of freedom in the lattice 
(\ie K\"ahler and complex structure moduli) than the Lie group, which leaves only the overall 
normalization of the lattice vectors unfixed. So when we speak of the lattice basis we mean the 
basis that captures the extra degrees of freedom of the orbifold and not the basis of the root 
lattice.}. 
The metric is conveniently parameterized via the lengths of the basis vectors and the angles 
between them: $\langle e_i, e_j\rangle= R_i R_j \cos\theta_{ij}$.
The form of the antisymmetric tensor $b$ is obtained in the same fashion, by solving
\eqn\b{Q^tb\,Q=b.}
By now we also know the number of untwisted K\"ahler and complex structure moduli: 
We count the number of independent deformations $d$ allowed by the solutions of $Q^tg\,Q=g$ 
and $Q^tb\,Q=b$. The orbifold has $d_b$ untwisted K\"ahler moduli and $\half(d_g-d_b)$ untwisted complex 
structure moduli.

To find the actual dependence of the K\"ahler and complex structure moduli on the degrees of freedom
following from the above described parameterization $g,C_4$ (or $b$) in the lattice basis, one has to go 
into a complex basis $\{z^i\}_{i=1,2,3}$, 
where the twist $Q$ acts diagonally on the complex coordinates, \ie
$\Theta\ :\ dz^i\ra e^{2\pi i v^i} dz^i$, with the eigenvalues $v^i$ introduced after \eqq \orbi.
To find these complex coordinates we make the ansatz 
\eqn\ansatz{dz^i=a^i_1\,dx^1+a^i_2\,dx^2+a^i_3\,dx^3+a^i_4\,dx^4+a^i_5\,dx^5+a^i_6\,dx^6}
and solve the equation $Q^t\,dz^i=e^{2\pi i v^i}\,dz^i$ which fixes $dz^i$ up to some constant factors. 
For convenience, we usually choose a normalization such that the first term comes with coefficient one. 
In addition, the ansatz \ansatz\ should yield a Hermitian metric, \ie we require the
identity: $ds^2=g_{ij}\ dx^i\otimes dx^j=g_{i\ov j}\ dz^i\otimes d\ov z^j$.
After having introduced the complex coordinates $dz^i$, which define the complex structure 
moduli $\Uc^i$, we write down the K\"ahler form 
$J=g_{i\ov j}\ dz^i\wedge d\ov z^j$ and the anti--symmetric $4$--form 
$C_4=\sum\limits_{i=1}^{h_{(2,2)}(X_6)} c^i\ d_i$
(or the $2$--form $B_2=b_{i\ov j}\ dz^i\wedge d\ov z^j\equiv \sum\limits_{i=1}^{h_{(1,1)}(X_6)} 
b^i\ \omega_i$ on the heterotic side). Here, the $\om_i$ represent a basis of  
twist--invariant $2$--forms of the real cohomology $H^2(X_6,\IZ)$, 
while the $d_i$ supply a twist--invariant basis for $H^4(X_6,\IZ)$.
On the heterotic side, the K\"ahler moduli $\Tc^i$ are defined via the pairing 
$J+i\,B=\sum\limits_{i=1}^{h_{(1,1)}(X_6)} {\cal T}^i\,\om_i$.
On the other hand, following the discussion after \eqq\ \CosetsKAE, in \tb we may define
the K\"ahler moduli $\Tc^i$:
\eqn\stringT{
\Tc^i=\int_{C_2^i} J +i\ \int_{C_4^i} C_4   \ \ \ ,\ \ \ i=1,\ldots,h_{(1,1)}(X_6)\ ,} 
with the $2$--cycle $C_2^i$ and the $4$--cycle $C_4^i$. The Poincar\'e dual $2$--form of $C_4^i$
is Hodge--dual to the Poincar\'e dual $4$--form of $C_2^i$. This is way both cycles 
$C_2^i, C_4^i$ carry the same index.
The first term $\int_{C_2^i} J$ of $\Tc^i$ describes the volume of the $2$--cycle $C_2^i$.
In subsection 2.5 we shall relate the latter to the volume of the $4$--cycle $C_4^i$, which
enters in the holomorphic moduli fields $T^i$ (\cf \eqq \NICEE).


As an illustrative example, we shall now work out the parameterization of the metric moduli
of a $\IZ_N\times \IZ_M$--orbifold.
All the $\IZ_N\times \IZ_M$--twists can be realized by automorphisms on either  $SU(3)$ or $G_2$--sublattices. For $\IZ_N\times \IZ_M$--orbifolds, we are looking for a complex structure that is compatible 
with the two different twists as well as the combination of the two. 
The constraints are more stringent than in a simple $\IZ_N$--twist, therefore the resulting lattices have less 
degrees of freedom. While the case  $\IZ_3\times \IZ_6$ 
is treated here, for  $\IZ_3\times \IZ_3,\,\IZ_2\times \IZ_6,\,\IZ_2\times \IZ_{6'}$ and $\IZ_6\times \IZ_6$ the reader is referred 
to appendix \appA. 
The $\IZ_3\times \IZ_6$--orbifold has the following two twists:
\eqn\twistactionththii{\eqalign{
Q_1:\ z^i& \lra e^{2\pi i v_1^i}\ z^i\ \ \ ,\ \ \ v_1^1=\fc{1}{3}\ ,\ v_1^2=0\ ,\ v_1^3=-\fc{1}{3}\ ,\cr
Q_2:\ z^i& \lra e^{2\pi i v_2^i}\ z^i\ \ \ ,\ \ \ v_2^1=0\ ,\ v_2^2=\fc{1}{6}\ ,\ v_2^3=-\fc{1}{6}\ .
}}
The combined twist $Q_3=Q_2Q_1$ is a $\IZ_{6-II}$--twist:
\eqn\twistcombii{
Q_3:\ z^i \lra e^{2\pi i v_3^i}\ z^i\ \ \ ,\ \ \ v_3^1=\fc{2}{6}\ ,\ v_3^2=\fc{1}{6}\ ,\ 
v_3^3=-\fc{3}{6}\ .}
We know that the $\IZ_{6-II}$--twist can live on $(SU(2))^2\times SU(3)\times G_2$, its 
action on the root lattice is worked out in the appendix \appA. 
A two-dimensional $\IZ_3$--twist 
lives on the $SU(3)$--lattice, while a two--dimensional $\IZ_6$--twist lives on the $G_2$--lattice. 
Knowing all this, it is possible to construct the following twists:
\eqn\twistactionththlii{\eqalign{
Q_1\, e_1&= e_2,\quad Q_1\,e_2=-e_1-e_2,\quad Q_1\,e_3=e_3,\quad Q_1\,e_4=e_4,\cr
Q_1\,e_5&=e_5+3\,e_6,\quad Q_1\,e_6=-e_5-2\,e_6,\cr
Q_2\, e_1&= e_1,\quad Q_2\,e_2=e_2,\quad Q_2\,e_3=2\,e_3+3\,e_4,\quad Q_2\,e_4=-e_3-e_4,\cr
Q_2\,e_5&=2\,e_5+3\,e_6,\quad Q_2\,e_6=-e_5-e_6.
}}
The twist on $e_5,\,e_6$ in $Q_1$ is minus the anti-twist of the usual Coxeter--twist on $G_2$, while all other twists are the usual $\IZ_2,\,\IZ_3$ and $\IZ_6$--twists on their respective lattices.
The twists reproduce the correct eigenvalues and the conditions $Q_1^3=1,\ Q_2^6=1$. 
The combined twist $Q_3$ has the form
\eqn\twicoii{\eqalign{
Q_3\, e_1&= e_2,\quad Q_3\,e_2=-e_1-e_2,\cr
Q_3\,e_3&=2\,e_3+3\,e_4,\quad Q_3\,e_4=-e_3-e_4,\cr
Q_3\,e_5&=-e_5,\quad Q_3\,e_6=-e_6,
}}
which is as just mentioned a $\IZ_{6-II}$--twist, namely the one on the lattice $SU(2)^2\times SU(3)\times G_2$. As before, we require the metric to be 
invariant under all three twists, i.e. we impose the three conditions $Q_i^Tg\,Q_i=g,\quad i=1,2,3$. 
This leads to the following solution:
\eqn\gii{g=\pmatrix{R_1^2&-\half R_1^2&0&0&0&0\cr
-\half R_1^2&R_1^2&0&0&0&0\cr
0&0&R_3^2&-\half R_3^2&0&0\cr
0&0&-\half R_3^2&{1\over 3} R_3^2&0&0\cr
0&0&0&0&R_5^2&-\half R_5^2\cr
0&0&0&0&-\half R_5^2&{1\over 3} R_5^2.}}
This corresponds exactly to the metric of $SU(3)\times (G_2)^2$ without any extra degrees of freedom.
The corresponding solution for $b$ has the form
\eqn\bii{b=\pmatrix{0&b_1&0&0&0&0\cr
-b_1&0&0&0&0&0\cr
0&0&0&b_3&0&0\cr
0&0&-b_3&0&0&0\cr
0&0&0&0&0&-b_5\cr
0&0&0&0&-b_5&0}.}
We have three K\"ahler moduli while the complex structure is completely fixed. We find the following complex coordinates:
\eqn\dzsii{\eqalign{
dz^1&=3^{1/4}\,(dx^1+e^{2\pi i/3}dx^2)\ ,\cr
dz^2&=dx^3+{1\over\sqrt3}\,e^{5\pi i/6}dx^4\ ,\cr
dz^3&=dx^5+{1\over\sqrt3}\,e^{-5\pi i/6}dx^6\ .}}
The invariant 2-forms in the real cohomology are simply $dx^1\wedge dx^2, \ dx^3\wedge dx^4$ and $dx^5\wedge dx^6$. 
The K\"ahler form in complex coordinates has the form
\eqn\kcplx{-i\,J={1\over\sqrt3}\,R_1^2\, dz^1\wedge d\ov z^1+R_3^2\, dz^1\wedge d\ov z^1+R_5^2\,dz^3\wedge d\ov z^3.}
Expressed in the real cohomology, it takes the form
\eqn\kreal{J={\sqrt3\over2}\,R_1^2\, dx^1\wedge dx^2+ {1\over2\sqrt3}\,R_3^2\, dx^3\wedge dx^4+{1\over2\sqrt3}\,R_5^2 \, dx^5\wedge dx^6\ .}
So via the pairing $J+i\,B={\cal T}^i\,\om_i$ in the real cohomology, we find the 
three K\"ahler moduli to take the following form
\eqn\ksii{\eqalign{
\Tc^1&={\sqrt3\over2}\,R_1^2+{i}\ c_1\ ,\cr
\Tc^2&={1\over2\sqrt3}\,R_3^2+{i}\ c_2\ ,\cr
\Tc^3&={1\over2\sqrt3}\,R_5^2+{i}\ c_3}}
after translating the heterotic components of $B_2$ into the respective components of the
Ramond $4$--form $C_4$, \ie $c_i\simeq b_i$.

We shall in the following derive the K\"ahler moduli and complex structures for the $\IZ_N$--orbifolds, which
allow for tadpole cancellation without introducing torsion or vector structure, 
\ie $\IZ_7,\,\IZ_{6-I},\,\IZ_{6-II}$ and $\IZ_{12-I}$, in appendix \appA.
The $\IZ_2\times \IZ_2$ and $\IZ_3$--orbifolds are treated in the next subsection.

\subsec{Parametrization of coset spaces with more than three K\"ahler moduli}

After having presented the general procedure of how to find the parameterization of 
the metric moduli in terms of the backgrounds $g,C_4$ (or $b$ in the heterotic case),
in this subsection we shall demonstrate this on  some selected orbifold examples
which have more than three (untwisted) K\"ahler moduli.
The simplest realization of the K\"ahler moduli spaces $\Mc_\K=\lf(\fc{SU(1,1)}{U(1)}\ri)^3,\ 
\Mc_\K=\fc{SU(2,2)}{SU(2)\times SU(2)\times U(1)}\times \lf(\fc{SU(1,1)}{U(1)}\ri)$ and 
$\Mc_\K=\fc{SU(3,3)}{SU(3)\times SU(3)\times U(1)}$, given in \eqq \cosets,  
are the toroidal orbifolds $T^6/\IZ_2\times \IZ_2$, $T^6/\IZ_4$ and $T^6/\IZ_3$, respectively.
Let us present the parameterization of their K\"ahler moduli in terms of their backgrounds $g,C_4$
for \tb orientifolds.

\br
$\underline{{\rm \it K\ddot{a}hler\ moduli\ space\ } \Mc_\K=\lf(\fc{SU(1,1)}{U(1)}\ri)^3}$
\br

This K\"ahler moduli space $\Mc_\K$ is realized \eg in the $\IZ_2\times\IZ_2$ \tb orientifold, 
with the twists $Q_1,Q_2$
\eqn\QQtwist{\eqalign{
&Q_1\ :\ e_{1,2}\lra-e_{1,2}\ \ \ ,\ \ \ e_{3,4}\lra-e_{4,5}\ \ \ ,\ \ \ e_{5,6}\lra e_{5,6}\ ,\cr
&Q_2\ :\ e_{1,2}\lra-e_{1,2}\ \ \ ,\ \ \ e_{3,4}\lra e_{4,5}\ \ \ ,\ \ \ e_{5,6}\lra -e_{5,6}}}
acting on the integral basis $\{e_i\}$ of the torus $T^6$. 
According to \twi, the twists $Q_1,Q_2$ 
only allow for a factorizable lattice, \ie $T^6$ being a direct product of three $2$--tori, \ie
$T^6=\otimes_{j=1}^3 T^{2,j}$, with the metrics $g^j$.
Each individual $2$--torus $T^{2,j}$ has one K\"ahler modulus $\Tc^j$
and  one complex structure modulus $\Uc^j$ describing the real parameters of the metric $g^j$. 
The complex structure moduli are given by: 
\eqn\CS{
\Uc^j=\fc{1}{g^j_{11}}\ (\ \sqrt{\det g^j}+i\ g^j_{12}\ )\ ,\ \ \ {\rm with:}\ \ \ 
g^j=\pmatrix{g_{11}^j&g_{12}^j\cr g_{12}^j&g_{22}^j}\ .}
Some of these moduli may be fixed through the orbifold group resulting in 
$h_{(2,1)}^{\rm untw.}$ complex structure  moduli $U^j$ entering \KAE.
The real part of the K\"ahler modulus $\Tc^j$ describes the size $\sqrt{\det g^j}$ 
of the subtorus $T^{2,j}$, \ie $\re(\Tc^j)=\sqrt{\det g^j}$.
The twist--invariant $4$--form $C_4$ is given by
\eqn\ffourform{
C_4=c^1\ dx^2\wedge dy^2\wedge dx^3\wedge dy^3+c^2\ dx^1\wedge dy^1\wedge dx^3\wedge dy^3
+c^3\ dx^1\wedge dy^1\wedge dx^2\wedge dy^2\ .}
The K\"ahler modulus $\Tc^j$ is complexified with the internal part of the
Ramond $4$--form $C_4$, \ie
$\im(\Tc^j)=\int\limits_{T^{2,k}\times T^{2,l}} C_4=c^j$.
Altogether, the three K\"ahler moduli $\Tc^j$ for the coset $\Mc_\Kc$ are:
\eqn\kaehler{
\Tc^j=i\ c^j+\sqrt{\det\ g^j}\ \ \ ,\ \ \ j=1,2,3\ .}
The \tb K\"ahler potential $K_\Kc$ is  (\cf \eqq \CosetsKAE):
\eqn\simply{
K_\K=-\ln Vol(X_6)=-\sum_{i=1}^3\ln\lf[\ \h(\Tc^i+\ov\Tc^i)\ \ri]\ .}
According to \eqq\ \NICEE, the moduli $\Tc$ have to be translated into the
holomorphic coordinates $T^j$. With $Vol(X_6)=\re(\Tc^1)\re(\Tc^2)\re(\Tc^3)$, we obtain \LustFI:
\eqn\OBTAIN{
T^j=i\ c^j+\re(\Tc^k)\ \re(\Tc^l)\ \ ,\ \ (j,k,l)\in \overline{(1,2,3)}\ .}
In terms of these holomorphic coordinates $T^j$ the K\"ahler potential \KAE\ becomes:
\eqn\Simply{
\kappa_4^{-2}\ K=-\ln(S+\ov S)-\sum_{i=1}^3\ln\lf[\ \h(T^i+\ov T^i)\ \ri]
-\sum_{j=1}^{h_{(2,1)}^{\rm untw.}}\ln(U^j+\ov U^j)\ .}

\br
$\underline{{\rm \it K\ddot{a}hler\ moduli\ space\ } 
\Mc_\K=\fc{SU(2,2)}{SU(2)\times SU(2)\times U(1)}\otimes\lf(\fc{SU(1,1)}{U(1)}\ri)}$
\br

Let us move on to the special K\"ahler moduli space $\Mc_\K$, which may be realized 
\eg in the $\IZ_4$--orbifold with the twist 
$v^i=\fc{1}{4}(1,1,-2)$. For concreteness we assume the twist to act block--diagonally within 
the six--torus $T^6$ and the latter to be factorizable into
$T^6=T^4\otimes T^2$. Then the $\IZ_2$--element of the twist entirely acts in the $2$--torus.
The complex structure modulus for the latter is defined as in \eqq \CS. The real part
of the K\"ahler modulus $\Tc^5$ of the $2$--torus $T^2$ describes the volume of
this torus. The complexified K\"ahler modulus $\Tc^5$ gives rise to 
the extra factor $\lf(\fc{SU(1,1)}{U(1)}\ri)$ of $\Mc_\K$ and will be given in a moment.

On the other hand, the parameterization of the four 
K\"ahler moduli $\Tc^{ij}\ ,\ i,j=1,2$
in terms of the lattice properties of $T^4$ need some more work.
According to \twi\ the orbifold twist $Q$, which acts through
\eqn\Qtwist{
Q\ :\ \ e_1\lra e_2\ \ \ ,\ \ \ e_2\lra-e_1\ \ \ ,\ \ \ 
e_3\lra e_4\ \ \ ,\ \ \ e_4\lra-e_3}
on the root basis $\{e_i\}$ of $T^4$ allows for the four metric components
\eqn\allowme{
g=\pmatrix{g_{11}&0&g_{13}&-g_{14}\cr
0&g_{11}&g_{14}&g_{13}\cr
g_{13}&g_{14}&g_{22}&0\cr
-g_{14}&g_{13}&0&g_{22}}\ .}
Following \ansatz\ , we introduce complex structures and the complex coordinates $z^1,z^2$:
\eqn\introducecomplex{
dz^1=\fc{1}{\sqrt 2}\ (dx^1+i\ dy^1)\ \ \ ,\ \ \ 
dz^2=\fc{1}{\sqrt 2}\ (dx^2+i\ dy^2)\ .}
W.r.t the latter, the
twist \Qtwist\ acts diagonally, \ie $\Theta\ : z^1\ra i z^1,\ z^2\ra i z^2$, with the 
twist eigenvalue $i$. 
In these complex coordinates the K\"ahler form $J=g_{i\ov j}\ dz^i\wedge d\ov z^{\ov j}$ 
becomes
\eqn\Kaehlerform{\eqalign{
-i\ J&=g_{11}\ dz^1\wedge d\ov z^1+g_{22}\ dz^2\wedge\ d\ov z^2+\re(\Tc^5)\ dz^3\wedge d\ov z^3\cr
&+(g_{13}-i\ g_{14})\ dz^1\wedge d\ov z^2+(g_{13}+i\ g_{14})\ dz^2\wedge d\ov z^1\ ,}}
or expanded w.r.t the integer cohomology $H^2(X_6,\IZ)$:
\eqn\kaehlerform{\eqalign{
J&=g_{11}\ dx^1\wedge dy^1+g_{22}\ dx^2\wedge dy^2+\re(\Tc^5)\ dx^3\wedge  dy^3\cr
&+g_{14}\ (dx^1\wedge dx^2+dy^1\wedge dy^2)+g_{13}\ (dx^1\wedge dy^2-dy^1\wedge dx^2)\ .}}
The $4$--form
$C_4$ may be expanded w.r.t. the twist--invariant elements of the integral cohomolgy $H^4(T^6,\IZ)$:
\eqn\fourform{\eqalign{
C_4&=c_1\ dx^2\wedge dy^2\wedge dx^3\wedge dy^3+c_2\ dx^1\wedge dy^1\wedge dx^3\wedge dy^3
+c_5\ dx^1\wedge dy^1\wedge dx^2\wedge dy^2\cr
&-\h\ c_3\ (dx^1\wedge dx^2\wedge dx^3\wedge dy^3+dy^1\wedge dy^2\wedge dx^3\wedge dy^3)\cr
&+\h\ c_4\ (dy^1\wedge dx^2\wedge dx^3\wedge dy^3-dx^1\wedge dy^2\wedge dx^3\wedge dy^3)\ ,}}
which becomes:
\eqn\Fourform{\eqalign{
C_4&=-c_1\ dz^2\wedge d\ov z^2\wedge dz^3\wedge d\ov z^3-
c_2\ dz^1\wedge d\ov z^1\wedge dz^3\wedge d\ov z^3
-c_5\ dz^1\wedge d\ov z^1\wedge dz^2\wedge d\ov z^2\cr
&+\h\ (c_4-ic_3)\ dz^1\wedge d\ov z^2\wedge dz^3\wedge d\ov z^3
-\h\ (c_4+ic_3)\ d\ov z^1\wedge dz^2\wedge dz^3\wedge d\ov z^3\ .}}
Thus, we are lead  to define the following K\"ahler moduli in \tb:
\eqn\conres{\eqalign{
\Tc^{1}&=g_{11}+i\ c_1\ \ \ ,\ \ \ \Tc^{2}=g_{22}+i\ c_2\ \ \ ,\ \ \ \Tc^5=\re(\Tc^5)+i\ c_5\ ,\cr
\Tc^{3}&=g_{13}+i\ c_4\ \ \ ,\ \ \ \Tc^{4}=g_{14}+i\ c_3\ .}}
The K\"ahler potential for the K\"ahler moduli $\Tc^j$ may be easily
obtained from the volume form:
\eqn\Kaehlerpot{\eqalign{
K_\K&=-\ln{\rm Vol}(T^6/\IZ_4)=-\ln \fc{1}{6}\ \int_{T^6} J\wedge J\wedge J+\ln 4\cr
&=-\ln\fc{1}{8}\ \lf\{(\Tc^5+\ov \Tc^5)\ 
\lf[(\Tc^1+\ov \Tc^1)(\Tc^2+\ov \Tc^2)-(\Tc^3+\ov \Tc^3)^2-(\Tc^4+\ov \Tc^4)^2\ri]\ri\}+\ln 4\ .}}
Note, that ${\rm Vol}(T^6/\IZ_4)=1/4\ {\rm Vol}(T^6)$. Besides, the piece in the bracket 
$\fc{1}{4}[\ldots]$
is the volume $\sqrt{\det g}=g_{11}g_{22}-g_{13}^2-g_{14}^2$ of the $4$--torus $T^4$. 
The holomorphic field--redefinition
\eqn\fieldred{
\tilde \Tc^{3}=\Tc^{3}+i\ \Tc^{4}=g_{13}+ig_{14}+ic_4-c_3\ \ \ ,\ \ \ 
\tilde \Tc^{4}=\Tc^{3}-i\ \Tc^{4}=g_{13}-ig_{14}+ic_4+c_3} 
would lead to the K\"ahler potential:
\eqn\otherK{\eqalign{
\tilde K_\K&=-\ln\lf\{(\Tc^5+\ov \Tc^5)\ \lf[(\Tc^{1}+\ov \Tc^{1})(\Tc^{2}+\ov \Tc^{2})-
(\tilde \Tc^{3}+\ov{\tilde\Tc}^{4})(\ov{\tilde \Tc}^{3}+{\tilde \Tc^{4}}) \ri]\ri\}\cr
&=-\ln(\Tc^5+\ov \Tc^5)-\ln\det(\Tc+\Tc^\dagger)\ \ \ ,\ \ \ 
\Tc=\pmatrix{\Tc^1&\tilde\Tc^3\cr \tilde \Tc^4&\Tc^2}\ .}}
This form of the K\"ahler potential reproduces the formula given in \CosetsKAE\ 
with the moduli fields $\Tc^i$ parameterized by the underlying string background.
The two different forms \Kaehlerpot\ and \otherK\ simply reflect 
the isomorphy $\fc{SU(2,2)}{SU(2)\times SU(2)\times U(1)}\simeq
\fc{SO(4,2)}{SO(4)\times SO(2)}$. 
Finally, with the relation \NICEE, we obtain the 
correct relations between the string--theoretical moduli fields $\Tc^i$ and
the field--theoretical fields $T^i$.
With
$$Vol(X_6)=\fc{1}{32}\ \lf\{(\Tc^5+\ov \Tc^5)\ 
\lf[(\Tc^1+\ov \Tc^1)(\Tc^2+\ov \Tc^2)-(\Tc^3+\ov \Tc^3)^2-(\Tc^4+\ov \Tc^4)^2\ri]\ri\} ,$$ 
we obtain the following field--theoretical moduli fields:
\eqn\fieldfinally{\eqalign{
T^1&=\fc{1}{32}\ (\Tc^2+\ov \Tc^2)\ (\Tc^5+\ov\Tc^5)+i\ c_1\ \ \ ,\ \ \ 
T^2=\fc{1}{32}\ (\Tc^1+\ov \Tc^1)\ (\Tc^5+\ov\Tc^5)+i\ c_2\ ,\cr
T^3&=-\fc{1}{16}\ (\Tc^3+\ov\Tc^3)\ (\Tc^5+\ov\Tc^5)+i\ c_3\ \ \ ,\ \ \ 
T^4=-\fc{1}{16}\ (\Tc^4+\ov \Tc^4)\ (\Tc^5+\ov\Tc^5)+i\ c_4\ ,\cr
T^5&=\fc{1}{32}\  \lf[(\Tc^1+\ov\Tc^1)\ (\Tc^2+\ov\Tc^2)-(\Tc^3+\ov \Tc^3)^2-(\Tc^4+\ov \Tc^4)^2\ri]+
i\ c_5\ .}}
The K\"ahler potential \KAE, written in terms of these fields, becomes: 
\eqn\fieldK{\eqalign{
\kappa_4^{-2}\ K&=-\ln(S+\ov S)+K_{CS}-\ln4\cr
&-\ln\lf\{(T^5+\ov T^5)\ \lf[(T^1+\ov T^1)(T^2+\ov T^2)-
\fc{1}{4}\ (T^3+\ov T^3)^2-\fc{1}{4}\ (T^4+\ov T^4)^2\ri]\ri\}\ .}}

\br
$\underline{{\rm \it  K\ddot{a}hler\ moduli\ space\ } 
\Mc_\K=\fc{SU(3,3)}{SU(3)\times SU(3)\times U(1)}}$
\br

The simplest realization of this  moduli space $\Mc_\K$
appears in the $\IZ_3$--orbifold with the block--diagonal orbifold action:
\eqn\QQQtwist{
Q\ :\ e_1\lra e_2\ ,\ e_2\lra-e_1-e_2\ \ \ ,\ \ \ e_3\lra e_4\ ,\ e_4\lra-e_3-e_4\ \ \ ,\ \ \ 
e_5\lra e_6\ ,\ e_6\lra-e_5-e_6\ .}
on the six roots $\{e_i\}$ describing the six--torus $T^6$.
According to \twi\ this twist allows for the metric background
\eqn\allowg{
g=\pmatrix{g_{11}&-\h\ g_{11}&g_{13}&g_{14}&g_{15}&g_{16}\cr
-\h\ g_{11}&g_{11}&-g_{13}-g_{14}&g_{13}&-g_{15}-g_{16}&g_{15}\cr
g_{13}&-g_{13}-g_{14}&g_{33}&-\h\ g_{33}&g_{35}&g_{36}\cr
g_{14}&g_{13}&-\h\ g_{33}&g_{33}&-g_{35}-g_{36}&g_{35}\cr
g_{15}&-g_{15}-g_{16}&g_{35}&-g_{35}-g_{36}&g_{55}&-\h\ g_{55}\cr
g_{16}&g_{15}&g_{36}&g_{35}&-\h\ g_{55}&g_{55}}\ ,}
characterized by nine metric parameters.
Following \ansatz\ we introduce complex structures and the 
complex coordinates $z^1,z^2,z^3$
\eqn\Introduce{
dz^1=\fc{1}{\sqrt 3}\ (dx^1+\rho\ dy^1)\ \ ,\ \ 
dz^2=\fc{1}{\sqrt 3}\ (dx^2+\rho\ dy^2)\ \ ,\ \ 
dz^3=\fc{1}{\sqrt 3}\ (dx^3+\rho\ dy^3).}
W.r.t. them, the twist \QQQtwist\ acts diagonally, \ie $\Theta\ :z^1\ra \rho z^1,\  
z^2\ra \rho z^2,\ z^3\ra \rho z^3$, with $\rho=e^{\fc{2\pi i}{3}}$.
W.r.t. these coordinates the K\"ahler form $J$ becomes:
\eqn\Kaehlerformm{\eqalign{
-i\ J&=\fc{3}{2}\ g_{11}\ dz^1\wedge d\ov z^1+\fc{3}{2}\ g_{33}\ dz^2\wedge d\ov z^2+
\fc{3}{2}\ g_{55}\ dz^3 \wedge d\ov z^3\cr
&-\sqrt 3i\ (\rho\ g_{13}-g_{14})\ dz^1\wedge d\ov z^2
-\sqrt 3i\ (\rho\ g_{15}-g_{16})\ dz^1\wedge d\ov z^3\cr
&-\sqrt 3i\ (\rho\ g_{35}-g_{36})\ dz^2\wedge d\ov z^3
+\sqrt 3i\ (\ov\rho\ g_{13}-g_{14})\ d z^2\wedge d\ov z^1 \cr
&+\sqrt 3i\ (\ov\rho\ g_{15}-g_{16})\ d z^3\wedge d\ov z^1 
+\sqrt 3i\ (\ov\rho\ g_{35}-g_{36})\ d z^3\wedge d\ov z^2\ .}}
Expressed w.r.t. the integer cohomology $H^2(X_6,\IZ)$ we obtain:
\eqn\realJ{\hskip-1cm\eqalign{
J&=\fc{\sqrt 3}{2}\ g_{11}\ dx^1\wedge dy^1+\fc{\sqrt 3}{2}\ g_{33}\ dx^2\wedge dy^2+
\fc{\sqrt 3}{2}\ g_{55}\ dx^3\wedge dy^3\cr
&-\fc{1}{\sqrt 3}\ \lf[(g_{13}+2g_{14})\ (dx^1\wedge dx^2-dy^1\wedge dx^2+dy^1\wedge dy^2)-
(2\ g_{13}+g_{14})\ (dx^1\wedge dy^2-dy^1\wedge dx^2)\ri]\cr
&-\fc{1}{\sqrt 3}\ \lf[(g_{15}+2g_{16})\ (dx^1\wedge dx^3-dy^1\wedge dx^3+dy^1\wedge dy^3)-
(2\ g_{15}+g_{16})\ (dx^1\wedge dy^3-dy^1\wedge dx^3)\ri]\cr
&-\fc{1}{\sqrt 3}\ \lf[(g_{35}+2g_{36})\ (dx^2\wedge dx^3-dy^2\wedge dx^3+dy^2\wedge dy^3)-
(2\ g_{35}+g_{36})\ (dx^2\wedge dy^3-dy^2\wedge dx^3)\ri].}}
Obviously, the $2$--forms in the bracket represent the nine twist--invariant two--forms of the
$\IZ_3$--orbifold.
The Ramond $4$--form
$C_4$ has nine independent parameters and 
may be expanded w.r.t. a twist--invariant basis of $H^4(X_6,\IZ)$:
\eqn\Fourformd{\eqalign{
C_4&=c_1\ dx^2\wedge dy^2\wedge dx^3\wedge dy^3+c_2\ dx^1\wedge dy^1\wedge dx^3\wedge dy^3
+c_3\ dx^1\wedge dy^1\wedge dx^2\wedge dy^2\cr
&+\fc{1}{3}\ (c_5-2c_4)\ (dx^1\wedge dx^2-dy^1\wedge dx^2+dy^1\wedge dy^2)\wedge  dx^3\wedge d y^3\cr
&+\fc{1}{3}\ (c_4-2c_5)\ (dx^1\wedge dy^2-dy^1\wedge dx^2)\wedge  dx^3\wedge d y^3\cr
&+\fc{1}{3}\ (c_7-2c_6)\ (dx^1\wedge dx^3-dy^1\wedge dx^3+dy^1\wedge dy^3)\wedge  dx^2\wedge d y^2\cr
&+\fc{1}{3}\ (c_6-2c_7)\ (dx^1\wedge dy^3-dy^1\wedge dx^3)\wedge  dx^2\wedge d y^2\cr
&+\fc{1}{3}\ (c_9-2c_8)\ (dx^2\wedge dx^3-dy^2\wedge dx^3+dy^2\wedge dy^3)\wedge  dx^1\wedge d y^1\cr
&+\fc{1}{3}\ (c_8-2c_9)\ (dx^2\wedge dy^3-dy^2\wedge dx^3)\wedge  dx^1\wedge d y^1\ .}}
This expansion has the property that the basis w.r.t. which the $4$--form $C_4$ is expanded,  
is Hodge--dual to the twist--invariant basis of $2$--forms appearing in the expansion 
of $B_2$, given in \eqq (D.5) of appendix \appE.
W.r.t.  $H^4(X_6,\IC)$, the expansion \Fourformd\ becomes:
\eqn\Fourformdc{\eqalign{
C_4&=-3\ c_1\ dz^2\wedge d\ov z^2\wedge dz^3\wedge d\ov z^3
-3\ c_2\ dz^1\wedge d\ov z^1\wedge dz^3\wedge d\ov z^3
-3\ c_3\ dz^1\wedge d\ov z^1\wedge dz^2\wedge d\ov z^2\cr
&-i\sqrt3\ (c_4+\rho\ c_5)\ dz^1\wedge d\ov z^2\wedge dz^3\wedge d\ov z^3
-i\sqrt3\ (c_4+\ov\rho\ c_5)\ d\ov z^1\wedge d z^2\wedge dz^3\wedge d\ov z^3\cr
&-i\sqrt3\ (c_6+\rho\ c_7)\ dz^1\wedge d\ov z^3\wedge dz^2\wedge d\ov z^2
-i\sqrt3\ (c_6+\ov\rho\ c_7)\ d\ov z^1\wedge d z^3\wedge dz^2\wedge d\ov z^2\cr
&-i\sqrt 3\ (c_8+\rho\ c_9)\ dz^2\wedge d\ov z^3\wedge dz^1\wedge d\ov z^1
-i\sqrt3\ (c_8+\ov\rho\ c_9)\ d\ov z^2\wedge d z^3\wedge dz^1\wedge d\ov z^1\ .}}
In the \tb orientifold we are discussing, the metric components of $J$ are complexified
with the respective components of $C_4$ and lead to:
\eqn\Ninemoduli{\eqalign{
\Tc^1&=\fc{\sqrt 3}{2}\ g_{11}+i\ c_1\ \ \ ,\ \ \ \Tc^2=\fc{\sqrt 3}{2}\ g_{33}+i\ c_2\ \ \ ,\ \ \ 
\Tc^3=\fc{\sqrt 3}{2}\ g_{55}+i\ c_3\ ,\cr
\Tc^4&=-\fc{1}{\sqrt 3}\ (g_{13}+2\ g_{14})+i\ c_4\ \ \ ,\ \ \ 
\Tc^5=\fc{1}{\sqrt 3}\ (2\ g_{13}+g_{14})+i\ c_5\ ,\cr
\Tc^6&=-\fc{1}{\sqrt 3}\ (g_{15}+2\ g_{16})+i\ c_6\ \ \ \ ,\ \ \ 
\Tc^7=\fc{1}{\sqrt 3}\ (2\ g_{15}+g_{16})+i\ c_7\ ,\cr
\Tc^8&=-\fc{1}{\sqrt 3}\ (g_{35}+2\ g_{36})+i\ c_8\ \ \ ,\ \ \ 
\Tc^9=\fc{1}{\sqrt 3}\ (2\ g_{35}+g_{36})+i\ c_9\ .}}
The K\"ahler potential for the K\"ahler moduli $\Tc^j$ may be easily
obtained from the volume form:
\eqn\Kaehlerpoti{
K_\K=-\ln \fc{1}{6}\ \int_{X_6} J\wedge J\wedge J=-\ln Vol(X_6)\ ,}
with 
\eqn\VOLU{
Vol(T^6/\IZ_3)=\fc{1}{3}\cdot \fc{1}{6}\ \Kc_{ijk}\ \re(\Tc^i)\ \re(\Tc^j)\ \re(\Tc^k)=\fc{1}{3}\ 
\sqrt{\det(g)}\ ,}
and the triple intersection numbers $\Kc_{ijk}$:
\eqn\internum{\eqalign{
&\Kc_{123}=1\ \ ,\ \ \Kc_{344}=-2\ \ ,\ \ \Kc_{345}=-1\ \ ,\ \ \Kc_{355}=-2\ \ ,\ \ \Kc_{266}=-2\ ,\cr
&\Kc_{267}=-1\ \ ,\ \ \Kc_{277}=-2\ \ ,\ \Kc_{468}=1\ \ ,\ \ \Kc_{568}=2\ \ ,\ \ \Kc_{478}=-1\ ,\cr
&\Kc_{578}=1\ \ ,\ \ \Kc_{188}=-2\ \ ,\ \Kc_{469}=2\ \ ,\ \ \Kc_{569}=1\ \ ,\ \ \Kc_{479}=1\ ,\cr
&\Kc_{579}=2\ \ ,\ \ \Kc_{189}=-1\ \ ,\ \ \Kc_{199}=-2\ .}}
Again, through a holmorphic field--redefinition we may pass to the  moduli fields :
\eqn\fieldred{\eqalign{
\tilde \Tc^4&=-i\ (\rho\ g_{13}-g_{14})+\fc{i}{\sqrt 3}\ (c_4+\rho\ c_5)\ ,\ 
\tilde \Tc^7=i\ (\ov\rho\ g_{13}-g_{14})+\fc{i}{\sqrt 3}\ (c_4+\ov\rho\ c_5)\ ,\cr
\tilde \Tc^5&=-i\ (\rho\ g_{15}-g_{16})+\fc{i}{\sqrt 3}\ (c_6+\rho\ c_7)\ ,\ 
\tilde \Tc^8=i\ (\ov\rho\ g_{15}-g_{16})+\fc{i}{\sqrt 3}\ (c_6+\ov\rho\ c_7)\ ,\cr
\tilde \Tc^6&=-i\ (\rho\ g_{35}-g_{36})+\fc{i}{\sqrt 3}\ (c_8+\rho\ c_9)\ ,\ 
\tilde \Tc^9=i\ (\ov\rho\ g_{35}-g_{36})+\fc{i}{\sqrt 3}\ (c_8+\ov\rho\ c_9)\ .}}
In terms of these moduli, the K\"ahler potential becomes:
\eqn\newK{
\tilde K_\Kc=-\ln\det(\Tc+\Tc^\dagger)\ \ \ ,\ \ \ \Tc=\pmatrix{\Tc^1&\tilde\Tc^4&\tilde\Tc^5\cr
\tilde\Tc^7&\Tc^2&\tilde\Tc^6\cr
\tilde\Tc^8&\tilde\Tc^9&\Tc^3}\ ,}
in agreement with the expression given in \CosetsKAE.
Finally, from \NICEE\ and \VOLU\ we may deduce the nine holomorphic moduli fields $T^i$
\eqn\correctth{
T^i=\h\ \Kc_{ijk}\  \re(\Tc^j)\ \re(\Tc^k)+i\ c_i\ ,}
with the intersection numbers given in \internum.
The full K\"ahler potential \KAE, written in terms of the fields \correctth, becomes 
\eqn\fieldKK{
\kappa_4^{-2}\ K=-\ln(S+\ov S)-\ln\lf[\ \fc{1}{8}\ \tilde\Kc_{ijk}\ (T^i+\ov T^i)\ (T^j+\ov T^j)\ 
(T^k+\ov T^k)\ \ri]\ ,}
with:
\eqn\internum{\eqalign{
&\tilde\Kc_{123}=\fc{1}{2}\ \ ,\ \ \tilde\Kc_{344}=-\fc{1}{3}\ \ ,\ \ \tilde\Kc_{345}=\fc{1}{6}\ \ ,\ \ 
\tilde\Kc_{355}=-\fc{1}{3}\ \ ,\ \ \tilde\Kc_{266}=-\fc{1}{3}\ ,\cr
&\tilde\Kc_{267}=\fc{1}{6}\ \ ,\ \ \tilde\Kc_{277}=-\fc{1}{3}\ \ ,\ \ \tilde\Kc_{568}=\fc{1}{6}\ \ ,\ \ 
\tilde\Kc_{478}=-\fc{1}{6}\ \ ,\ \ \tilde\Kc_{188}=-\fc{1}{3}\ ,\cr
&\tilde\Kc_{469}=\fc{1}{6}\ \ ,\ \ \tilde\Kc_{569}=-\fc{1}{6}\ \ ,\ \ \tilde\Kc_{579}=\fc{1}{6}\ \ ,\ \ 
\tilde\Kc_{189}=\fc{1}{6}\ \ ,\ \ \tilde\Kc_{199}=-\fc{1}{3}\ .}}

\subsec{String--theoretical K\"ahler moduli $\Tc^i$ vs.  field--theoretical fields $T^i$}

The imaginary part of the modulus $\Tc^i$, defined in \eqq \stringT, follows from the integral 
$\im(\Tc^i)=\int_{C^i} C_4$ of the Ramond $4$--form over a certain $4$--cycle $C^i$. 
A $D7$--brane has the world--volume
Chern--Simons coupling $\int\ C_4\wedge F\wedge F$. 
Hence, a $D7$--brane wrapped around this $4$--cycle
$C_i$ gives rise to the $CP$--odd gauge term $\int\ \im(\Tc^i)\ F\wedge F$ in $D=4$.
On the other hand, the real parts of the moduli $\Tc^i$, 
which derive from the underlying string background (\cf Subsection 2.3)
do not yet  properly fit into complex scalars $T^i$ of N=1 chiral 
multiplets in field theory.
According to the previous discussion, the real part $\re(T^i)$ of those scalar fields 
has to describe 
the gauge coupling of a $D7$--brane, which is wrapped around the  four--cycle $C_i$.
This coupling is measured by the volume of this $4$--cycle $C_i$.
More precisely, from the Born--Infeld action $e^{-\phi_{10}}\ \int\ d^8\xi\ 
\det(g+2\pi\ap F)^{1/2}$ 
we derive the $CP$--even gauge--coupling $\re(T^i):=e^{-\phi_{10}}\ \int_{C_i} d^4\xi\ 
\det(g)^{1/2}$.
In order for the $D7$--brane to respect $1/2$ of the supersymmetry
of the bulk theory, which is N=2 in $D=4$, the internal $4$--cycle $C_i$ the $D7$--brane
is wrapped on has to fulfill the calibration condition \Ruben:
\eqn\calibration{
e^{-\phi_{10}}\ \int_{C_i} d^4\xi\  \det(g)^{1/2}=\h\ \int_{C_i}\ J\wedge J\ .}
Note, that the r.h.s. just describes the volume of the $4$--cycle $C_i$.
Hence, the real part of the correct holomorphic modulus $T^i$ is:
\eqn\calibrationi{
\re(T^i):=e^{-\phi_{10}}\ \int_{C_i} d^4\xi\  \det(g)^{1/2}=\h\ 
\int_{C_i}\ J\wedge J\ .}
More precisely, with $\omega_i$ the Poincar\'e dual
$2$--form of the $4$--cycle $C_i$, we have:
$\int\limits_{C_i}\ J\wedge J=\int\limits_{X_6} \omega_i \wedge J\wedge J$. 
With $J=\sum\limits_{j=1}^{h_{(1,1)}}\re(\Tc^j)\ \omega_j$ we may write
\eqn\NICE{
\re(T^i):=\fc{1}{6}\ \fc{\p}{\p \re(\Tc^i)}\ \int_{X_6}\ J\wedge J\wedge J=
\fc{\p}{\p \re(\Tc^i)}\ Vol(X_6)\ ,}
which proves \NICEE.
Furthermore, with $\int_{X_6} J\wedge J\wedge J=\Kc_{ijk}\ \re(\Tc^i)\ \re(\Tc^j)\ \re(\Tc^k)$, and 
$\Kc_{ijk}$ the intersection form of the CYM $X_6$, we may also write
\eqn\ALSONICE{
\re(T^i):=\h\ \Kc_{ijk}\ \re(\Tc^j)\ \re(\Tc^k)\ ,}
which gives the volume of the $4$--cycle (in the string frame).

For the case when the K\"ahler moduli space is described by the coset space
$\lf(\fc{SU(1,1)}{U(1)}\ri)^3$ we have worked out 
the field--theoretical moduli fields $T^i$ in \eqq \OBTAIN, for the coset
$\fc{SU(2,2)}{SU(2)\times SU(2)\times U(1)}\times \fc{SU(1,1)}{U(1)}$,  
in \eqq \fieldfinally, and for the coset
$\fc{SU(3,3)}{SU(3)\times SU(3)\times U(1)}$ in \eqq  \correctth.

\subsec{Three--form flux $G_3$ in $\IZ_N$ and $\IZ_{N}\times \IZ_M$--orbifolds}

Let us now give non--vanishing vevs to some of the (untwisted) 
flux components $H_{ijk}$ and $F_{ijk}$, with $F_3=dC_2$, $H_3=dB_2$. 
The two $3$--forms $F_3,H_3$ are organized in the $SL(2,\IZ)_S$ covariant field:
\eqn\fluxcomb{
G_3=F_3+i\ S\ H_3\ .}
On the torus $T^6$, we would have 20+20 independent internal
components for $H_{ijk}$ and $F_{ijk}$. 
However, only a portion of them is invariant under the orbifold group $\Gamma$.
More precisely, of the $20$ complex (untwisted) components comprising the flux $G_3$, only 
$2h^{\rm untw.}_{(2,1)}(X_6)+2$ survive the orbifold twist. 
The orientifold action $\Om(-1)^{F_L}I_6$ producing $O3$--planes does not give rise to any 
further restrictions. If the orbifold group $\Gamma$ contains 
$\IZ_2$--elements $\th$ which leave the $j$--th complex plane fixed, we also 
encounter $O7_j$--planes transverse to the $j$--th plane. Since $I_2^j=I_6\th$, the 
orientifold generator $\Om(-1)^{F_L}I_2^j$ does not put further restrictions on the 
$2h^{\rm untw.}_{(2,1)}(X_6)+2$ twist invariant  components.
Hence, the allowed flux components are most conveniently found in the complex basis, in which
the orbifold group $\Gamma$ acts diagonally. In the following, we shall
concentrate on the orientifold/orbifolds $T^6/(\Gamma+\Gamma\Om I_6)$, with $\Gamma$ being
one of the orbifold twists $\IZ_N$ or $\IZ_N\times\IZ_M$ encountered above. 
Note that $O7$--planes appear in the case that
the orbifold twist $\Gamma$ is of even order.

The most general $3$--form flux $G_3$ on $T^6$ has 20 components, which appear 
in the expansion
\eqn\GC{{1\over{(2\pi)^2\alpha'}}\ G_3=\sum_{i=0}^3 (A^i\omega_{A_i}+B^i\omega_{B_i})+
\sum_{j=1}^6(C^j\omega_{C_j}+D^j\omega_{D_j})}
w.r.t. the complex $3$--form cohomology  
$H^3=H^{(3,0)}\oplus H^{(2,1)}\oplus H^{(1,2)}\oplus H^{(0,3)}$:
\eqn\cplxz{\eqalign{
\om_{A_0}&=d z^1\wedge dz^2\wedge d z^3\ \ \ ,\ \ \om_{B_0}=d\ov z^1\wedge d \ov z^2\wedge d \ov z^3\cr
 \om_{A_1}&=d\ov z^1\wedge
dz^2\wedge dz^3\ \ \ ,\ \ \ \om_{B_1}= dz^1\wedge d\ov z^2\wedge d\ov z^3\cr
\om_{A_2}&=dz^1\wedge
d\ov z^2\wedge dz^3\ \ \ ,\ \ \ \om_{B_2}=d\ov z^1\wedge dz^2\wedge d\ov z^3\cr
\om_{A_3}&=dz^1\wedge dz^2\wedge d\ov z^3\ \ \ ,\ \ \ \om_{B_3}= d\ov z^1\wedge d\ov z^2\wedge dz^3\cr
\om_{C_1}&=dz^1\wedge d\ov z^1\wedge dz^2\ \ \ ,\ \ \om_{D_1}=dz^1\wedge d \ov z^1\wedge d \ov z^2\cr
 \om_{C_2}&=dz^1\wedge
d\ov z^1\wedge dz^3\ \ \ ,\ \ \ \om_{D_2}= dz^1\wedge d\ov z^1\wedge d\ov z^3\cr
\om_{C_3}&=dz^1\wedge
d z^2\wedge d\ov z^2\ \ \ ,\ \ \ \om_{D_3}=d\ov z^1\wedge dz^2\wedge d\ov z^2\cr
\om_{C_4}&=dz^2\wedge d\ov z^2\wedge d z^3\ \ \ ,\ \ \ \om_{D_4}= 
d z^2\wedge d\ov z^2\wedge d\ov z^3\cr 
\om_{C_5}&=dz^1\wedge
dz^3\wedge d\ov z^3\ \ \ ,\ \ \ \om_{D_5}=d\ov z^1\wedge dz^3\wedge d\ov z^3\cr
\om_{C_6}&=dz^2\wedge dz^3\wedge d\ov z^3\ \ \ ,\ \ \ \om_{D_6}= 
d\ov z^2\wedge d z^3\wedge d\ov z^3\ .}}
The $\om_{A_i}$,  $\om_{B_i}$ correspond to flux components 
with all one--forms coming from different planes, 
while the $\om_{C_i}$, $\om_{D_i}$ are flux components 
with two one--forms coming from the same plane. 
The latter we have just written down for completeness, as they are projected out in 
all orbifolds. 
In the $\IZ_2\times \IZ_2$ orientifold/orbifold, which allows for the largest number of 
(untwisted) fluxes \LustFI, all $\om_{A_i}$ and $\om_{B_i}$ remain, whereas in most other orbifolds 
only $\om_{A_0}$ and $\om_{B_0}$ survive.
In appendix \appB\ we list the 3--form flux components, which are invariant under
the respective orbifold action $\Gamma$.
That the $(0,3)$ and $(3,0)$-flux always
survive is quite clear, as the $(3,0)$-flux corresponds to the Calabi-Yau
3-form $\Omega$, which is always present, and the $(0,3)$-flux to its
conjugate.

While in the form \GC, the cohomology structure of $G_3$ is manifest, 
in order to impose the flux quantization on $G_3$, \ie
\eqn\fluxqu{
\fc{1}{(2\pi)^2\ap}\int_{C_3} F_3 \in n_0\ \IZ\ \ \ ,\ \ \ 
\fc{1}{(2\pi)^2\ap}\int_{C_3} H_3 \in n_0\ \IZ\ ,}
with some integer $n_0$ (depending on the orbifold group $\Gamma$)  
to be specified later, one has to transform the forms \cplxz\
into a real basis of the following 20 elements
\eqn\realbase{
\eqalign{\alpha_0&=dx^1 \wedge dx^2  \wedge dx^3\ \ \ ,\ \ \ 
\beta^0=dy^1 \wedge dy^2 \wedge dy^3\ ,\cr
\alpha_1&=dy^1 \wedge dx^2  \wedge dx^3\ \ \ ,\ \ \ \beta^1=-dx^1 \wedge dy^2\wedge dy^3\ ,\cr
\alpha_2&=dx^1 \wedge dy^2  \wedge dx^3\ \ \ ,\ \ \ \beta^2=-dy^1 \wedge dx^2 \wedge dy^3\ ,\cr
\alpha_3&=dx^1 \wedge dx^2  \wedge dy^3\ \ \ ,\ \ \ \beta^3=-dy^1 \wedge dy^2 \wedge dx^3\ ,\cr
\gamma_1&=dx^1 \wedge dy^1  \wedge dx^2\ \ \ ,\ \ \ \delta^1=-dy^2 \wedge dx^3 \wedge dy^3\ ,\cr
\gamma_2&=dx^1 \wedge dy^1  \wedge dx^3\ \ \ ,\ \ \ \delta^2=-dx^2 \wedge dy^2 \wedge dy^3\ ,\cr
\gamma_3&=dx^1 \wedge dx^2  \wedge dy^2\ \ \ ,\ \ \ \delta^3=-dy^1 \wedge dx^3\wedge dy^3\ ,\cr
\gamma_4&=dx^2 \wedge dy^2  \wedge dx^3\ \ \ ,\ \ \ 
\delta^4=-dx^1 \wedge dy^1 \wedge dy^3\ ,\cr
\gamma_5&=dx^1 \wedge dx^3  \wedge dy^3\ \ \ ,\ \ \ \delta^5=-dy^1 \wedge dx^2 \wedge dy^2\ ,\cr
\gamma_6&=dx^2 \wedge dx^3  \wedge dy^3\ \ \ ,\ \ \ \delta^6=-dx^1 \wedge dy^1\wedge dy^2\ ,}}
with the six real periodic coordinates $x^i,y^i$ on the torus $T^6$, \ie 
$x^i\cong x^i+1$ and $y^i\cong y^i+1$. The basis \realbase\ has the property $\int_{X_6} 
\alpha_i \wedge \beta^j=\delta^j_i,\ \int_{X_6} \gamma_i \wedge \delta^j=\delta^j_i$,
with the choice of orientation $\int_{X_6}dx^1\wedge dx^2\wedge dx^3\wedge dy^1\wedge dy^2\wedge dy^3=1$.
In real notation, the flux has the form:
\eqn\GR{{1\over{(2\pi)^2\alpha'}}{G_3}=\sum_{i=0}^{3} \lf[(a^i+iS
c^i)\alpha_i+(b_i+iS d_i)\beta^i\ri]+\sum_{j=1}^{6} \lf[(e^j+iS
g^j)\gamma_j+(f_j+iS h_j)\delta^j\ri]\ .}
In this basis, the  $SL(2,\IZ)_S$--covariance of $G_3$ is manifest. 
The coefficients $a^i,\ b_i,\ e^i,\ f_i$ refer to the Ramond part of $G_3$, whereas the
coefficients $c^i,\ d_i,\ g^i,\ h_i$ refer to the Neveu-Schwarz part.

To pass from the complex basis \cplxz\ to the real basis \realbase, one introduces complex
structures, \ie the complex coordinates (\cf \ansatz):
\eqn\achieved{
dz^j=\sum_{i=1}^3\ \rho^j_i\ dx^i+ \tau^j_i\ dy^i\ \ \ ,\ \ \ j=1,2,3\ .}
Most of the parameters $\rho^j_i$ and $\tau^j_i$ are fixed through the orbifold twist $\Gamma$,
with  only those remaining undetermined, which correspond to the $\IZ_2$--elements of $\Gamma$.
The latter are eventually fixed through the flux quantization condition (\cf appendix \appB).
As we shall see in a moment, the  specific values of the constants $\rho^j_i$ and $\tau^j_i$
are relevant for finding flux solutions.

Let us briefly comment on the integers $n_0$, introduced in the flux quantization conditions
\fluxqu.
It has been pointed out in {\it Ref.} \FP, that there are subtleties for toroidal orientifolds  
due to additional $3$--cycles, which are not present in the covering space $T^6$.
If some integers are odd, additional discrete flux has to be
turned on in order to meet the quantization rule for those $3$--cycles.
We may bypass these problems in the $\IZ_N$ ($\IZ_N\times \IZ_M$)--orientifolds, 
if we choose the quantization numbers to be multiples of $n_0=2N$ ($n_0=2NM$)  
and do not allow for discrete flux at the orientifold planes \threeref\BLT\CU\FontCY.
Note, that for $h_{(2,1)}^{\rm twist.}\neq 0$, 
in addition to the untwisted flux components $H_{ijk}$ and $F_{ijk}$ there may
be also $NSNS$-- and $RR$--flux components from the twisted sector. We do not
consider them here. It is assumed, that their quantization rules freeze the blowing up 
moduli at the orbifold singularities.

To illustrate the above procedure,  we shall discuss the $\IZ_{6-II}$ orbifold with the 
lattice $(SU(2))^2\times SU(3)\times G_2$ and present the fluxes compatible with the complex 
structures of this orbifold.
We will again present only one example, while the other orbifolds  are treated in appendix \appB.

At this level, no supersymmetry conditions are imposed. 
Imposing further conditions will fix $S$ and the complex structure moduli 
(in case they are present in the particular orbifold) and/or constrain the coefficients 
$a_i,\,b_i,\,c_i,\,d_i$ which are real integers.

The $\IZ_{6-II}$ orbifold on the 
lattice $(SU(2))^2\times SU(3)\times G_2$ is a case with one complex structure modulus $\Uc^3$ left 
unfixed, therefore the flux takes the form\foot{In the remainder of this section and 
in the two appendices \appA, \appB\ the 
real and imaginary parts of the complex structure moduli $\Uc$ are exchanged compared to the 
previous setup.}
$${1\over (2\pi)^2\alpha'}\,G_3=A_0\,\om_{A_0}+A_3\,\om_{A_3}+B_0\,\om_{B_0}+B_3\,\om_{B_3}.$$
The $(3,0)$--form on this orbifold takes the form\foot{Contrary to the normalization of $dz^3$ 
used in the context of the linear $\sigma$--model we do not use the 
prefactor $1/\sqrt{{\rm Im}\,{\cal U}^3}$ here in order to obtain a holomorphic superpotential.}
\eqn\threezerosixi{\eqalign{
\om_{A_0}=&{1\over3}\{ 3\,\alpha_0+\sqrt3\,e^{5\pi i/6}\,\alpha_1+3\,e^{2\pi i/3}\,\alpha_2\cr
&+{\cal U}^3\,[\, 3\,\alpha_3-i(\sqrt3\,\beta_0+3\,e^{\pi i/6}\,\beta_1+\sqrt3\,e^{2\pi i/6}\,
\beta_2)]+{i\sqrt3}\beta_3\,\}.
}}
The one $(2,1)$--form surviving the twist takes the form
\eqn\twoone{\eqalign{
\om_{A_3}=&{1\over3}\{ 3\, \alpha_0+\sqrt3\,e^{5\pi i/6}\,\alpha_1+3\,e^{2\pi i/3}\,\alpha_2\cr
&+\ov{\cal U}^3\,[\,3\, \alpha_3-i(\sqrt3\,\beta_0+3\,e^{\pi i/6}\,\beta_1+\sqrt3\,e^{2\pi i/6}\,
\beta_2)]+{i\sqrt3}\,\beta_3\,\}.
}}
$\om_{B_0}$ and $\om_{B_3}$ are the complex conjugates of the above.
For the complex coefficients we find
\eqn\cocoeffi{\eqalign{
A_0=&{1\over2\,{{\rm Im}\,{\cal U}^3}}\left\{\,e^{2\pi i/12}b_0-ib_2+iS\,(e^{2\pi i/12}d_0-i\,d_2)
\right.\cr
&\left.+\ov{\cal U}^3\,\left[\,{1\over\sqrt3}\,e^{2\pi i/6}\,a_0+{1\over\sqrt3}\,a_2+
iS\,(-{\sqrt3}\,e^{2\pi i/6}\,c_0+{1\over\sqrt3}\,c_2)\right]\right\},\cr
B_0=&{1\over2\,{{\rm Im}\,{\cal U}^3}}\left\{\,e^{-2\pi i/12}b_0+ib_2+iS\,(e^{-2\pi i/12}\, 
d_0+i\,d_2)\right.\cr
&\left.+{\cal U}^3\,\left[\,{1\over\sqrt3}\,e^{-2\pi i/6}\,a_0+{1\over\sqrt3}\,a_2+iS\,
({\sqrt3}\,e^{-2\pi i/6}\,c_0+{1\over\sqrt3}\,c_2)\right]\right\},\cr
A_3=&{1\over2\,{{\rm Im}\,{\cal U}^3}}\left\{\,e^{-10\pi i/12}b_0+ib_2+iS\,(e^{-10\pi i/12}\, 
d_0+i\,d_2)\right.\cr
&\left.-{\cal U}^3\,\left[\,{1\over\sqrt3}\,e^{2\pi i/6}\,a_0+{1\over\sqrt3}\,a_2+iS\,
({\sqrt3}\,e^{2\pi i/6}\,c_0+{1\over\sqrt3}\,c_2)\right]\right\},\cr
B_3=&{1\over2\,{{\rm Im}\,{\cal U}^3}}\left\{\,e^{10\pi i/12}b_0-ib_2+iS\,(e^{10\pi i/12}\, 
d_0-i\,d_2)\right.\cr
&\left.-\ov{\cal U}^3\,\left[\,{1\over\sqrt3}\,e^{-2\pi i/6}\,a_0+{1\over\sqrt3}\,a_2+iS\,
(\sqrt3\,e^{-2\pi i/6}\,c_0+{1\over\sqrt3}\,c_2)\right]\right\}\ .}}
Note that the normalization of the 3--forms is 
$\int \om_{A_0}\wedge\om_{B_0}=2i\,{\rm Im}\,{\cal U}^3$.
Expressed in the real 3--forms, the flux takes the form
\eqn\realtwelveii{\eqalign{
{1\over (2\pi)^2\alpha'}\,G_3=&(a^0+iS\,c^0)\,\alpha_0+{1\over 3}\,(-a^0+a^2-iS\,
(c^0-c^2))\,\alpha_1\cr
&+(a^2+iS\,c^2)\,\alpha_2+(-b_0+2\,b_2+iS\,(-d_0+2\,d_2))\,\alpha_3+(b_0+iS\,d_0)\,\beta^0\cr
&+(b_0+b_2+iS\,(d_0+d_2))\,\beta^1+(b_2+iS\,d_2)\,\beta^2+
{1\over3}\,(a^0+2\,a^2+iS\,(c^0+2\,c^2))\,\beta^3]\ .}}

\newsec{Vacuum structure of orientifolds in the orbifold limits}

In this section, we investigate  the vacuum structure of \tb orientifold compactifications
in their orbifold limits. 
The discussion is based on the following effective N=1
superpotential
\eqn\superpot{
 W=W_{\rm flux}(S,U^j)+W_{\rm np}(T^i)\ ,}
with:
\eqn\WITH{\eqalign{
W_{\rm flux}(S,U^j)&=\fc{\lambda}{(2\pi)^2\alpha '}\int_{X_6}  G_3\wedge \Omega\ ,\cr
W_{\rm np}(T^i)&=\sum_{i=1}^{h_{(1,1)}(X_6)}  
{g^i} \, e^{-h^iT^i}\ \ \ , \ \ \ g^i \in  \IC,\ h^i\in 
\IR^+\ .}}
The first term is the perturbative contribution to the superpotential
due to non-vanishing $3$--form fluxes \TV, 
and it depends on the dilaton field $S$ and, if present,
also on the untwisted complex structure moduli $U^j$
(with the normalization $\kappa_{10}^{-2}=\fc{\lambda}{(2\pi)^2\alpha '}$).
The second term is
of non-perturbative nature and depends on the untwisted K\"ahler moduli
$T^i$. At this stage, we do not discuss any further the possible
microscopic origin of the non--perturbative superpotential, but are only 
interested in a rather phenomenological way in its effect on the
ground state structure of the orbifold models. Later in section 4, we shall 
provide some more details on the non--perturbative effects which may
cause such a non-perturbative superpotential, namely wrapped Euclidean
$D3$-branes and/or gaugino condensates. As a result, we will see that a
non--perturbative superpotential of this type may emerge
if also effects from blowing up modes of the orbifold singularities
are taken into account. Hence this superpotential that depends only
on the untwisted moduli $T^i$ can be regarded as a certain truncation
of a more complete non-perturbative superpotential.
This and several other aspects will be also the subject of a forthcoming
publication \future.

The vacua of the effective N=1 supergravity theory are determined
by the associated scalar potential \WessCP
\eqn\scalarpot{
V=e^{\kappa^2_4 K}\Biggl(|D_{S}W|^2+\sum_{i=1}^{h_{(1,1)}(X_6)}|D_{T^i}W|^2+
\sum_{j=1}^{h_{(2,1)}(X_6)}|D_{U^j}W|^2-3\ |W|^2\Biggr)\ ,}
with the K\"ahler potential for the fields $S,T^j,U^j$.
During the process of minimizing $V$, the following two aspects will
become important:
first, the supersymmetry conditions $D_{S,T^i,U^j}W=0$ will imply that
generic
supersymmetric AdS ground states are described by
fluxes which are not anymore $ISD$ with only
$G_{2,1}$ components, but rather will include $G_{0,3}$ and
also all $IASD$ (imaginary
anti self-dual) types as well. 
The second issue concerns the stability of the obtained 
extrema after imposing the supersymmetry conditions.
As stated already in the introduction, in the framework of
the KKLT scenario, one has to require the absence of any tachyonic
scalar fields, i.e. the (mass)$^2$ of all scalars must be positive.
This means that all eigenvalues of the scalar field mass matrix
${\partial ^2V\over\partial \phi_\alpha\partial \ov\phi_\beta}$ 
($\phi_\alpha,\phi_\beta=S,U^j,T^i$) must be
positive. As we will see, this requirement 
can be only satisfied by those orbifolds which contain untwisted
complex structure
moduli $U^j$. In this way, we derive
some severe constraints
on which orbifolds can lead to stable vacua.
This result is contrasted by the
procedure originally applied in KKLT, where first the dilaton field $S$ and
the complex structure moduli were integrated out by solving
the flux supersymmetry conditions
$D_SW_{\rm flux}=D_{U^j}W_{\rm flux}=0$
using $ISD$ $(2,1)$-- or $(0,3)$--fluxes, and then plugging the obtained
values for $S$ and $U^j$ back into $W$. This leads to a constant term $W_0$.
However,  the integrating-out procedure is in addition only
consistent, if the masses of the integrated-out fields $S$ and
$U^j$ are heavy compared to the K\"ahler moduli $T^i$. Otherwise,
the results on the vacuum structure and especially what concerns
the stability problems are misleading and cannot be trusted anymore.

This problem has been emphasized and thoroughly discussed recently in {\it Ref.} \ChoiSX.
In this section, we want to generalize this discussion into several directions.
First,  we discuss under what conditions  stable minima may be found if all moduli are 
minimized at
once without first integrating out the complex structure moduli. This way, in subsection 3.5 
we find 
a stable minimum for the case $h_{(1,1)}^{\rm untw.}=3$ and $h_{(2,1)}^{\rm untw.}=1$.
On the other hand, in {\it Ref.} \ChoiSX\ it is has been proven that this case would not lead to 
a stable minimum, if the complex structure modulus was integrated out first.
Secondly, in subsections 3.3 and
3.4, we shall investigate the KKLT scenario in toroidal orbifolds for  more than 
one K\"ahler modulus and more general K\"ahler potentials (\cf \CosetsKAE) at fixed complex
structure modulus. We find, that in those cases no stable minimum is possible generalizing the
one K\"ahler modulus case discussed in \ChoiSX. This result rules out 
all toroidal orbifold limits with only K\"ahler moduli for a KKLT scenario, as \eg the 
$\IZ_7$--orbifold.
In fact, going beyond the orbifold limit in {\it Ref.} \future,   
we will give the criteria under which a KKLT scenario may
be possible with only K\"ahler moduli.
Furthermore, in subsection 3.6, we find a more general effective superpotential
(compared to the ones discussed in \ChoiSX)   
after integrating out several complex structure moduli.
Finally, the conditions and solutions for the extrema are presented.

\subsec{Supersymmetry conditions}

In this subsection, we shall study the SUSY conditions for the $\IZ_2\times \IZ_2$ orientifold, 
with $h_{(1,1)}^{\rm untw.}=h_{(2,1)}^{\rm untw.}=3$. They read
\eqn\Di{
D_i {W}\equiv\partial_i {W}+\kappa_4^2\ {W}\ \partial_i K=0 \ ,\  \quad i=S,U^i,T^i}
and allow us to explore the Hodge structure of the flux $G_3$ in the supersymmetric case. 
The K\"ahler
potential for the dilaton $S$ and K\"ahler moduli $T^i$ is given in \Simply, while for the complex
structure moduli $U^i$ it may be read off from \CosetsKAE. With the superpotential \superpot\
the conditions \Di\ lead to: 
\eqn\susyeq{\eqalign{
D_{T^i}{W}&=0 \Longrightarrow \fc{\lambda}{(2\pi)^2\alpha '}\int 
 G_3\wedge\Omega =-\left(T^i+\bar T^i\right) {g^i} h^ie^{-h^iT^i}-\sum_{j=1}^3  
{g^j} e^{-h^jT^j}\ , \quad i=1,2,3 \ ,\cr
D_S W&=0 \Longrightarrow \fc{\lambda}{(2\pi)^2\alpha '}
\int \bar G_3 \wedge \Omega=-\sum_{i=1}^3 {g^i} e^{-h^iT^i} \ ,\cr
D_{U^i}{W}&=0\Longrightarrow \fc{\lambda}{(2\pi)^2\alpha '}
\int  G_3\wedge\omega_{Ai} =-\sum_{j=1}^3 {g^j} e^{-h^jT^j} \ , \quad i=1,2,3 \ .}}
After writing $G_3$ in the complex basis (\cf \eqq \GC)
\eqn
\cbasis{
\fc{1}{(2\pi)^2\alpha '}\ G_3=\sum_{i=0}^3\left(A^i\omega_{Ai}+B^i\omega_{Bi}\right) \ ,}
where $\omega_{A0}$ is a $(3,0)$-form, $\omega_{Ai}$ are $(2,1)$-forms, 
$\omega_{B0}$ is a $(0,3)$-form and  $\omega_{Bi}$ are $(1,2)$-forms, 
we obtain from \susyeq:
\eqn\abloes{\eqalign{
B^0&=-\fc{1}{\lambda\prod\limits_{k=1}^3\left(U^k+\bar U^k\right)}\ \left[\left(T^i+\bar T^i\right) 
 {g^i} h^i e^{-hT^i}+\sum_{j=1}^3  {g^j} e^{-h^jT^j}\right] \ , \qquad i=1,2,3 \ ,\cr
A^0&=-\fc{1}{\lambda}\ \fc{\sum\limits_{i=1}^3 {g^i} e^{-h^i\bar T^i}}{\prod\limits_{j=1}^3
\left(U^j+\bar U^j\right)}\ \ \ ,\ \ \ 
B^i=-\fc{1}{\lambda}\ \fc{\sum\limits_{k=1}^3 {g^k} e^{-h^k T^k}}{\prod\limits_{j=1}^3
\left(U^j+\bar U^j\right)}\ , \qquad i=1,2,3 \ .}}
Here we have used $\int_{X_6} \omega_{A_0}\wedge\omega_{B_0}=\int_{X_6} \omega_{B_i}\wedge\omega_{A_i}=\prod\limits_{k=1}^3 (U^k+\ov U^k)$.
We see that in the presence of the non--perturbative term 
the $(1,2),(0,3)$ and $(3,0)$--components of the flux are no longer vanishing.
Next, with the formula \WessCP
\eqn\FTERMS{
\ov F^{\ov I}=e^{\kappa_4^2/2\ K}\ K^{\ov I J}\ (\p_J W+\kappa_4^2\ W\ \p_J K)}
we present  the $F$-terms:
\eqn\fterms{\eqalign{
\bar F^{\bar S}=&\left(S+\bar S\right)^{\fc12}\prod_{i=1}^3\left(T^i+\bar T^i\right)^{-\fc12}
\prod_{j=1}^3\left(U^j+\bar U^j\right)^{-\fc12}\kappa^2_4\left(\fc{\lambda}{(2\pi)^2\alpha '}
\int \bar G_3\wedge \Omega +\sum_{k=1}^3 {g^k} e^{-h^kT^k}\right)\ ,\cr
\bar F^{\bar U ^i}=&\left(S+\bar S\right)^{-\fc12}\left(U^i+\bar U^i\right)^{\fc12}
\left(U^j+\bar U^j\right)^{-\fc12}\left(U^k+\bar U^k\right)^{-\fc12}\prod_{l=1}^3
\left(T^l+\bar T^l\right)^{-\fc12}\times\cr
&\times \kappa_4^2\ \left(\fc{\lambda}{(2\pi)^2\alpha '}\int  G_3\wedge \omega_{Ai} 
+\sum_{m=1}^3 {g^m} e^{-h^mT^m}\right)\ ,\cr
\bar F^{\bar T ^i}=&\left(S+\bar S\right)^{-\fc12}\left(T^i+\bar T^i\right)^{\fc12}
\left(T^j+\bar T^j\right)^{-\fc12}\left(T^k+\bar T^k\right)^{-\fc12}\prod_{l=1}^3
\left(U^l+\bar U^l\right)^{-\fc12}\times\cr
&\times\kappa_4^2\ 
\left[W_{\rm flux}+\left(T^i+\bar T^i\right)\  {g^i} h^i\ e^{-h^iT^i}\right]\ , 
\quad i\neq j \neq k \ .}}
With \WessCP
\eqn\POTENT{
V=K_{I\ov J}\ F^I\ \ov F^{\ov J}-3\ e^{\kappa_4^2\ K}\ \kappa_4^2\ |W|^2}
the potential becomes:
\eqn\pot{\eqalign{
{V}=&\kappa^2_4 \Bigg(|S+\bar S|\prod_{j=1}^3(T^j+\bar T^j)\prod_{k=1}^3(U^k+\bar U^k)\Bigg)^{-1}
\times \cr
&\times\Bigg\{\sum_{i=1}^3\left|W_{\rm flux}+(T^i+\bar T^i)\ 
{g^i} h^i\ e^{-h^iT^i}\right|^2+
\left|\ \fc{\lambda}{(2\pi)^2\alpha '}\int \bar G_3\wedge \Omega+\sum_{l=1}^3\ 
{g^l}\ e^{-h^lT^l}\ \right|^2\cr
&+\sum_{l=1}^3\left|\ \fc{\lambda}{(2\pi)^2\alpha '}\int G_3\wedge \omega_{Al}+\sum_{m=1}^3 
{g^m} \ e^{-h^mT^m}\ \right|^2-3\ |{W}|^2\Bigg\} \ .}}
Using \cbasis\ we can rewrite the potential as:
\eqn\porreal{\eqalign{
{V}=&\kappa^2_4\left(\left|S+\bar S\right|\prod_{i=1}^3\left|T^i+\bar T^i\right|\prod_{j=1}^3
\left|U^j+\bar U^j\right|\right)^{-1} \Bigg\{-3\left|\ B^0\lambda\prod_{l=1}^3(U^l+\bar U^l)+
\sum_{l=1}^3  {g^l} e^{-h^lT^l}\ \right|^2\cr
&+\sum_{k=1}^3\left|\ B^0\lambda\ \prod_{l=1}^3(U^l+\bar U^l)+\sum_{l=1}^3 \ {g^l}\ e^{-h^lT^l}+
(T^k+\bar T^k) {g^k} h^ke^{-h^kT^k}\ \right|^2\cr
&+\left|\ \lambda\ \prod_{n=1}^3(U^n+\bar U ^n)\bar A^0+\sum _{p=1}^3 \ {g^p}\ 
e^{-h^pT^p}\ \right|^2\cr
&+\sum_{r=1}^3\left|\ \lambda\ \prod_{n=1}^3(U^n+\bar U ^n)B^r+\sum _{p=1}^3\  
{g^p}\ e^{-h^pT^p}\ \right|^2\Bigg\}.}}
In the supersymmetric case, \ie $F^S=F^{U^j}=F^{T^i}=0$, the potential reduces to:
\eqn\potsusy{
V_0=-3\ \kappa_4^2\ {\left|\ B^0\lambda\ 
\prod\limits_{l=1}^3(U^l+\bar U^l)+\sum\limits_{l=1}^3  {g^l}\  e^{-h^lT^l}\ \right|^2 \over 
\left|S+\bar S\right|\prod\limits_{i=1}^3\left|T^i+\bar T^i\right|\prod\limits_{j=1}^3
\left|U^j+\bar U^j\right|}\ .}
Next, we plug the superpotential \superpot\ (\cf also \LustFI\ for $W_{\rm flux}$)
\eqn\super{
\eqalign{
W=& \left( a^0+iSc^0\right)U^1U^2U^3-\{ \left( a^1+iSc^1\right)U^2U^3+ \left( a^2+iSc^2\right)U^1U^3+ 
\left( a^3+iSc^3\right)U^1U^2\}\cr
	&-\sum_{i=3}^{3} 
\left( b_i+iSd_i\right)U^i- \left( b_0+iSd_0\right)+ \sum_i{g^i} e^{-h^iT^i}}}
into {\it Eqs.}  \susyeq.
The equations become:
\eqn\system{\eqalign{
0=&\bar U^1\bar U^2\bar U^3 \left( a^0+iSc^0\right)-\sum_{i\neq j\neq k} \left( a^i+iSc^i\right)\bar U^j\bar U^k- \left( b_0+iSd_0\right)\cr
&-\sum_{i=1}^3 \left( b_i+iSd_i\right)\bar U^i+\sum_{i=1}^3{g^i}e^{-h^i\bar T^i} \ ,\cr
0=&U^1U^2U^3 \left( a^0+iSc^0\right)-\sum_{i\neq j\neq k} \left( a^i+iSc^i\right)U^jU^k- \left( b_0+iSd_0\right)\cr
&-\sum_{i=1}^3 \left( b_i+iSd_i\right)U^i+\sum_{j=1}^3g^j\ 
e^{-h^jT^j}+{g^i}h^i \left( T^i+\bar T^i\right)e^{-h^iT^i}\ , i=1,2,3 \ ,\cr
0=&\bar U^1U^2U^3 \left( a^0+iSc_0\right)-\{ \left( a^1+iSc^1\right)U^2U^3+ \left( a^2+iSc^2\right)\bar U^1U^3+ \left( a^3+iSc^3\right)\bar U^1U^2\}\cr
&- \left( b_0+iSd_0\right)-\{ \left( b_1+iSd_1\right)\bar U^1+ \left( b_2+iSd_2\right)U^2+ \left( b_3+iSd_3\right)U^3\}+\sum_{i=1}^3{g^i}e^{-h^iT^i} \ ,\cr
0=&U^1\bar U^2U^3 \left( a^0+iSc^0\right)-\{ \left( a^1+iSc^1\right)\bar U^2U^3+ \left( a^2+iSc^2\right)U^1U^3+ \left( a^3+iSc^3\right)U^1\bar U^2\}\cr
&- \left( b_0+iSd_o\right)-\{ \left( b_1+iSd_1\right)U^1+ \left( b_2+iSd_2\right)\bar U^2+ \left( b_3+iSd_3\right)U^3\}+\sum_{i=1}^3{g^i}e^{-h^iT^i} \ ,\cr
0=&U^1U^2\bar U^3 \left( a^0+iSc^0\right)-\{ \left( a^1+iSc^1\right)U^2\bar U^3+ \left( a^2+iSc^2\right)U^1\bar U^3+ \left( a^3+iSc^3\right)U^1U^2\}\cr
&- \left( b_0+iSd_0\right)-\{ \left( b_1+iSd_1\right)U^1+ \left( b_2+iSd_2\right)U^2+ \left( b_3+iSd_3\right)\bar U^3\}+\sum_{i=1}^3{g^i}e^{-h^iT^i} \ .
}}
These are the equations to be satisfied at the supersymmetric point of the moduli space.

\subsec{Orientifolds without complex structure modulus}

Let us now discuss the vacuum structure of orientifold compactifications
without any complex structure moduli, \ie  $h_{(2,1)}=0$. So the moduli fields which
we want to determine by the supersymmetry conditions are the dilaton $S$
and the K\"ahler moduli $T^i$ ($i=1,\dots ,h^{\rm untw.}_{(1,1)}$).
Since $\omega_{A_0}$ and $\omega_{B_0}$ are the only non-trivial
$3$-forms, the flux $G_3$, expressed in the complex basis, reads:
\eqn\gthree{
{1\over (2\pi)^2\alpha'}~G_3=G_{(3,0)}+G_{(0,3)}=A^0(S)~\omega_{A_0}+
B^0(S)~\omega_{B_0}\, .}
$B^0(S)$ is a linear function in $S$ with complex coefficients $B_1^0$, $B_2^0$:
\eqn\anull{
B^0(S)=B_1^0-iS\,B_2^0\, .}
The precise form of the $B_K^0$ ($K=1,2$) depends on the considered orbifold, as we will
discuss in the following. However, the other flux coefficient $A^0(S)$ is  not anymore
an independent function, but it is given as
\eqn\bnull{
A^0(S)=\bar B_1^0+iS\,\bar B_2^0\, .}
The flux superpotential which contains the contribution from the
$G_3$ flux as well as the non-perturbative K\"ahler moduli dependent term,
is given in \eqq \superpot.
Inserting $G_3$ of \eqq \gthree, ${W}$ becomes:
\eqn\supog{
{W}=\lambda\ (B_1^0-iS\,B_2^0)+\sum_{i=1}^3g~e^{-h^iT^i}\, .}
This superpotential is of the same structure as the superpotential
discussed in \ChoiSX\ (see section 3.1 in that paper).
The main difference to the superpotential of \ChoiSX\
is that here, the coefficients $B_1^0$ and $B_2^0$ have a microscopic
explanation in terms of $3$--form flux quantum numbers. It follows that
these coefficients are integer-valued. Hence the flux quantization will
put some additional constraints on the allowed solutions of the supersymmetry
equations.

Let us  consider in more detail the $\IZ_3\times \IZ_3$ orbifold.
Here the complex flux coefficients read (see appendix \appBvii):
\eqn\ai{
B_1^0={1\over \sqrt 3}\ (i\,a^1+e^{-5\pi i/6}\,b_1)\, ,\quad
B_2^0={1\over \sqrt3}\ (i\,c^1+e^{-5\pi i/6}\,d_1)
\, ,\quad a^1,b_1,c^1,d_1\in \IZ\, .}
In order to determine the exact form of the 
flux part of the superpotential, we also
need the prefactor $\lambda$. For the $\IZ_3\times \IZ_3$ orbifold
it takes the value $\lambda=i\sqrt 3$ \FontCY.

Now we may determine the solutions of the two supersymmetry
conditions $D_T{W}=0$ and $D_S{W}=0$. We may essentially follow
the procedure outlined in \ChoiSX.
We shall consider the simplified case where all K\"ahler moduli $T^i$ are
identified, \ie $T^i=T$, and also $h^i=h$.
Now observe that via a field redefinition in $T$, namely a constant shift
in ${\rm Im}~T$,
the coefficient $g$ can always be chosen to be real.
Similarly one can shift ${\rm Im}~S$, such that
$i\sqrt 3 B_1^0$ is real. So we choose $b_1=0$ in eq.\ai.
For simplicity we also choose $i\sqrt 3 B_2^0$ to be real.
Taking all this into account, the superpotential \supog\ becomes:
\eqn\supoga{
{W}=-a^1+\fc{\sqrt{3}}{2}d_1\ S+3g~e^{-hT}\, , \quad a^1,d_1\in\IZ\ .}
As in \ChoiSX, we may restrict the analysis to the case where the
moduli $S$ and $T$ are purely real, i.e. $T=t$ and $S=s$.
Then the two supersymmetry conditions provide the following two
constraints on $s$ and $t$:
\eqn\susynilles{
a^1=g~e^{-ht}(ht+3)\, ,\quad {d_1\over a^1}=-{2ht\over\sqrt{3} s(ht+3)}\, .}
Since $e^{-ht}(ht+3)\leq3$, it follows that the first equation has only
solutions for integer values of $a^1$, if the parameter
$|g|\geq 1/3$. In fact due
to charge quantization, for any given $|g|\geq 1/3$,
this equation has a finite number of allowed solutions
(for $|g|=1/3$ the solution occurs at $t=0$).
Specifically, the first equation
possesses solutions in $t$ for the following values
of the flux $a^1$:
\eqn\asol{
a^1=1,\dots ,[g']\, ,\quad g'=3g\ .}
Here we have assumed that $g>0$, otherwise $a^1<0$.
Finally, after  having solved the first constraint in \susynilles\
which fixes the modulus $t$,
the second equation does not put any further conditions on the
allowed fluxes, it possesses precisely one solution in $s$ for any given
choice of $a^1,c^1$.
Let us assume that $g$ is very large, $|g|>>|a^1|$.
Then the supersymmetry condition is solved for very large $t$.
Furthermore, if we insist on weak string coupling, \ie large $s$,
we have to demand that
$|a^1|>>|c^1|$.

As discussed in \ChoiSX, the above solutions of the two supersymmetry
conditions do not correspond to stable supersymmetric vacua,
but the supersymmetric point rather is a saddle point with instabilities
along the moduli and axionic directions.
Hence, we like to proceed to consider  orbifolds with 
more than one K\"ahler modulus and/or  complex
structure moduli in order to see whether stable supersymmetric
ground states now become possible.

\subsec{Orientifolds with three K\"ahler moduli and fixed complex structure moduli}

After having discussed the case of one K\"ahler modulus in the previous subsection, we 
now shall move on to the three K\"ahler moduli case. This case captures  \eg the
$\IZ_7$--orbifold.
We start with the following ansatz for the superpotential \superpot
\eqn\ansatzW{
W={\alpha_1}+{\alpha_2}\ S+\sum_{j=1}^3 g^j e^{-{h^j} T^j}\ ,}
with complex coefficients ${\alpha_1}=B_1^0$,\ ${\alpha_2}=-iB_2^0$,\ $g^j$ and ${h^j}>0$.
With the K\"ahler potential 
\eqn\Kaehl{
\kappa_4^2\ K=-\ln(S+\ov S)-\sum_{j=1}^3\ln(T^j+\ov T^j)}
for the closed string moduli sector
we derive the following $F$--terms:
\eqn\Fterms{\eqalign{
-(S+\ov S)^{-1/2}\ \prod\limits_{i=1}^{3}(T^i+\ov T^i)^{1/2} \ \ov F^S&=
\ {\alpha_1}-{\alpha_2} 
\ov S+\sum_{j=1}^3{g^j}\ e^{-{h^j}\ T^j}\ ,\cr
-\fc{(S+\ov S)^{1/2}\ (T^i+\ov T^i)^{1/2}\ (T^k+\ov T^k)^{1/2}}{(T^j+\ov T^j)^{1/2}}\  
\ov F^{T^j}&={h^j}\ {g^j}\ (T^j+\ov T^j)\ e^{-{h^j}\ T^j}
+{\alpha_1}+{\alpha_2}\ S\cr
&+\sum_{j=1}^3{g^j}\ e^{-{h^j}\ T^j}\ \ \ ,\ \ \ (i,j,k)=\overline{(1,2,3)}\ ,}}
and similarly for their complex conjugate $F^{T^j}$ and $F^S$. 
Demanding $F^S=0=F^{T^j}$ leads to the following relations:
\eqn\relations{
{\alpha_1}={\alpha_2}\ \lf(\ov S+\sum_{j=1}^3\fc{S+\ov S}{{h^j}\ \left(T^j+\ov T^j\right)}\ri)
\ \ \ ,\ \ \ 
{g^j}=-\fc{{\alpha_2}\ e^{{h^j}\ T^j}}{{h^j}}\ \fc{S+\ov S}{T^j+\ov T^j}\ ,\ j=1,2,3\ ,}
and their complex conjugate. These relations have to be obeyed at the extremum of the
potential. In principle, the point $(S_0,T_0^j)$ of the extremum may be determined
from these relations \relations. 
It is straightforward to calculate the scalar potential $V(S,T^j)$.
At the extremum $(S_0,T_0^j)$, its value is given by
\eqn\EXTREM{
V_0=-3\ \fc{|{\alpha_2}|^2\ (S_0+\ov S_0)}{\prod\limits_{j=1}^3 (T^j_0+\ov T^j_0)}\ .}
To determine the kind of extremum, we have to calculate the second derivatives
of the potential $V(S,T^j)$ w.r.t. the moduli fields.
It is convenient to introduce $S=s_1+is_2$ and $T^j=t_1^j+it_2^j$.
W.r.t. the parameters $s_i,t_i^j$ we find the following identities for the mixed
derivatives:
\eqn\obtainrel{
\fc{\p^2 V}{\p s_1\ \p t_2^j}=\fc{\p^2 V}{\p t_1^j\ \p s_2}=\fc{\p^2 V}{\p s_1\ \p s_2}=
\fc{\p^2 V}{\p t_1^k\ \p t_2^l}=0\ .}
On the other hand, the non--vanishing components  of the Hessian $H=\pmatrix{H_1&0\cr
0&H_2}$ are arranged in a block--form with two $4\times 4$ matrices $H_1$ and $H_2$, 
with their determinants given by:
\eqn\hessian{\eqalign{
\det H_1&=-\fc{s_1^2\ |{\alpha_2}|^8}{512\ (t_1^1t_1^2t_1^3)^6}
\lf(\ 2 + {h^1}\ {h^2}\ {h^3}\ t_1^1\ t_1^2\ t_1^3- \sum_{j=1}^3 {h^j}\ t_1^j\ \ri)\cr  
&\times\lf(\ 5 + 16\ {h^1}\ {h^2}\ {h^3}\ t_1^1\ t_1^2\ t_1^3 + 
8\ \sum_{j=1}^3 {h^j}\ t_1^j  + 6\ \sum_{i\neq j}h^i\ {h^j}\ t_1^i\ t_1^j\ \ri)\ ,\cr
\det H_2&=-\fc{{h^1}\ {h^2}\ {h^3}\ s_1^2\ |{\alpha_2}|^8}{512\ (t_1^1t_1^2t_1^3)^5}\ 
\lf(\ 27 + 16\ {h^1}\ {h^2}\ {h^3}\ t_1^1\ t_1^2\ t_1^3 
-6\ \sum_{i\neq j}h^i\ {h^j}\ t_1^i\ t_1^j\ \ri)\ .}}
The latter may become positive in a certain region of the parameter space $h^i\ t_1^i$.
In order for $H_1$ and $H_2$ to be positive definite, also
their subdeterminants have to be positive, \ie $H_{11}>0,\ H_{11}H_{22}-H_{12}^2>0$ and 
$\det\pmatrix{
H_{11}&H_{12}&H_{13}\cr
H_{12}&H_{22}&H_{23}\cr
H_{13}&H_{23}&H_{33}}>0$.
However, we find 
$$(H_1)_{11}(H_1)_{22}-(H_1)_{12}^2=-\fc{1}{32 (t_1^1)^4(t_1^2)^2(t_1^3)^2}\ 
|{\alpha_2}|^4\ (4+5{h^1}\ t_1^1)<0\ ,$$ 
and 
$$(H_2)_{11}(H_2)_{22}-(H_2)_{12}^2=-
\fc{3}{32 (t_1^1)^3(t^2_1)^2(t_1^3)^2}\ |{\alpha_2}|^4\ {h^1}\ <0$$ 
and conclude that the extremum $(S_0,T_0^j)$ is no minimum.

Hence a KKLT scenario is not possible in the $\IZ_7$--orbifold with only untwisted
K\"ahler moduli. This generalizes the results of  \ChoiSX\ for one K\"ahler modulus 
to the three K\"ahler moduli case.

\subsec{Orientifolds with five K\"ahler moduli and fixed complex structure moduli}

After having found this negative result for three K\"ahler moduli arising from orbifolds
with a diagonal Hermitian complex metric, we shall now investigate the case
with non--diagonal metric. We want to discuss the K\"ahler moduli space 
$\fc{SU(2,2)}{SU(2)\times SU(2)\times U(1)}\times \fc{SU(1,1)}{U(1)}$ with five
K\"ahler moduli. This moduli space has been parameterized in subsection 2.4 and is relevant
to the $\IZ_{6-I}$--orientifolds.
In \eqq \fieldK\ the K\"ahler potential has been given.

In order to keep formulae short we shall stick to the case $T^1=T^2$ and $T^3=T^4$,
though we have performed the analysis for all five K\"ahler moduli kept arbitrary.
So we start with the following ansatz for the superpotential \superpot
\eqn\ansatzWW{
W=\al_1+\al_2\ S+g^1 e^{-{h^1} T^1}+g^3 e^{-{h^3} T^3}+g^5 e^{-{h^5} T^5}\ ,}
with complex coefficients ${\alpha_1}$,\ ${\alpha_2}$,\ ${g^j}$ and $h^j>0$.
With the K\"ahler potential \fieldK\
\eqn\Kaehll{
\kappa_4^2\ K=-\ln(S+\ov S)-\ln(T^5+\ov T^5)-\ln\lf[(T^1+\ov T^1)^2-(T^3+\ov T^3)^2\ri]}
we calculate the $F$--terms:
\eqn\Ftermss{\eqalign{
\fc{Y^{1/2}}{(S+\ov S)^{1/2}}\ \ov F^{\ov S}&=-\al_1+\al_2\ \ov S-\sum_{j=1,3,5} g^j\ 
e^{-h^j T^j}\ ,\cr
(S+\ov S)^{1/2}\ Y^{1/2}\ \ov F^{\ov T^1}&=-\al_1-\al_2\ S-
\sum_{j=1,3,5} g^j\ e^{-h^j T^j}\cr
&-\h\ g^1h^1\ e^{-h^1 T^1}\ [(T^1+\ov T^1)^2+(T^3+\ov T^3)^2]\cr
&-g^3\ h^3\ e^{-h^3 T^3}\ (T^1+\ov T^1)(T^2+\ov T^2)\ ,\cr
(S+\ov S)^{1/2}\ Y^{1/2}\ \ov F^{\ov T^3}&=-\al_1-\al_2\ S-
\sum_{j=1,3,5} g^j\ e^{-h^j T^j}\cr
&-g^1h^1\ e^{-h^1 T^1}\ (T^1+\ov T^1)(T^2+\ov T^2)\cr
&-\h\ g^3\ h^3\ e^{-h^3 T^3}\ [(T^1+\ov T^1)^2+(T^3+\ov T^3)^2]\ ,\cr
\fc{(S+\ov S)^{1/2}\ Y^{1/2}}{(T^5+\ov T^5)}\ \ov F^{\ov T^5}&=
-\al_1-\al_2\ S-g^5h^5\ e^{-h^5 T^5}\ (T^5+\ov T^5)-\sum_{j=1,3,5} g^j\ e^{-h^j T^j}\ ,\cr}}
with $Y=(T^5+\ov T^5)\ [(T^1+\ov T^1)^2-(T^3+\ov T^3)^2]$.
Demanding $F^S=0=F^{T^j}$ leads to the following relations
\eqn\relationss{\eqalign{
\al_1&=\al_2\lf(\ov S+\fc{S+\ov S}{h^5\ (T^5+\ov T^5)}+
\fc{2(S+\ov S)\ (T^1+\ov T^1)}{h^1\ [(T^1+\ov T^1)^2-(T^3+\ov T^3)^2]}-
\fc{2(S+\ov S)\ (T^3+\ov T^3)}{h^3\ [(T^1+\ov T^1)^2-(T^3+\ov T^3)^2]}\ri),\cr
g^1&=-2\ \fc{\al_2\ e^{h^1 T^1}}{h^1}\ \fc{(S+\ov S)\ (T^1+\ov T^1)}
{(T^1+\ov T^1)^2-(T^3+\ov T^3)^2}\ \ \ ,\ \ \ 
g^3=2\ \fc{\al_2\ e^{h^3 T^3}}{h^3}\ \fc{(S+\ov S)\ (T^3+\ov T^3)}
{(T^1+\ov T^1)^2-(T^3+\ov T^3)^2}\ ,\cr
g^5&=-\fc{\al_2\ e^{h^5\ T^5}}{h^5}\ \fc{S+\ov S}{T^5+\ov T^5}\cr}}
to be satisfied at the extremum.
At the extremum $(S_0,T_0^j)$ the scalar potential takes the value:
\eqn\EXTREMUMM{
V_0=-3\ |\al_2|^2\ \fc{(S_0+\ov S_0)}{(T^5_0+\ov T^5_0)\ 
[(T^1_0+\ov T^1_0)^2-(T^3_0+\ov T^3_0)^2]}\ .}
The scalar masses $m_S,m_{T^j}$ are encoded in the Hessian of the scalar potential
$V(S,T^1,T^3,T^5)$. After introducing $S=s_1+is_2$ and $T^j=t_1^j+it_2^j$ 
we calculate the Hessian $H$ of the potential w.r.t. to the eight real variables 
$s_1,s_2,t_1^j,t_2^j\ ,\ j=1,3,5$. 
This gives an $8 \times 8$ matrix, which has to be positive definite in
order to guarantee for positive scalar masses. This means, that all upper left matrices
of $H$ must have positive determinants.
Similarly as before, we determine:
\eqn\obtainrell{
\fc{\p^2 V}{\p s_1\ \p t_2^j}=\fc{\p^2 V}{\p t_1^j\ \p s_2}=\fc{\p^2 V}{\p s_1\ \p s_2}=
\fc{\p^2 V}{\p t_1^k\ \p t_2^l}=0\ .}
Hence, the Hessian decomposes into two $4\times 4$ blocks, which individually have to
be positive definite.
However, at the extremum $(S_0,T^j_0)$ we find the following:
\eqn\findfollowing{\eqalign{
\fc{\p^2 V}{\p s_2^2}&=\fc{|\al_2|^2}{8}\ \fc{1}{s_1t_1^5\ [(t_1^1)^2-(t_1^3)^2]}>0\ ,\cr
\det\pmatrix{\fc{\p^2 V}{\p (s_2)^2}&\fc{\p^2 V}{\p s_2 \p t_2^1}\cr
                        \fc{\p^2 V}{\p t_2^1 \p s_2 }&\fc{\p^2 V}{\p (t_2^1)^2}}&=
 -\fc{|\al_2|^4\ h^1t_1^1}{16}\ \fc{(3+2\ h^1t^1_1)}{(t_1^5)^2\ [(t_1^1)^2-(t_1^3)^2]^3}<0\ ,\cr
\det\pmatrix{\fc{\p^2 V}{\p (s_2)^2}&\fc{\p^2 V}{\p s_2 \p t_2^1}&\fc{\p^2 V}{\p s_2 \p t_2^3}\cr
      \fc{\p^2 V}{\p t_2^1 \p s_2 }&\fc{\p^2 V}{\p (t_2^1)^2}&\fc{\p^2 V}{\p t_2^1 \p t_2^3}\cr
      \fc{\p^2 V}{\p t_2^3 \p s_2 }&\fc{\p^2 V}{\p t_2^3 \p t_2^1}&\fc{\p^2 V}{\p (t_2^3)^2}}&=
 -\fc{|\al_2|^6\ h^1h^3\ s_1\ t_1^1\ t_1^3}{32}\ \fc{(3+2\ h^1t^1_1)\ (3+2\ h^3t^3_1)}
{(t_1^5)^3[(t_1^1)^2-(t_1^3)^2]^5}<0\ .}}
Since $(t_1^1)^2-(t_1^3)^2>0$  (\cf subsection 2.4) and $h^j>0$, we conclude, that there 
is no parameter range $g^j,h^j$, for which our exremum becomes a minimum.
This continues to hold for  all five K\"ahler moduli $T^j$ kept arbitrary.
In that case in the Hessian we encounter {\it e.g.} the submatix
\eqn\encounter{
\det\pmatrix{\fc{\p^2 V}{\p (s_2)^2}&\fc{\p^2 V}{\p s_2 \p t_2^1}\cr
                        \fc{\p^2 V}{\p t_2^1 \p s_2 }&\fc{\p^2 V}{\p (t_2^1)^2}}=
 -\fc{3\ |\al_2|^4\ h^1t_1^2}{32}\ 
\fc{h^1t^1_1}{(t_1^5)^2\ (t_1^1t_1^2-t_1^3t_1^4)^3}<0\ ,}
which is negative--definite for $t_1^1t_1^2-t_1^3t_1^4>0$ (\cf subsection 2.4).

Furthermore, due to the similarity of the coset structure of the nine K\"ahler moduli case 
$\fc{SU(3,3)}{SU(3)\times SU(3)\times U(1)}$ to the above case,
it is expected that no stable minimum can be found either.

\subsec{Orientifolds with one untwisted complex structure modulus}

Now consider orientifolds with one untwisted complex structure modulus, labelled by $U^3$.
The main issue will be to solve the supersymmetry
conditions, taking into account the flux quantization, and to see if
in contrast  to the previous case there are stable vacua.
The relevant  3-forms are the $(3,0)$--form  $\omega_{A_0}$ and
one $(2,1)$--form $\omega_{A_3}$ plus their conjugate $(0,3)$ and $(1,2)$--forms $\omega_{B_0}$
and $\omega_{B_3}$.
In terms of these complex $3$--forms, the flux $G_3$ may be expanded as:
\eqn\gthreeb{\eqalign{
{1\over (2\pi)^2\alpha'}~G_3= & ~G_{(3,0)}+G_{(2,1)}+G_{(0,3)}+G_{(1,2)}=\cr
= & ~ A^0(S,U^3)~\omega_{A_0}+
A^3(S,U^3)~\omega_{A_3}+
B^0(S,U^3)~\omega_{B_0}+B^3(S,U^3)~\omega_{B_3}\ .}}
Now,  the complex coefficients take the form
\eqn\anull{
B^0(S)=B_1^0(U^3)-iB_2^0(U^3)~S\, , \quad B^3(S)=B_1^3(U^3)-iB_2^3(U^3)~S\ ,}
where the $B^0(U^3),\ B^3(U^3)$ each contain a constant term
and a term linear in $U^3$. All together they
comprise eight real integer valued flux parameters, whose
explicit forms depend on the individual orientifold under investigation (see later). 
Using this $3$--form flux, the  superpotential \superpot\ may  be written as
\eqn\supoga{
{W}=\lambda \ \big[B_1^0(U^3)-i\ B_2^0(U^3)\ S\big]+\sum\limits_{i=1}^3{g^i}e^{-h^iT^i}\ ,}
which for convenience we parameterize as:
\eqn\supogb{
{W}=\alpha_0+\alpha_1\ U^3+\alpha_2\ S+\alpha_3\ SU^3+\sum_{i=1}^3{g^i}e^{-h^iT^i}\, ,
\quad \alpha_i\in{\bf R}\ . }

In the following, we consider first the situation, where in the first step
 the complex structure modulus $U^3$ is integrated out; this leads
to an effective superpotential $W_{\rm eff}(S,T)$. 
In the second step, the 
supersymmetry conditions $D_TW_{\rm eff}(S,T)=D_SW_{\rm eff}(S,T)=0$
are imposed for the effective 
superpotential $W_{\rm eff}(S,T)$.
As pointed out in {\it Ref.} \ChoiSX, this procedure is valid as
long as the vacuum has the property that the complex structure   
moduli $U^i$ are much heavier than the fields $S$ and $T_i$.
Assuming that this assumption indeed holds, we consider the supersymmetry
condition for  $U^3$,
\eqn\supoga{
D_{U^3}{W}=\alpha_1+\alpha_3S-{\alpha_0+\alpha_1\ U^3+\alpha_2\ S+\alpha_3\ SU^3+
\sum\limits_{i=1}^3{g^i}e^{-h^iT^i}\over U^3+\bar U^3}=0\, ,}
and plug back its solution for $U^3$ into the superpotential.
This results in the following effective superpotential that now
depends only on $S$ and $T_i$ (for real $U^3$):
\eqn\supoga{
W_{\rm eff}(S,T)=2\ \Big(\alpha_0+\alpha_2\ S+\sum\limits_{i=1}^3{g^i}e^{-h^iT^i}\Big)\, .}
We see that this effective superpotential is again a linear function in $S$.
In fact, it is precisely of the same structure as the superpotential \supog\
of the previous section without complex structure modulus.
Hence all conclusions 
about  the vacuum structure with respect to $S$ and $T$ 
are unchanged. In particular, the supersymmetric stationary points
in $S$ and $T$ are not stable ground states with a positive definite moduli
mass matrix. This result has already been obtained in \ChoiSX.

Alternatively, we can also determine the solutions
of all supersymmetry conditions $D_{U^3}{W}=D_{S}{W}=
D_{T}{W}=0$ at the same time without first integrating out $U^3$.
For simplicity we consider the isotropic case $T:=T^1=T^2=T^3$,
$h_1=h_2=h_3$ and real flux parameters $\alpha_i$. We write the moduli fields 
as $T=t+i\tau$, $S=s+i\sigma$ and $U^3=u_3+i \nu$. 
To make the calculation clear, we confine ourselves 
to the supersymmetric point with $\sigma=0$, $\nu=0$, $\tau=0$.
%
The constraints which have to be fulfilled at the supersymmetric point become: 
\eqn\solSU{\eqalign{
s=&-{1\over \alpha_2}\left(\alpha_0+(3+ht){g} e^{-ht}\right) \ , \cr
u_3=&-{\alpha_2\over \alpha_3}\left(\alpha_0+3{g} e^{-ht}\over \alpha_0+(3+ht){g} 
e^{-ht}\right) \ , \cr
\alpha_1=&{\alpha_3\left(\alpha_0+{g} e^{-ht}(3+ht)\right)^2\over \alpha_2\ 
\left(\alpha_0+3{g} e^{-ht}\right)} \ .}}
Here, $s$ and $u_3$ are the real parts of the dilaton and complex structure moduli respectively, 
and should be positive. From the above constraints we see that this excludes 
some values for $\alpha_0$, $\alpha_2$ and $\alpha_3$. 
If we allow for $t$ every positive value, the situation is  simple. One has two possibilities 
\eqn\awerte{
\alpha_0\geq 0 \ , \quad \alpha_2<0 \ , \quad \alpha_3>0 }
and
\eqn\bwerte{
\alpha_0<-(3+ht)\ g\ e^{-ht} \ , \quad \alpha_2>0 \ , \quad \alpha_3<0 \ .}
In the same way as in the previous section we compute the potential and then calculate 
the second derivatives at the supersymmetric point. This means that we plug the constraints \solSU\ into 
the matrix of second derivatives. The resulting six-dimensional matrix is of block diagonal 
form (two blocks $3\times 3$).
The condition for the supersymmetric point to be a minimum is that the diagonal blocks should 
be positive definite. 
This requirement may be translated into the statement that the determinants associated 
with all upper--left submatrices are positive. We abbreviate the sub--determinants of the upper block
by  $a_{11}$, $a_{22}$, $a_{33}$ and those of the 
lower block by $a_{44}$, $a_{55}$, $a_{66}$. They are
\eqn\hauptminoren{\eqalign{
a_{11} =&{\alpha_3^3\over 8\alpha_2^2t^3(\alpha_0+3 {g} ~e^{-ht})^3}
 \Big(\alpha_0+(3+ht) {g} ~e^{-ht}\Big)^2\times\cr
 &\times\Big(2\alpha_0^2+2\alpha_0(6+ht) {g} ~e^{-ht}+ {g} ^2e^{-2ht}\big(18+6ht+h^2t^2\big)\Big) \ , 
\cr
a_{22}=&   {\alpha_3^4\over64t^6(\alpha_0+3 {g} ~e^{-ht})^4}
  \Big(2\alpha_0^2+\alpha_0(12+ht) {g} ~e^{-ht}+ {g} ^2e^{-2ht}\big(18+3ht-h^2t^2\big)\Big) \times \cr
 &\times  \Big(2\alpha_0^2+3\alpha_0(4+ht) {g} ~e^{-ht}+3 {g} ^2e^{-2ht}\big(6+3ht+h^2t^2\big)\Big) \ ,
\cr 
a_{33}=  & {3\alpha_3^5h^2 {g} ^2e^{-2ht}\over 512 t^9(\alpha_0+3 {g} ~e^{-ht})^5} 
  \Big(2\alpha_0^2+\alpha_0(12+ht) {g} ~e^{-ht}+ {g} ^2e^{-2ht}\big(18+3ht-h^2t^2\big)\Big) \times \cr
 &\times  \Big(\alpha_0(1+2ht)+3 {g} ~e^{-ht}(1+ht)\Big)\Big(2\alpha_0(2+ht)+3 {g} ~e^{-ht}(4+3ht+h^2t^2)\Big) \ ,
\cr
a_{44}=& {\alpha_3^3\over 8\alpha_2^2t^3(\alpha_0+3 {g} ~e^{-ht})^3}
 \Big(\alpha_0+(3+ht) {g} ~e^{-ht}\Big)^2\times\cr
 &\times \Big(2\alpha_0^2+2\alpha_0(6+ht) {g} ~e^{-ht}+ {g} ^2e^{-2ht}\big(18+6ht+h^2t^2\big)\Big)  \ ,
\cr
 a_{55}=& {\alpha_3^4\over 64 t^6(\alpha_0+3 {g} ~e^{-ht})^3}
  \Big(2\alpha_0+3(2h+t) {g} ~e^{-ht}\Big) \times \cr
 &\times \Big(2\alpha_0^2+\alpha_0(12+ht){g} e^{-ht}+ {g} ^2e^{-2ht}(18+3ht+2h^2t^2)\Big) \ ,
\cr
  a_{66}=&{3\alpha_3^5h^3 {g} ^2e^{-2ht}\over 512 t^8(\alpha_0+3 {g} ~e^{-ht})^4} 
  \Big(4\alpha_0^2+2\alpha_0(9+2ht) {g} ~e^{-ht}+(18+3ht-3h^2t^2) {g} ^2e^{-2ht}\Big)\times\cr
  &\times \Big(\alpha_0(3+2ht)+(9+6ht+2h^2t^2) {g} ~e^{-ht}\Big) \ .}}
To analyze these minors we have to distinguish the two cases \awerte\ und \bwerte.

In the first case \awerte, the conditions for the positivity of the minors are
\eqn\condition{\eqalign{
2\alpha_0^2&+\alpha_0(12+ht){g}~e^{-ht}+\big(18+3ht-h^2t^2\big){g}^2e^{-2ht}>0\cr
4\alpha_0^2&+2\alpha_0(9+2ht){g}~e^{-ht}+(18+3ht-3h^2t^2){g}^2e^{-2ht}>0\ .}}
In the case for  vanishing $\alpha_0$ we
obtain $ht<3$. In  other cases  the term $\alpha_0^2$  is  dominant for large $t$ and 
\condition\ is true. For the small $t$, the values of $a_{22}$, $a_{33}$ and $a_{66}$ could 
be negative.
However, this depends on the values of ${g}$ and $h$. 

In the second case \bwerte, the conditions are the same \condition, with the difference 
that $\alpha_0<-(3+ht){g} e^{-ht}$. It means  all minors are positive for large $t$ as in 
the previous case.

To conclude,  stable minima do exist for orbifolds with one complex
structure modulus.
In addition, we see that there is a discrepancy between whether we integrate out the complex structure modulus or not. 
The reason for this discrepancy is that the complex structure modulus
is not heavy and therefore is not allowed to be simply integrated out.

Finally, we give an example which falls into class \awerte\ of the solutions.
This example is $Z_{6-II}$ on $(SU(2))^2\times SU(3)\times G_2$. The 
superpotential is given by:
\eqn\superpotexample{\eqalign{
W=&- ie^{-\pi i/ 6}b_0+b_2 +S\lf(ie^{-\pi i / 6}d_0-d_2\ri) \cr 
    & - { U}^3\lf[{i\over\sqrt 3}e^{-\pi i / 3}a_0+{i\over \sqrt 3}a_2
    -S\lf(\sqrt3\,e^{-2\pi i/6}c_0+{1\over\sqrt 3} c_2\ri)\ri] +{g} e^{-hT}\ .}}
We choose $b_0=d_0=c_0=0$ and $a_2=-{1\over 2}a_0$. Further $a_0$, $b_2$, $c_2$ 
and $d_2$ should be positive. In this case, we obtain a superpotential of the form \supogb.

\subsec{Orientifolds with three untwisted complex structure moduli}

In the $\IZ_2\times \IZ_2$ orientifold, we have three untwisted complex structure moduli $U^i$
undetermined. In this case, all
$\om_{A_i}$ and $\om_{B_i}$ survive, and the (primitive) $3$-form flux takes the following form:
\eqn\GCa{{1\over{(2\pi)^2\alpha'}}\ G_3
=\sum_{i=0}^3 \big[\ A^i(S,U^i)\ \omega_{A_i}+B^i(S,U^i)\ \omega_{B_i}\ \big] \ .}
The corresponding superpotential \superpot\ becomes:
\eqn\supogd{
{W}=\lambda\ \big[\ B_1^0(U^i)-i\ B_2^0(U^i)\ S\ \big] +\sum_{i=1}^3{g^i}\ e^{-h^iT^i}\ .}
The coefficients $B_{1,2}^0$ are
determined by $16$ integer valued flux quantum numbers.
The supersymmetry conditions for this superpotential
with seven moduli fields and $16$ flux quantum numbers have been  given in \system.
However, it is very involved to solve them in a closed form. 
Therefore, we reduce the number of fields and
parameters by setting two of the complex structure moduli equal to each other, 
\eg $U^1=U^2$.
Then the superpotential is somewhat simpler and may be parameterized 
by eight integer valued fluxes ${\alpha}_j$ ($j=0,\dots ,7$)
in the following way:
\eqn\supoge{
{W}={\alpha}_0+{\alpha}_1\ U^1+{\alpha}_2\ U^3+{\alpha}_3\ S+{\alpha}_4\ SU^1+{\alpha}_5\ SU^3+
{\alpha}_6\ U^1U^3+{\alpha}_7\ SU^1U^3+\sum_{i=1}^3{g^i}e^{-h^iT^i}\ .}
In this case, the effective K\"ahler potential is given by (\cf section 2):
\eqn\kpnew{
\kappa^2_4  K =-\ln(S+\bar S)-2\ln(U^1+\bar U^1)-\ln(U^3+\bar U^3)
-\sum_{i=1}^3 \ln(T^i+\bar T^i)\ .}
In order to determine the vacuum structure of
this class of models we will first use the integrating out procedure
for all three complex structure moduli, assuming that they are heavy
compared to $S$ and $T^i$.
Again, the aim of this investigation is to see, whether there
are stable supersymmetric vacua with positive definite mass matrix
in $S$ and $T$ or not.
Hence, we consider the two supersymmetry conditions $D_{U^{1,2}}{W}=0$.
Their solution becomes (for real $U^i$):
\def\ss#1{{\scriptstyle{#1}}}
\eqn\soluu{\eqalign{
\ss{U^1=}&\ss{\fc{
-2\ ({\alpha}_2+{\alpha}_5S)\ \lf({\alpha}_0+{\alpha}_3\ S+\sum\limits_{i=1}^{3} {g^i}\ e^{-h^iT^i}\ri) }{
 {\alpha}_0{\alpha}_6+{\alpha}_1({\alpha}_2+{\alpha}_5S)
 +S\big[({\alpha}_2{\alpha}_4+\alpha_0{\alpha}_7+{\alpha}_4{\alpha}_5S+{\alpha}_3({\alpha}_6+{\alpha}_7S)\big]+
({\alpha}_6+{\alpha}_7 S)\sum\limits_{j=1}^{3} {g}^je^{-h^jT^j} }\ ,}\cr
\ss{U^3=}&
\ss{-{{\alpha}_0+{\alpha}_3S+\sum\limits_{i=1}^{3} g^i\ e^{-h^iT^i} \over {\alpha}_2+{\alpha}_5S}\ .}}}
We can now insert this solution back into \eqq \supoge. This way
we derive the following effective superpotential:
\eqn\weffab{\eqalign{
W_{\rm eff} & (S,T^i)=
\Biggl\lbrace
2\Big({\alpha}_0+{\alpha}_3S+\sum_{i=1}^3 {g^i} e^{-h^iT^i}\Big)\Big[-{\alpha}_1({\alpha}_2+{\alpha}_5S)+{\alpha}_0({\alpha}_6+{\alpha}_7S)
+iS\big({\alpha}_2{\alpha}_4 \cr &
+{\alpha}_4{\alpha}_5S -{\alpha}_3({\alpha}_6+{\alpha}_7S)\big) +({\alpha}_6+{\alpha}_7S)\sum_{k=1}^3{g}^ke^{-h^kT^k}\Big]\Biggr\rbrace /
\Biggl\lbrace \sum_{j=1}^3 g^je^{-h^jT^j}({\alpha}_6+{\alpha}_7S) 
\cr &
+{\alpha}_0{\alpha}_6+{\alpha}_1({\alpha}_2+{\alpha}_5S)+S\ \Big[{\alpha}_2{\alpha}_4+{\alpha}_0{\alpha}_7+{\alpha}_4{\alpha}_5S+{\alpha}_3({\alpha}_6+{\alpha}_7S)\Big]
\Biggr\rbrace\, .}}
The numerator is a polynomial of third degree in $S$ and second degree in denominator. 

To apply the analysis of \ChoiSX\ requires to compute the ratio
${S W_{\rm eff}(S)''\over W_{\rm eff}(S)'}$ and to analyze, if its value 
is bigger than one. However, this analysis assumes a superpotential of the form
$W_{\rm eff}(S)+\sum\limits_{i=1}^3 {g^i} e^{-h^iT^i}$. 
Obviously, our effective superpotential \weffab\ is not of this form.
This would only be achieved for a special choice of the coefficients ${\alpha}_i$. The condition 
on ${\alpha}_i$ for  the numerator and denominator be divisible without remainder is:
\eqn\aicond{
({\alpha}_1+{\alpha}_4S)\ ({\alpha}_2+{\alpha}_5 S)=0 \ .}
After inserting this condition the effective superpotential \weffab\ becomes
\eqn\supspecial{
W_{\rm eff}=2\ \Big({\alpha}_0+{\alpha}_3 S+\sum_{i=1}^3 {g^i} e^{-h^iT^i}\Big) \ .}
This is again the already analyzed case of the previous section, in which there 
is no stable minimum.

\subsec{Cubic superpotential}

We consider the case of three complex structure moduli ($U^i$) and three K\"ahler moduli ($T^i$). 
To make the calculations simple, we assume $U:=U^1=U^2=U^3$, $T:=T^1=T^2=T^3$.
The superpotential and the K\"ahler potential have the following form:
\eqn\superpotdreiu{
W={\alpha}_0+{\alpha}_1 U+{\alpha}_2 (U)^2 +{\alpha}_3 (U)^3 +S({\alpha}_4+{\alpha}_5 U+{\alpha}_6 (U)^2 +{\alpha}_7 (U)^3) + 3 g e^{-hT} \ ,
}
\eqn\kahlerpotdreiu{
K=-\ln(S+\bar S)-3 \ln(U+\bar U)-3 \ln(T+\bar T) \ , \quad g,{\alpha}_i,h \in {\bf R}\ , \ h \ {\rm positive} \ .
}
We rewrite $U$ and $T$ using the real basis: $U=u+i\nu$, $T=t+i \tau$ and compute
the supersymmetry conditions ($D_{U}W=D_TW=0$) at the point of vanishing $\nu$ and $\tau$:
\eqn\eqdreiu{\eqalign{
{\alpha}_0&=g e^{-ht}(-3+2ht)+u({\alpha}_5 s+ {\alpha}_2u+2{\alpha}_6su+2{\alpha}_3u^2+3{\alpha}_7su^2))\ , \cr
{\alpha}_1&=-\fc{1}{u}3ght e^{-ht}-{\alpha}_5s-u(2{\alpha}_2+2{\alpha}_6s+3{\alpha}_3u+{\alpha}_7su)\ , \cr
{\alpha}_4&=-\fc{1}{s}\left(ght e^{-ht}+su({\alpha}_5+{\alpha}_6u+{\alpha}_7u^2)\right) \ .
}}
As in the previous cases, we compute the scalar potential and its Hessian at the supersymmetric points. 
This means that we calculate the second derivatives of the potential and eliminate ${\alpha}_0$, ${\alpha}_1$, ${\alpha}_4$ by using \eqdreiu . 
It is irrelevant which of the parameters or fields are eliminated through \eqdreiu. We choose this particular combination 
by the criterion of simplicity of the later analysis.

The Hessian should have positive eigenvalues at the minimum or equivalently, its upper-left submatrices should be positive definite. 
The determinants of the upper-left submatrices are of the form
\eqn\eqsubdetdreiu{\eqalign{
a_{11}=&a_{44}=\fc{1}{48st^3u^3}\Big(9u^4(4a^2_3+8{\alpha}_3{\alpha}_7s+7{\alpha}_7^2s^2)+u^3(24{\alpha}_2{\alpha}_3+24{\alpha}_2{\alpha}_7s+24{\alpha}_3{\alpha}_6s+60{\alpha}_6{\alpha}_7s^2)\cr
&+u^2(4{\alpha}_2^2+8{\alpha}_2{\alpha}_6s+16{\alpha}_6^2s^2+18{\alpha}_5{\alpha}_7s^2)+12{\alpha}_5{\alpha}_6s^2u+3{\alpha}_5^2s^2\Big)+{\cal{O}}\left(e^{-ht}\right)\ ,\cr
a_{22}=&a_{55}={\left({\alpha}_5+2{\alpha}_6u+3{\alpha}_7u^2\right)^2\over768t^4u^4}+{\cal{O}}\left(e^{-ht}\right)\ ,\cr
a_{33}=&\fc{g^2h^2e^{-2ht}(2+ht)(1+2ht)({\alpha}_5+u(2{\alpha}_6+3{\alpha}_7u))^4}{8192st^9u^7}+{\cal{O}}\left(e^{-3ht}\right)\ ,\cr
a_{66}=&\fc{g^2h^3e^{-2ht}(3+2ht)({\alpha}_5+u(2{\alpha}_6+3{\alpha}_7u))^4}{8192st^8u^7}+
{\cal{O}}\left(e^{-3ht}\right)\ .}}
For positive flux parameters ${\alpha}_2,{\alpha}_3,{\alpha}_5,{\alpha}_6,{\alpha}_7$ and in the region of large $t$ ($t^{-4}>>e^{-ht}$), all sub-determinants are positive. 
So in the case of three complex structure moduli, there is some region for which there is a supersymmetric minimum.

\subsec{Open string moduli and soft--supersymmetry breaking terms}

So far, we have not yet discussed the open string moduli accounting for $D3/D7$--brane
positions, Wilson line moduli and matter fields. The $D7$--brane moduli $\phi_{7,j}$ 
are fixed through $W_{\rm flux}$ in the case of $ISD$--fluxes  with 
their respective scalars 
acquiring soft--supersymmetry breaking masses $\tilde m_{\phi_{7,j}}$. The latter 
have been calculated in \fourref\LustFI\CIUii\LustDN\FontCX. Generically, the 
locations of the $D7$--branes are fixed to those of the $O7$--planes.

After including the non--perturbative superpotential $W_{\rm np}$,
supersymmetry is restored at the minimum and the scalar masses take the generic form:
\eqn\softmasses{
(m_{\phi_{7,j}})^2=-2\ \kappa_4^6\ e^{\kappa_4^2 \hat K}\ |W|^2\   
G_{\phi_{7,j}\ov \phi_{7,j}}=\fc{2}{3}\ \kappa_4^4\ V_{0}\ G_{\phi_{7,j}\ov \phi_{7,j}}\ .}
Here, $V_0$ is the value of the potential at the minimum (\cf \eqq \potsusy)
and $G_{\phi_{7,j}\ov \phi_{7,j}}$ is the metric  of the fields $\phi_{7,j}$.
Since supersymmetry is restored, these masses correspond to some sort of 
effective $\mu$--terms in the superpotential \superpot.
Nevertheless, the soft--supersymmetry breaking mass terms $\tilde m_{\phi_{7,j}}$ 
are important quantities as they represent certain coefficients of a 
low--energy supergravity expansion of the $D$--brane dynamics 
(encoded in the Born--Infeld action) coupled to non--vanishing $3$--form fluxes. 
Moreover, as we shall argue in subsection 4.2, 
they are relevant for the discussion of possible non--perturbative contributions  
$W_{\rm np}$ to the superpotential.
For the $\IZ_2\times\IZ_2$ orientifold, the relation between $m_{\phi_{7,j}}$ and 
$\tilde m_{\phi_{7,j}}$ may be determined by the $F$--terms, given in subsection 4.1.
{\it E.g.} the gaugino mass $m_{g,j}$ of a $D7$--brane, which is transversal to the 
$j$--th internal complex $2$--plane becomes
\eqn\Newgaugino{
m_{g,j}=\tilde m_{g,j}\ \lf[\ 1+\fc{(T^j+\ov T^j)\ g^jh^j\ e^{-h^j T^j}}{W_{\rm flux}}\ 
\ri]\ ,}
with $\tilde m_{g,j}$ the gaugino mass before introducing the non--perturbative superpotential
$W_{\rm np}$. 
Hence, \eg for all K\"ahler moduli $T^j$ and the parameters $h^j,g^j$ equal, we find:
\eqn\RATIO{
\fc{m_{g,j}}{m_{g,k}}=\fc{\tilde m_{g,j}}{\tilde m_{g,k}}\ .}
Hence, the ratio of gaugino masses after inclusion of $W_{\rm np}$ is given by the same
ratio as without $W_{\rm np}$.  Strictly speaking, this ratio is only valid for the case
$F^M\neq 0$. Similar relations may be found for other soft--masses.
Eventually, soft--supersymmetry breaking terms after the KKLT uplift, \ie after inclusion of
the anti $D3$--branes, have been determined recently in {\it Ref.} \HPN.

The \tb orientifolds, introduced in section 2,  imply $64$ $O3$--branes with 
negative $D3$--brane charges. 
At the orbifold point, no extra contribution comes 
from the induced $D3$--brane charge of wrapped $D7/O7$--stacks
(without world--volume $2$--form flux on the $D7$). 
Hence tadpoles must be cancelled by adding $D3$--branes or/and some amount
of flux $N_{flux}$.
The original KKLT proposal is, that only $ISD$--flux should be turned on in order to have
a reliable supergravity approximation at least before the uplift.
Since $ISD$--fluxes do not freeze the positions of $D3$--branes, 
the tadpoles should be cancelled by pure flux, \ie $N_{flux}$=64.
It has been shown in {\it Ref.} \MS, that $ISD$--fluxes  with $N_{flux}=64$ 
are possible in the $\IZ_2\times \IZ_2$ orientifold.


\newsec{Non--perturbative superpotential, moduli stabilization and resolved orbifolds}

\subsec{The non-perturbative superpotential}

We will first present some general facts 
about the non-perturbative superpotential due to wrapped Euclidean
$D3$--brane instantons.
As an alternative but related mechanism to
generate a non--perturbative superpotential we also discuss
gaugino
condensation
in the effective N=1 gauge theory that lives on $D7$--branes wrapped on internal
4-cycles. As we will see,  this possibility of getting a non-perturbative
superpotential is largely influenced by the $3$--form flux turned on.
Hence there is a very non-trivial interplay between the
tree--level superpotential due to $3$--form flux and the non--perturbative
superpotential due to $D3$-brane instantons and/or gaugino condensation.
Some useful papers on this
subject include {\it Refs.}
\sevenref\DenefDM\RobbinsHX\GorlichQM\TripathyHV\KalloshGS\SaulinaVE\DenefMM.

We start with some type $IIB$ orientifold on a 6-dimensional CY space $M_3$
with some possible $3$--form flux $G_3$ through some 3-cycles $\Sigma$.
Assume that
$M_3$ possesses a complex 2-dimensional divisor, called ${\cal O}_i$.
The relevant question
is whether in \eqq \superpot\ non--perturbative effects contribute to a non-vanishing 
superpotential of the form
\eqn\supo{
W_{\rm np}\sim g_i~e^{-h_iV_i}\ ,}
where $V_i$ denotes the volume of ${\cal O}_i$.
The constant $h_i$ in the exponent depends on which mechanism, $D3$--brane instantons
or gaugino condensation, is responsible
for the generation of $W_{\rm np}$.
The most important question is, if, for any given ${\cal O}_i$,
$g_i$ is zero or not. (The even more difficult problem is
what the value of $g_i$ is, and whether it depends on the complex structure
moduli, which is expected on general grounds.)
This question can be answered in three different, but 
related ways:

\vskip0.2cm\noindent
{\sl (i) Euclidean $D3$-brane instantons in type $IIB$} 

\vskip0.2cm
\noindent
We consider possible contributions to $W_{\rm np}$ from 
Euclidean $D3$--branes wrapped on ${\cal O}_i$. In order to decide whether
one gets a contribution to $W_{\rm np}$,
one has to count the number of fermionic zero modes, $n_f$, on the world volume
theory of the $D3$--branes. According to \DenefDM, $n_f$ must be 2 in order
for $g_i$ to be non-vanishing.
The actual value for $n_f$ crucially depends on the number of
supersymmetries which are supported by ${\cal O}_i$.
However, this number also depends on the $3$--form flux $G_3$ 
turned on. Hence it is
useful to define the number of fermionic zero modes 
in the presence of $3$--form flux by the number $n_f(G_3)$.
The actual requirement for a non-vanishing superpotential now is that
$n_f(G_3)=2$.

In a N=1 supersymmetric Yang-Mills gauge theory,
the contribution from a single $D3$-instanton to the superpotential
is given by the following expression:
\eqn\supoins{
W_{\rm np}\sim \Lambda^{3b}\, .
}
$b$ is the $\beta$-function coefficient of the corresponding group
(see below), and $\Lambda$ is the dynamical scale
of the gauge theory:
\eqn\lam{
\Lambda^3= e^{-{8\pi^2\over bg^2}}\, .
}
After relating  the gauge coupling to the volume (\cf subsection 4.3 for a 
definition of the gauge coupling) of ${\cal O}_i$,
\eqn\gvol{
{4\pi\over g^2}=V_i\ ,}
one obtains
\eqn\supoinsa{
W_{\rm np}\sim \, g_i~e^{-2\pi V_i}\ .}
So $h_i=2\pi$.

\vskip0.2cm\noindent
{\sl (ii) Euclidean $M5$-brane instantons in $F$-theory}

\vskip0.2cm
\noindent
A second description is given in terms of $F$--theory on a four--manifold
$M_4$. Here we are considering Euclidean $M5$--branes wrapped around  a 
six-dimensional divisor $D_i$ inside $M_4$.
$M_4$ is an elliptic $T^2$ fibration over some six--dimensional $IIB$ base
$B_6$. 
The condition for a non-vanishing superpotential in $F$--theory now is
given by the requirement that the arithmetic genus $\chi_{D_i}=1$ 
\WittenBN, where $\chi_{D_i}$ is given by the following formula:
\eqn\chid{
\chi_{D_i}=h_{(0,0)}(D_i)-h_{(0,1)}(D_i)+h_{(0,2)}(D_i)-h_{(0,3)}(D_i)\, .}

Just like in the $D3$--brane case,
the number of fermionic zero modes, the arithmetic genus can
be ``changed'' in the presence of 4-form flux $G_4$ in $F$--theory. This
gives rise to an effective arithmetic genus 
$\chi_{D_i}(G_4)$. Therefore it is not necessary that the geometric
arithmetic genus 
$\chi_{D_i}$
is one, but rather the requirement for a non-vanishing
superpotential becomes 
$\chi_{D_i}(G_4)=1$.
Whether this condition is fulfilled depends again on the number
of supersymmetries preserved by the flux. In particular,
if the flux preserves N=2 supersymmetry, then in
general $\chi_{D_i}(G_4)\neq 1$. On the other hand if $G_4$ preserves
 N=1, one can expect the above condition to be satisfied.

\vskip0.2cm\noindent
{\sl (iii) $D7$-branes in type $IIB$ orientifolds and gaugino condensation}

\vskip0.2cm
\noindent
Instead of wrapping $D3$--branes on ${\cal O}_i$ 
respectively $5$-branes on $D_i$, we may also consider
a stack of $N$ $D7$--branes which fill the space-time and 
wrapped around four--dimensional divisors ${\cal O}_i$. 
In general orientifold compactifications, the existence of
the $D7$--branes is in fact forced by the tadpole
cancellation conditions, namely in order
to cancel the Ramond charge of the orientifold planes.
Now consider the open string spectrum on
the $D7$--branes. 
It is in general given by an effective N=1 supersymmetric
$SU(N)$ gauge theory with some additional matter fields.
In the following we will study the conditions, under which gaugino condensation
will take place in the effective N=1 gauge theory on the
$D7$--branes and will lead to a contribution to $W_{\rm np}$.

First consider pure N=1 Yang--Mills gauge theory with gauge group
$G$ without any matter fields.
Gaugino condensation generates a nonperturbative superpotential
\eqn\supogaugino{
W_{\rm np}\sim \Lambda^{3}=e^{-{8\pi^2\over bg^2}}\ ,}
With eq.\gvol\ we then get
\eqn\supoinsa{
W_{\rm np}\sim \, g_i~e^{-{2\pi V_i\over b}}\, ,}
and hence $h_i=2\pi/b$ for pure  SQCD.

Now consider  N=1 SQCD with gauge group $G=SU(N_C)$ and with
$N_F$ 
matter fields $Q$, $\tilde Q$ in the
fundamental plus anti--fundamental representations
$N_F(\underline N_C\oplus \underline{\bar N_C})$.
For $N_F<N_C$, there is a dynamically generated
superpotential (for a review see \eg \IntriligatorAU)
\eqn\suposeiberg{
W_{\rm np}=(N_c-N_f)\ \biggl({\Lambda^{3N_C-N_F}\over\det Q\tilde Q}\biggr)^{1/(N_c-N_f)}\, .}
Here, $b=3N_C-N_F$ is the N=1 $\beta$--function coefficient of
SQCD. 
The vacuum expectation values of the meson superfields $M\sim Q\tilde Q$
break the gauge group $SU(N_C)$ to the non-Abelian subgroup
$SU(N_C-N_F)$.
If $N_F=N_C-1$, the superpotential is generated by gauge instantons.
On the other hand,
 the superpotential arises due to the gaugino 
condensation in the unbroken $SU(N_C-N_F)$ gauge group.
Therefore the gaugino condensate is determined by the scale of the
unbroken gauge group,
$\langle \lambda\lambda\rangle\sim \Lambda^3_{N_C-N_F}$, where the
scale $\Lambda_{N_C-N_F}$ of the low-energy $SU(N_C-N_F)$ gauge theory
can be associated to the scale $\Lambda$ of the high-energy
gauge theory as $\Lambda^{3(N_C-N_F)}_{N_C-N_F}=\Lambda^{3N_C-N_F}/\det M$.
This precisely yields the effective superpotential eq.\suposeiberg.
Finally, for $N_F\geq N_C$ there is no dynamically generated superpotential
of this type.
 
It is also important to know what is happening if the squarks fields $Q$ and $\tilde Q$
get a mass. 
Consider the case where we have $N_F'$ matter fields with some
common mass $m$.
If we are interested in the effective low energy field theory,
i.e. for energies much smaller than the masses of the squark fields, they decouple
and the low energy theory is $SU(N_C)$ with $N_F-N_F'$ flavors.
For this theory the previous discussion applies. So in the case where all
squark fields are heavy, one is back to gaugino condensation in pure SQCD.

An interesting class of relevant models are 
 N=1 quiver gauge theories with gauge group
$G=SU(N_C)\times SU(N_C')$ and bifundamental matter
fields in the representations 
$(\underline N_C,{\overline{\underline N_C'}})+{\rm h.c.}$
(plus possibly massless adjoint fields).
These models generically appear as effective gauge theories
of intersecting $D7$-branes.
One way to analyze the dynamics of these field theories is
to consider $SU(N_C')$ as the flavor
group of the $SU(N_C)$ gauge group and vice versa.
Then the $\beta$-function coefficient for  $SU(N_C)$
becomes $b=3N_C-N_C'$, and analogously for $SU(N_C')$.
For $N_C'>N_C$, gaugino condensation in the $SU(N_C)$
part is excluded, but is possible in the $SU(N_C')$
gauge theory. However, following our previous discussion,
for  $N_C'=N_C$ it will not be possible in either of the
two gauge group factors.

Next let us briefly discuss what will happen if one adds
also $\tilde N_F$ massless adjoint chiral matter fields $X$. 
The N=1 $\beta$-function coefficient for 
$G=SU(N_C)$ becomes $b=(3-\tilde N_F)N_C-N_F$.
Obviously, for $\tilde N_F\geq 3$ asymptotic freedom
is lost, and there will be no non-perturbative superpotential.
However also the remaining cases $\tilde N_F=1,2$ are problematic.
For $\tilde N_F=1$ and $N_F=0$ without tree level superpotential,
the theory becomes  N=2 supersymmetric. This theory
is known to possess a moduli space in the Coulomb branch
without dynamical superpotential. Adding a tree level 
superpotential $W_{\rm tree}\sim {\rm Tr}X^k+\dots$ might 
change this conclusion, but 
the discussion now becomes more model dependent.
Similarly for $\tilde N_F=2$ with a tree level superpotential
it is not possible to make very specific statements.

In conclusion we have seen that the possibility
of having a non-perturbative superpotential
is indeed rather model dependent. Therefore we like
to simplify the discussion and want to formulate
a few simple conditions, which are sufficient but
not always necessary for a non-perturbative superpotential
(later we will discuss how these rules can by implemented
by the $D7$-branes in orientifold models):

\vskip0.2cm
\noindent 
{\sl 1.) For
a non-vanishing $W_{\rm np}$
we require that all fundamental matter fields, if present, 
become massive.} This conditions
implies some strong conditions on the world volume theory
of the $D7$-branes. Massless bi-fundamental matter
fields typically arise as open string states localized at
the intersection loci of two $D7$-branes. The number of those
states depends on the geometrical intersection number and
also on the open string 2-form magnetic fluxes
on the  $D7$-branes. 
So we will essentially require that those $D7$-branes
which are responsible for gaugino condensation in the
hidden gauge sectors do not intersect each other.

\vskip0.2cm
\noindent 
{\sl 2.) For
a non-vanishing $W_{\rm np}$
we require that all adjoint matter fields, if present, become massive.}
As we will discuss
later, the number of massless adjoint chiral
multiplets is a $3$--form flux dependent quantity.
This means in particular, that possibly after turning on some
$3$--form flux,

\noindent 
a) there are no
massless adjoint scalar fields associated to the positions of the
$D7$-branes in the space transversal to  ${\cal O}_i$
(called {\sl position fields $\phi_7$} in the
following),

\noindent
b) there are no massless adjoint scalar fields living on the $D7$-branes
being associated to Wilson lines (called {\sl Wilson line fields $\phi_7'$ } 
in the following).

\vskip0.3cm\noindent

\subsec{$M_3=T^6/(\IZ_N\times \IZ_M)$}

These are the models which are of main interest in our paper.
The simplest model of this class is the $\IZ_2$ orientifold
with $M_3=T^6$. 
This model was recently
considered in \TripathyHV\ and some time ago originally
in     \KST .
In this case, the $F$--theory lift is
described by $M_4=(T^4\times K3_2)/Z_2$.
All type $IIB$ divisors are simply $T^4$.
In $F$--theory, we get $D_i=T^4\times T^2$.
Since 
$\chi_{T^4\times T^2}=0$,
a
non-perturbative superpotential due to Euclidean
5-branes is excluded.
We briefly compare this result with the 
spectrum of the  type $IIB$ orientifold.
Since $T^4$ supports 16 supercharges, it simply follows that
$n_f=16$, i.e. no superpotential.\foot{It
 is perhaps interesting to note that the 
geometric arithmetic genus of $T^4\times T^2$,
expressed in terms of the Hodge numbers of $T^4$ can be written as
\eqn\divfc{
\chi_{T^4\times T^2}=h_{(0,0)}(T^4)-h_{(0,1)}(T^4)+h_{(0,2)}(T^4)=0
\, .}
The actual numbers $h_{(0,1)}(T^4)=2$ and  $h_{(0,2)}(T^4)=1$
perfectly agree with the number of massless adjoint fields on the
$D7$-branes.}

The spectrum can be again changed by turning on $3$--form flux. 
As shown  in \TripathyHV, the number of fermionic zero modes 
on the $D3$-branes is reduced by a particular (2,1) $3$--form flux
$G_{2,1}$ from 16 to $n_f(G_3)=4$. 
So we see
that turning on $G_{2,1}$ 
is not sufficient to generate $W_{\rm np}$.
However, one expects that by turning  on additional
 $3$--form fluxes, the number of
fermionic zero modes is further reduced, and a non-perturbative
superpotential may eventually become possible.

Now we will turn to the general
$T^6/(\IZ_N\times \IZ_M)$ orientifolds. 
Many of these orientifolds require $D7$-branes
for tadpole cancellation. 
Since the $\IZ_N\times \IZ_M$ 
orbifold group now has a non-trivial action on the
six-dimensional space,
the four-dimensional divisors ${\cal O}_i$
split into two classes:
(a) untwisted divisors $D_i$
and (b) twisted, i.e. exceptional divisors 
$E_\alpha$.
The untwisted divisors are in one-to-one correspondence with the
untwisted 4-cycles of the orbifold. The number of
linearly independent untwisted divisors is precisely given by
the untwisted Hodge number $h_{(1,1)}^{\rm untw.}$ of the orbifold, \ie 
the number of untwisted K\"ahler moduli. On the other hand, the twisted
divisors correspond to the twisted, blown-up 4-cycles of the orbifolds, which
arise after resolving the orbifold singularities. Hence the number 
of linearly independent twisted divisors
is given by the number of twisted K\"ahler moduli on the orbifold, namely by
$h_{(1,1)}^{\rm twist.}$. The actual  values for $h_{(1,1)}^{\rm untw.}$, $h_{(1,1)}^{\rm twist.}$
in each of the orientifold examples can be found in Table 1 and 2.
According to our previous discussion, a gaugino condensate will
form if there are no massless fundamental and no massless adjoint
matter fields on the world volume theories of the $D7$-branes.
This means that the respective divisors do not intersect each other and
hence there are no massless bi-fundamental fields, and that
the adjoint matter fields are absent or massive due to $3$--form fluxes.
In the following we will now consider
each case separately:

\vskip0.2cm\noindent
{\sl (a) Untwisted divisors in the orbifold limit}
\vskip0.2cm

In the unresolved orbifolds, the untwisted divisors always have  
topology $D_i\sim T^4$. 
The untwisted open string states on
the $D7$-branes wrapped on $T^4$ are always given by 
N=4 super Yang-Mills multiplets of
the gauge group $G$. It contains 1 adjoint
chiral position field $\phi_7$ and 2   adjoint
Wilson line fields
$\phi_7'$. In addition, if we consider $N_C^i$
and $N_C^j$ $D7$-branes wrapped
on two untwisted divisors $D_i$ and $D_j$, there are massless
bi-fundamental matter fields. 
More specifically, the three $D_i$ are defined by the algebraic
conditions $z^i=a^i={\rm const}$.
Therefore they mutually intersect each other on the (complex)
one-dimensional subspaces $z^i=a^i$, $z^j=a^j$ ($i\neq j$).
The corresponding triple intersection number is just one.
(Here we do not consider possible 2--form flux on the $D7$-branes.)
The twisted open string spectrum on this common intersection locus
can be easily determined
and is  given by massless bi-fundamental matter fields in
the representations 
$(\underline N_C^i,{\overline{\underline N_C^j}})+{\rm h.c.}$
of the gauge group $G=U(N_C^i)\times U(N_C^j)$.
Similarly, if we consider models with more than three
K\"ahler moduli, like the $\IZ_3$ and $\IZ_{6-I}$ orientifolds,
the additional untwisted divisors are defined by equations like
$D_4=z_1+z_2={\rm const}$. These divisors mutually intersect 
with all other $D_i$. (However the triple intersection number
vanishes for some combinations, \eg the triple intersection number
$D_1\cdot D_2\cdot D_4=0$).
Therefore we again get massless bi-fundamental matter fields
on the intersection of all these untwisted divisors. 

So the presence of the massless adjoint fields $\phi_7$ and $\phi_7'$
and also the possible bi-fundamental matter fields due to $D7$-brane
intersections will hamper the formation of a gaugino condensate.
Hence
the important question is again, which $3$--form fluxes make
the fields $\phi_7$ massive, and which $3$--form fluxes are responsible
for the masses of the fields $\phi_7'$.
More precisely, in order
to get a non-vanishing superpotential, those $3$--form fluxes have
to be turned on in such a way that the N=4 supersymmetry
on the world volume of the $D7$-branes is broken to N=1 supersymmetry.
Then the untwisted open string spectrum contains only one massless adjoint
N=1 vector multiplet and no further adjoint chiral multiplets.
This should be possible by turning on two kinds of possible $3$--form
fluxes.
First, $G_{2,1}$ generically gives a mass to  $\phi_7$.
This is the effective, supersymmetric $\mu$-term which is of
the following form \LustBD:
\eqn\muterm{
\mu_{\phi_7\phi_7}\sim\int\bar G_3\wedge\omega_A\, .}
Next consider $3$--form fluxes which can 
give a mass also to the Wilson line fields.
This kind of $3$--form fluxes were considered in \LustFI\ and in \LustDN\
in the context of softly supersymmetry breaking mass parameters.
There, it was shown that  the $(0,3)$--form flux $G_{0,3}$ indeed generates 
a mass for the adjoint Wilson line 
scalar fields $\phi_7'$.
(In addition, the 
(3,0)-form flux $G_{3,0}$
flux  will also contribute to the masses of the
fields $\phi_7$.)
The specific mass term for the
two adjoint matter fields
has following form \doubref\LustFI\LustDN\ (in case of
vanishing 2-form flux on the $D7$-branes):
\eqn\massterm{
(m_{\phi_7'\phi_7'}^1)^2\ ,\ (m_{\phi_7'\phi_7'}^2)^2
\sim \lf|\int\bar G_3\wedge\Omega\,\ri|^2\, .}
More precisely, at the extremum $V_{0}$ of the 
scalar potential, the soft--masses are given by \eqq \softmasses.
One might worry that $G_{(0,3)}$ completely breaks all
supersymmetries on the $D7$-branes. This is indeed the case without $W_{\rm np}$.
However, as we have seen in the previous chapter, in the context of
full moduli stabilization with non-perturbative superpotential, supersymmetric
AdS--vacua are possible in the presence of these more general $3$--form fluxes.
So in conclusion, without being more specific, it is quite reasonable
to assume that after turning on specific  $3$--form flux components
$G_{(0,3)}$, $G_{(3,0)}$ and, if allowed by the orbifold twist, also
$G_{(2,1)}$, $G_{(1,2)}$, all unwanted adjoint matter fields become massive.

Next we turn to the problem of the massless matter fields
in bi-fundamental representations
due to $D7$-brane intersections. As explained before, they are always present
if more than one stack of $D7$-branes 
are wrapped on the untwisted divisors $D_i$ in the unresolved orbifold
space. However after the resolution of the orbifold
singularities some of the divisors $D_i$ do not intersect anymore.
In physical terms this means that the twisted blowing-up modulus
fields acts as a mass parameter for the bi-fundamental matter
fields. 
Therefore the bi-fundamental matter fields from open strings
spanned between $D7$-branes around these two divisors
became massive.
A short discussion on resolved orbifolds will be given in section (4.3).
Many more details based on a toric description of the
divisors will be presented in \future.

\vskip0.2cm\noindent
{\sl (b) Twisted divisors}
\vskip0.2cm

The twisted divisors $E_\alpha$ 
correspond to the twisted 4-cycles on the orientifold
and receive a finite volume after blowing up the orbifold
singularities. Generically they cannot be moved inside the
compact space. The orbifold group generically reduces
the number of supersymmetries from N=4  down to
N=1. Therefore the twisted divisors are expected to contribute to
the superpotential without turning on additional $3$--form
fluxes. In $F$--theory language it means that the arithmetic genus
of the twisted divisors is generically equal to one.
This can be also seen from the $D7$-brane perspective.
Wrapping $D7$-branes around twisted divisors means that we
are considering open strings at orbifold singularities.
After blowing up the twisted 4-cycles, the corresponding 
open string gauge theory on the $D7$-branes generically is that of pure
N=1 Yang-Mills. Hence gaugino condensation will take place.
If one wants to avoid massless chiral multiplets in the
bi-fundamental representation of the gauge group, one has to choose disjoint
twisted divisors which do not intersect.
The intersection pattern of the $E_\alpha$ can be again best read
off from the toric diagrams, \cf {\it Ref.} \future.
It can be seen that there exist several possibilities
for twisted divisors $E_\alpha$ that do not mutually intersect.

As a specific model with exceptional divisors that
lead to a non-vanishing superpotential the $T^6/({\bf Z}_2\times {\bf Z}_2)$
orientifold was recently investigated in full detail in \DenefMM.
This model allows for an $F$--theory description in terms of an elliptic
Calabi-Yau four-fold. The superpotential arises due to
48 $D3$-branes which are wrapped around all 48 linearly independent
exceptional divisors. 
In addition, this model also contains 12 stacks of 8 $D7$-branes
which are wrapped on 12 disjoint divisors that are linear combinations
of the 3 untwisted divisors and 12 particular twisted divisors.
In this model, the $D7$-branes cancel the Ramond charge of the 12 orientifold
O7-planes in a local way. It follows that the open string spectrum
is given by pure N=1 Yang-Mills with gauge group
$G=SO(8)^{12}$.

Let us try to 
summarize the discussion of this section:
in N=1 orientifolds, Euclidean $D3$-instantons
and/or gaugino condensation on $D7$-branes generically do generate 
a non-perturbative superpotential. 
For $D3$/$D7$-branes being wrapped around untwisted
divisors one needs certain 3--form fluxes for a non-vanishing superpotential.
Moreover if one wants to avoid massless matter fields in bi-fundamental
representations due to $D$--brane intersections, one has to resolve
some or all orbifold singularities.
If the branes are wrapped around twisted divisors, all conditions
for a non-vanishing superpotential are in general met without any further
fluxes.
We are confident that the conditions for a non-vanishing
$W_{\rm np}$ can be satisfied in many or if not all of the
considered orientifold spaces, although we do not provide here
concrete, full fledged string models with all tadpole
conditions satisfied. 
(An nice concrete example is the model considered in
\DenefMM.)
Nevertheless we assume that 
the total non-perturbative superpotential can 
be written in the following form:
\eqn\wnpfinal{
W_{\rm np}=\sum_{j=1}^{n_3^{\rm untw.}}e^{-2\pi V_j^{\rm untw.}}
+
\sum_{\alpha=1}^{n_3^{\rm twist.}}e^{-2\pi V_{\alpha}^{\rm twist.}}
+
\sum_{j'=1}^{n_7^{\rm untw.}}e^{-{2\pi V_{j'}^{\rm untw.}\over b_{j'}(G)}}
+
\sum_{\alpha'=1}^{n_7^{\rm twist.}}e^{-{2\pi V_{\alpha'}^{\rm twist.}\over b_{\alpha'}(G)}}
\, .}
The first two terms denote the non-vanishing contributions of the
Euclidean $D3$-instantons wrapped around $n_3^{\rm untw.}$ untwisted divisors
and $n_3^{\rm twist.}$ twisted divisors with volumes  $V_j^{\rm untw.}$
or  $V_{\alpha}^{\rm twist.}$, respectively.
The last two terms describe the contribution from the gaugino
condensation on different stacks of $D7$-branes, which are wrapped
around $n_7^{\rm untw.}$ untwisted divisors
and $n_7^{\rm twist.}$ twisted divisors with volumes  $V_{j'}^{\rm untw.}$
or  $V_{\alpha'}^{\rm twist.}$, respectively.
The numbers  $b_{j'}(G)$, $b_{\alpha'}(G)$ are the suitably normalized
$\beta$-function coefficients of the associated effective gauge theories.
For pure N=1 Yang-Mills, the $\beta$-functions coefficients
are just given by the quadratic Casimir invariants $c_2(G)$ of the
gauge group $G$.
It is instructive to compare this superpotential with the
one obtained
in the ${\bf Z}_2\times {\bf Z}_2$ model of \DenefMM.
Here, there are
48 twisted divisors contributing to the $D3$-brane part and
12 twisted divisors associated to the $D7$-branes; setting all
twisted volumes $V_{\alpha}^{\rm twist.}$
and also all volumes $V_{\alpha'}^{\rm twist.}$ equal to each other,
the superpotential \wnpfinal\ reduces to (see eq.(6.17) in \DenefMM)
\eqn\wdenef{
W_{\rm np}=48~e^{-2\pi V_{\alpha}^{\rm twist.}}
+12~e^{-{2\pi V_{\alpha'}^{\rm twist.}\over 6}}
\, .}

\subsec{Moduli Stabilization in the resolved $\IZ_2\times \IZ_2$ orientifold revisited}

It has been pointed out in {\it Ref.} \ChoiSX\ and demonstrated in detail in 
subsection 3.6 for the case $h_{(1,1)}^{\rm untw.}=3,\ h_{(2,1)}^{\rm untw.}=1$,
that first integrating out the dilaton field $S$ and the complex structure moduli $U^j$
is not necessary to find a stable supersymmetric minimum w.r.t. the K\"ahler moduli $T^i$.
In fact, in the recent work \DenefMM, it has been shown that a stable supersymmetric
minimum is possible in the resolved $(T^2)^3/\IZ_2\times \IZ_2$ orientifold.
More concretely, the latter \tb compactification with $h_{(1,1)}=51,\ h_{(2,1)}=3$
allows for supersymmetric minima with all moduli fixed {\it at the same time}, \ie
without integrating out first $U^j$ and $S$. 
This result is interesting as the K\"ahler potential
involves a non--trivial dependence on the K\"ahler moduli which is quite different from the
cases discussed in section 3. Moreover, both $ISD$ and $IASD$ fluxes are turned on.

In this subsection, we shall investigate some aspects of the stabilization of \DenefMM.
With $r_i$ being the volume of the three two--tori $T^{2,i}$ and $t_{i\alpha,j\beta}$
the size of the $48$ exceptional divisors, the K\"ahler potential for the K\"ahler moduli
reads \DenefMM:
\eqn\kaehlerres{
K=-2\ \ln(V)=-2\ \ln(r^3-24 rt^2+48t^3)\ }
on the locus of moduli space, which maximally respects the symmetries between the divisors, \ie
$r=r_i,\ t_{i\alpha,j\beta}=t$. The parameters $r,t$ refer to the string moduli fields
(\cf the discussion in subsection 2.5). Hence,  with 
$\re T^1=e^{-\phi_{10}}\ \lf(\fc{1}{3}\fc{\p V}{\p r}+\fc{1}{6}\ \fc{\p V}{\p t}\ri)$ and 
$\re T^2=-\fc{1}{48}\ e^{-\phi_{10}}\ \fc{\p V}{\p t}$ 
following from the relation \NICEE, we determine the field--theoretical moduli to be:
\eqn\FTM{
T^1=e^{-\phi_{10}}\ (r^2-8rt+16t^2)+i \int_{C_1} C_4\ \ \ ,\ \ \ 
T^2=e^{-\phi_{10}}\ (rt-3 t^2) +i \int_{C_2} C_4\ .}
This allows us to express the K\"ahler potential \kaehlerres\ with the 
holomorphic coordinates:
\eqn\Kaehlerres{\eqalign{
K_{K}&=6\ln 2+2\ln(T^2+\ov T^2)-\ln\lf\{(T^1+\ov T^1)+ 2\ (T^2+\ov T^2)\ri.\cr
&\lf.-(T^1+\ov T^1)^{1/2}\ [T^1+\ov T^1+4\ (T^2+\ov T^2)]^{1/2}\ri\}\cr
&-2\ln\lf\{(T^1+\ov T^1)^2+16\ (T^1+\ov T^1)\ (T^2+\ov T^2)+32\ (T^2+\ov T^2)^2\ri.\cr
&+(T^1+\ov T^1)^{3/2}\ [(T^1+\ov T^1)+4\ (T^2+\ov T^2)]^{1/2}\cr
&\lf.+8\ (T^2+\ov T^2)\ (T^1+\ov T^1)^{1/2}\ [(T^1+\ov T^1)+4\ (T^2+\ov T^2)]^{1/2}\ri\}\ .}}
The K\"ahler potential for the dilaton and three (untwisted) complex structure moduli $U^j$,
with $U^j=U$, is given by (\cf section 2)
\eqn\givenCS{
K_{CS}=-\ln(S+\ov S)-3\ \ln(U+\ov U)\ .}
With these K\"ahler potentials, the tree--level superpotential $W_{\rm flux}(S,U^j)$ 
of \superpot\ and the non--perturbative superpotential \wdenef, 
with $V^{\rm twist.} _\alpha=T^2$ and $V_{\alpha'}^{\rm twist.}=T^1$, \ie the full superpotential 
\eqn\considerW{
W=W_{\rm flux}(S,U^j)+12\ \Lambda\ e^{-2\pi\ \fc{T^1}{6}}+48\ \Lambda\ e^{-2\pi\ T^2}\ ,}
a miminum is found at the values
$$T^1=5.84\ \ ,\ \ T^2=1.18\ \ ,\ \ U=0.84+0.46\ i\ \ ,\ \ S=3.64+7\ i\ ,$$
with $-\Lambda=e^{-K_{CS}/2}=5.87$. Note, that the factor $\Lambda$ has been introduced\foot{
In fact, in practice \DenefMM, one may first 
look for a minimum in the $T^1,T^2$--sector for constant $S,U^j$ and $W_{\rm flux}(S,U^j)$,
then promote this minimum to a minimum in the full moduli space $S,U,T^1,T^2$ 
through introducing the K\"ahler gauge. Of course, the values for $T^1,T^2$ do not change
this way.}
in \DenefMM\ as ``K\"ahler gauge''. 
Though the fluxes turned on are non--supersymmetric, supersymmetry is restored
after including the non--perturbative potential \wdenef\ and all moduli are stabilized at 
the above values. Indeed, we have checked that the mass matrix for the scalar fields 
$S,U,T^1,T^2$ is positive definite, with the following eight positive eigenvalues:
\eqn\poseigen{\eqalign{
\lf\{0.00010817\ ,\ 0.000128819\ ,\ 0.00030082\ ,\ 0.000354985\ ,\ri. \cr
\lf. 0.0105129\ ,\ 0.0112272\ ,\ 0.0925661\ ,\ 0.096739\ri\}\ .}}
Like the orbifold example discussed in subsection 3.6, this represents  a case where
the combined minimization of the moduli $S,U,T^1,T^2$ at one stroke leads to a stable 
minimum.
As one can see from the eigenvalues, the masses of the scalars of the dilaton and complex
structure moduli are by a factor of one hundred larger than those of the K\"ahler moduli.

The K\"ahler potentials for the blown up orbifold compactifications (\eg \eqq \Kaehlerres\ for 
$\IZ_2\times\IZ_2$ in the symmetric resolution)
take a quite different form from the ones for the cosets in \CosetsKAE, relevant to the 
unresolved orbifolds.
Hence the discussion about finding
stable minima in the context of orbifold limits of orientifolds in section 3 
has to be redone for those
more involved K\"ahler potentials \future.

One may now ask the question whether with a more complicated K\"ahler potential such
as \Kaehlerres, first integrating out
the complex structure modulus is  possible, \ie whether a minimum may still be found.
To check this, let us start with the superpotential
\eqn\startW{
W=\al_1+\al_2\ S+12\ e^{-2\pi\ \fc{T^1}{6}}+48\ e^{-2\pi\ T^2}\ \ ,\ \ \al_i\in \IC}
and the K\"ahler potential \Kaehlerres\ for the K\"ahler moduli and 
$K_S=-\ln(S+\ov S)$ for the dilaton field. This may be a situation which appears after 
integrating out one complex structure modulus 
or it may mimic the situation of resolved orbifolds without complex structure moduli.
To minimize the potential, we proceed like in the previous section: At the extremum for
which $F_S,F_{T^1},F_{T^2}=0$, we solve for the parameters $\al_1$ and $\al_2$, obtaining
a non--trivial relation between the parameters $\al_1,\al_2$ and the moduli $S,T^1,T^2$.
However, after plugging these relations into the mass matrix, we find that the latter
has negative eigenvalues. Hence, there is in general no minimum for this setup with
one complex structure modulus integrated out, though
stable minima may be found within the K\"ahler moduli sector $T^1,T^2$, while keeping
$S$ fixed (\cf the previous footnote).  On the other hand to discuss the situation in the above
discussed $\IZ_2\times \IZ_2$ orientifold one has to integrate out three complex structure moduli.
The results, discussed in subsection 3.7, should be combined with the above non--perturbative
superpotential.

To conclude, while for orbifolds with complex structure moduli it is always possible to find
a stable minimum both at the orbifold point and after resolving, 
orbifolds without complex structure moduli do not seem to allow for moduli stabilization
both at the orbifold point and after their resolution.
This issue among others will be discussed in more detail in \future.


\subsec{Outlook: Resolved Orbifolds}

As the analysis in section 3 has shown, all the those orbifolds with {\it no} complex structure moduli do {\it not} allow a stable vacuum. This leaves only the $\IZ_{6-II}$--orbifolds on the different lattices, the $\IZ_2\times \IZ_2$-- and the $\IZ_2\times \IZ_6$--orbifold in the game. But can these receive contributions to the non-perturbative superpotential at the orbifold point? 

From $D3$--brane instantons, it must be checked whether any of the divisors present fulfill the criterion $\chi=1$. Exceptional divisors, resulting from blowing up the orbifold singularities  are likely to  contribute, as they are fixed at the locations of the singularities and therefore have no moduli.

Let us now consider contributions originating from gaugino--condensates. 
$\IZ_2\times\IZ_2$ and $\IZ_2\times \IZ_6$ have each 12 $O7$--planes using the standard orientifold involution, sitting at the planes corresponding to $z^i=0,\half,\half\tau,\half(1+\tau),\ z^j, z^k={\rm free}$, where $\tau$ is the modular parameter of the torus corresponding to $z^i$. Canceling the tadpoles locally by placing the $D7$--branes on top of the $O7$--planes, we end up having bifundamental matter, as the divisors wrapped by the $D7$--branes intersect. This means that we do not have a pure SYM--theory and therefore no contribution to the non-perturbative superpotential can arise. Resolving the singularities via blow-ups cures this problem, as the arising exceptional divisors change the intersection pattern of the divisors such that the divisors which are wrapped by the $D7$--branes no longer intersect. That we get gaugino condensates for $\IZ_2\times \IZ_2$ has been shown in \DenefMM. For  $\IZ_2\times \IZ_6$, the question whether there is adjoint matter must be settled first. 
The  $\IZ_{6-II}$--orbifolds have four stacks of $O7$--planes with the exception of the $SU(2)\times SU(6)$--lattice which has one stack of $O7$--planes. For these cases, the question of existence of gaugino condensates at the orbifold point remains to be studied in detail.

When resolving the orbifold singularities, all orbifolds except for the prime orbifolds $\IZ_3,\,\IZ_7$ and $\IZ_3\times \IZ_3$ which do not allow $O7$--planes are possible candidates for moduli stabilization via gaugino--condensates. 
It is more difficult to settle the case for contributions coming from $D3$--brane instantons, as an analysis of the $F$-theory lift of the model seems inevitable, yet for many of the orbifolds in question, no $F$-theory lift is known to exist.
At any rate, orbifold models without complex structure moduli do not 
exhibit a stable uplift even in the resolved case \future.

We will address the questions outlined above in {\it Ref.} \future. 
The first steps will be the local analysis of the models at the location of the orbifold singularities via the methods of toric geometry. In a next step, the local patches must be glued together to form the compact model. An orientifold--action must be determined, and after the divisors which are wrapped by the $D7$--branes have been identified and their topology analyzed, it is possible to ascertain the existence of gaugino--condensates. 

As an illustration, we present here the toric diagrams of the blow-ups of the local models for $\IZ_{6-II}$ and $\IZ_2\times \IZ_6$. Near the location of a fixed point, the orbifold $T^6/\IZ_N$ looks like ${\bf C}^3/\IZ_N$, irrespective of the lattice the orbifold lives on. The lattice becomes important when the local patches are glued together to form the compact model, as for each lattice, the configuration of the singularities turns out to be different.
\fig{Toric diagrams of the resolutions of ${\bf C}^3/\IZ_{6-II}$ and dual graphs}{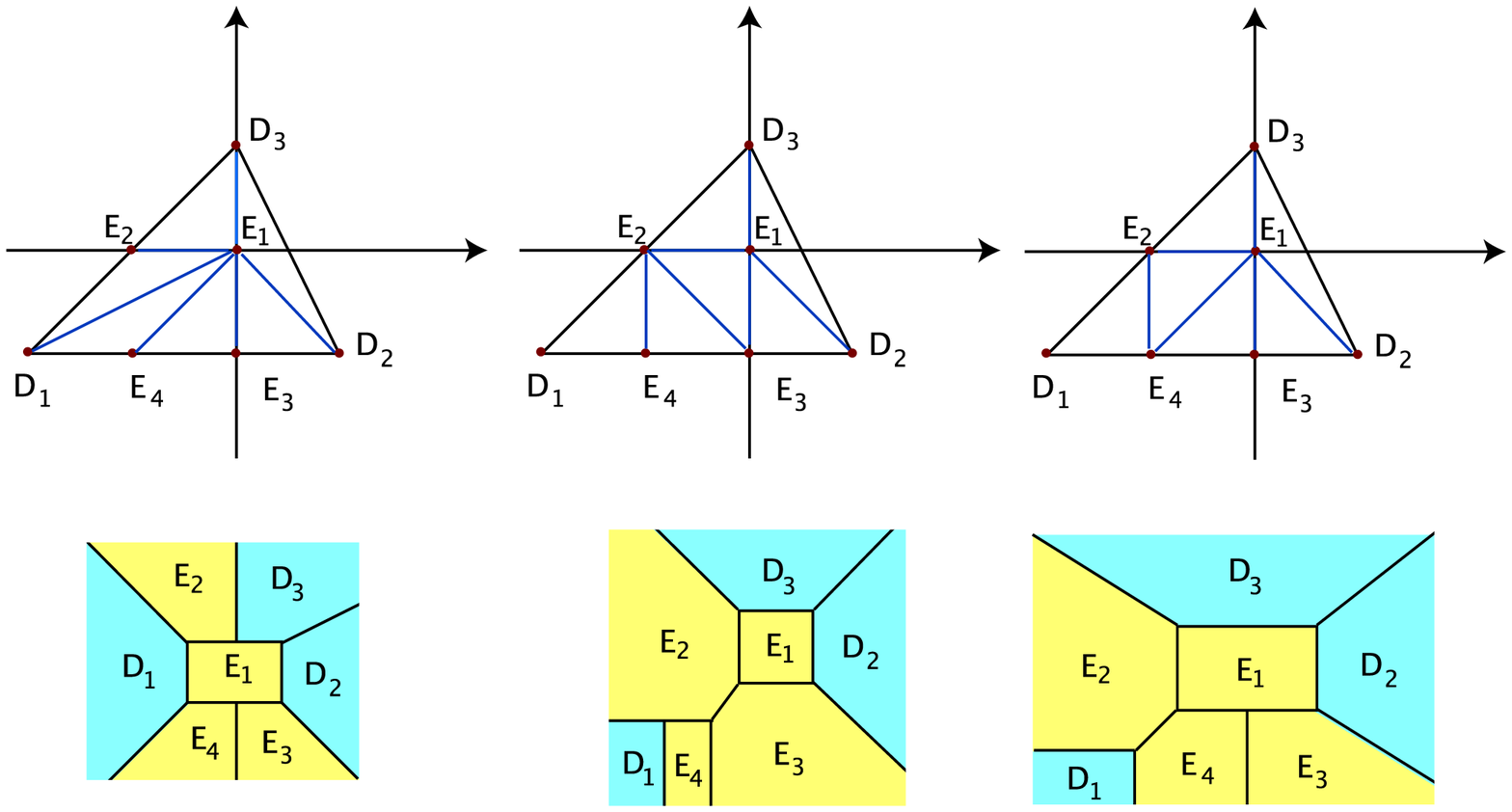}{16truecm}

Figure 1 shows three of the five possible resolutions of ${\bf C}^3/\IZ_{6-II}$. The other two can be obtained by taking case b) and flopping the curve $E_1\cdot E_2$ to $E_3\cdot D_2$ and taking case c) and flopping the curve $E_1\cdot E_3$ to $D_3\cdot E_4$. The $D_i$ denote the divisors inherited from the unresolved geometry, whereas the $E_i$ denote the exceptional divisors. There is one compact exceptional divisor, namely $E_1$. The diagrams in the bottom row are the dual graphs, where faces have become vertices and vice versa.

Figure 2 shows one of the 156 possible resolutions\foot{The number of possible triangulations was  obtained using the package TOPCOM.} of ${\bf C}^3/\IZ_2\times \IZ_6$. Here, we have two compact exceptional divisors, namely $E_7$ and $E_8$.
\fig{Toric diagram of one of the resolutions of ${\bf C}^3/\IZ_{2}\times\IZ_6$ and dual graph}{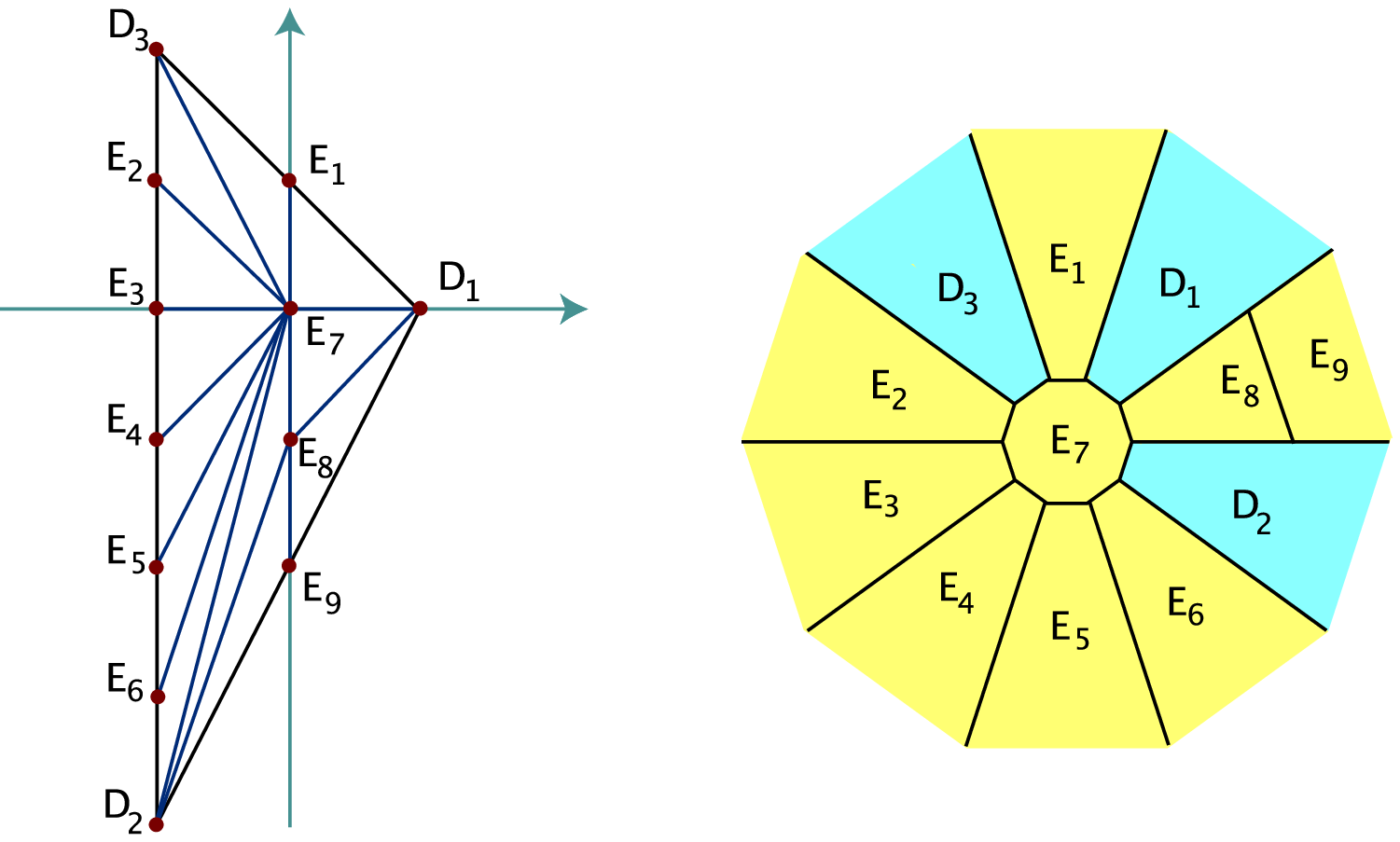}{16truecm}

\newsec{Conclusions}
This paper deals with moduli stabilization in type $IIB$ orientifolds \`a la
KKLT, i.e. with tree--level $3$--form flux superpotential plus
non-perturbative superpotential from $D3$-instantons and/or gaugino
condensation.
The main emphasis of the work are orientifold compactifications
in their various orbifold limits.
We show that it is indeed possible to find stable (i.e. tachyon free),
supersymmetric AdS-minima with stabilized dilaton, untwisted K\"ahler $T^i$ and
untwisted complex structure moduli  $U^j$, as long as the 
geometrical orbifold group still allows
for the existence of untwisted complex structure moduli fields.
On the other hand, if the orbifold group action already freezes all
complex structure moduli, then the 
scalar potential in the $S,T^i$-sector is such that
the (mass)$^2$ matrix for these fields contains negative eigenvalues.
We also point out some problems with the integrating out procedure
of complex structure moduli, namely we investigate
cases, where integrating out the
$U^j$ leads to non-stable AdS-vacua in the remaining $S,T^i$ potential,
whereas the minimization of the full $S,T^i,U^j$ potential does not
suffer from any instabilities.

In the paper we give a complete classification of
the untwisted moduli spaces of all type $IIB$ orientifolds,
together with the flux quantizations of the Ramond and NS $3$--form fluxes on
the orbifold spaces. We also provide the correct definitions of the
untwisted K\"ahler moduli fields for type $IIB$ orientifolds, including also
the orbifolds with more than three K\"ahler moduli $T^i$. 
However we would like to emphasize 
that we do not claim in this paper that we have constructed 
full-fledged type $IIB$ orientifold models
with {\sl all} moduli  fixed. The reasoning for this caveat is two-fold:
\vskip0.1cm
\noindent
(i) We mainly stick to the orbifold limit. Hence we do not analyze the
stability properties of the blow-up (i.e. twisted) moduli, which are
nevertheless still present.

\vskip0.1cm
\noindent
(ii) We do not discuss in great detail  the 
form and the microscopic origin of the non-perturbative superpotential,
but  just parameterize it 
by a
sum of exponential functions given 
in terms
of the untwisted K\"ahler moduli. Again, the blowing up
of the orbifold singularities will be very important for a more
complete analysis.

\vskip0.1cm
\noindent
Both these points as well as
other aspects will be addressed in a forthcoming paper \future.
In this work we just remark on some properties of the 
potential including blowing-up moduli
of the concrete type orientifold, which was recently
constructed in \DenefMM.
Specifically, we extend the work of \DenefMM\ 
by writing the K\"ahler potential and the
superpotential in terms of the correct 
supergravity variables
for the K\"ahler moduli fields. In this way, a K\"ahler potential is
obtained which differs quite a bit from the standard K\"ahler potentials
which are used in string compactifications on orbifold spaces.

\vskip20pt

\noindent{\bf Acknowledgements:} We are grateful to R. Blumenhagen,
G. Curio, B. Florea, 
W. Lerche, P. Mayr,
E. Scheidegger and P. Tripathy 
for useful discussions. This work is supported in part by the
Deutsche Forschungsgemeinschaft as well as by the
EU-RTN network {\sl Constituents, Fundamental Forces and Symmetries
of the Universe} (MRTN-CT-2004-005104). S.R. and W.S.
thank the university of Munich for hospitality.

\vskip20pt
\goodbreak


\appendix\appA{Complex structures of $\IZ_N$-- and $\IZ_N\times\IZ_M$--orbifolds}

In this appendix, we will derive the complex structures and K\"ahler moduli of all the 
$\IZ_N$-- and $\IZ_N\times\IZ_M$--orbifolds listed in tables 1 and 2 which were not treated 
already in the main text, i.e. 
$\IZ_{6-I},\,\IZ_{6-II},\, \IZ_7,\,\IZ_{12-I}$, $\IZ_3\times \IZ_3, \ \IZ_2\times \IZ_6,\ 
\IZ_2\times \IZ_{6'}$ and $\IZ_6\times \IZ_6$.

Throughout the whole section, we shall work with the heterotic anti--symmetric $2$--form $B_2$.
We have already explained in section 2, that the $2$--form components entering
the K\"ahler moduli have to be replaced by the
corresponding part of the Ramond $4$--form in the \tb orientifolds we are discussing.
This map is obvious due to Poincar\'e duality on $X_6$.
 

\subsec{$\IZ_{6-I}$--orbifold}

The action of the $\IZ_{6-I}$--twist on the complex coordinates is as follows:
\eqn\twistactionsixi{
z^i \lra e^{2\pi i v^i}\ z^i\ \ \ ,\ \ \ v^1=\fc{1}{6}\ ,\ v^2=\fc{1}{6}\ ,\ v^3=-\fc{2}{6}\ .}

\noindent{\it $G_2^2\times SU(3)$--lattice}
\vskip0.5cm

On the root lattice of $G_2^2\times SU(3)$, the twist $Q$ has the following action (see discussion in section 3):
\eqn\twistactionii{\eqalign{
Q\ e_1&=2\,e_1+3\,e_2\ \ \ ,\ \ \ Q\ e_2=-e_1 -e_2\ ,\cr
Q\ e_3&=2\,e_3+3\,e_4\ \ \ ,\ \ \ Q\ e_4=-e_3-\ e_4\ ,\cr
Q\ e_5&=e_6\ \ \ ,\ \ \ Q\ e_6=-e_5-e_6\ .}}
The twist $Q$ allows for five independent real deformations of the metric $g$
and five real deformations  of the
anti--symmetric tensor $b$. As before, these results follow again from solving the equations
$Q^tgQ=g$ and $Q^tbQ=b$:
\eqn\metricsixi{
g=\pmatrix{R_1^2&-{1\over 2}R_1^2&R_1R_3\cos\theta_{13}&y&0&0\cr
-{1\over 2}R_1^2&{1\over 3}R_1^2&{1\over\sqrt3}R_1R_3\cos\theta_{23} &{1\over 3}R_1R_3\cos\theta_{13}&0&0\cr
R_1R_3\cos\theta_{13}&{1\over\sqrt3}R_1R_3\cos\theta_{23}&R_3^2&-{1\over 2}R_3^2&0&0\cr
y&{1\over3}R_1R_3\cos\theta_{13}&-{1\over 2}R_3^2&{1\over 3} R_3^2&0&0\cr
0&0&0&0&R_5^2&-\half R_5^2\cr
0&0&0&0&-\half R_5^2&R_5^2},
}
with $y=-{1\over3}(3R_1R_3\cos\theta_{13}-\sqrt3 R_1R_3\cos\theta_{23})$ and the five real parameters $R_1^2,\ R_3^2,\ R_5^2,\ \theta_{13},\ \theta_{23}$. For the choice $\theta_{13}=\theta_{23}=\pi/2$, we get the metric one obtains from the Cartan matrices back. For $b$ we find
\eqn\bfieldsixi{
b=\pmatrix{0&b_1&3\,b_2&-3b_2-b_3&0&0\cr
-b_1&0&b_3&b_2&0&0\cr
-3\,b_2&-b_3&0&b_4&0&0\cr
3b_2+b_3&-b_2&-b_4&0&0&0\cr
0&0&0&0&0&b_5\cr
0&0&0&0&-b_5&0}
}
with the five real parameters $b_1,\ b_2,\ b_3,\ b_4,\ b_5$. We see that we get 5 untwisted K\"ahler moduli in this orbifold, while the complex structure is completely fixed.

It is possible to fix the complex structure up to a few constants by choosing a suitable ansatz such that it repects the twist without the use of the lattice vectors:

\eqn\cplxzsixi{\eqalign{
dz^1=&a\,(-(1+e^{2\pi i/6})\,dx^1+dx^2)+b\,(-(1+e^{2\pi i/6})\,dx^3+dx^4),\cr
dz^2=&c\,(-(1+e^{2\pi i/6})\,dx^1+dx^2)+d\,(-(1+e^{2\pi i/6})dx^3+dx^4),\cr
dz^3=&e\,(e^{-2\pi i/3}dx^5+dx^6).
}}
$a,\,b,\,c,\,d$ and $e$ are constants left unfixed by the twist alone. In the following, we will choose $a,\,d,\,e$ such that $dx^1,\,dx^3,\,dx^5$ have the coefficient one and set $b=c=0$, so the complex structure takes the following form:
\eqn\cplxzsixisim{\eqalign{
dz^1=&dx^1+{1\over \sqrt3}\,e^{5\pi i/6}\,dx^2,\cr
dz^2=&dx^3+{1\over \sqrt3}\,e^{5\pi i/6}\,dx^4,\cr
dz^3=&3^{1/4}\,(dx^5+e^{2\pi i/3}\,dx^6).
}}
Now we proceed as outlined in section 2. From the metric \metricsixi\ we can easily read off the K\"ahler form, which expressed in the complex coordinates \cplxzsixisim\ reads
\eqn\Kaehlersixicplx{\eqalign{
-i\,J=&\,R_1^2\ dz^1\wedge d\ov z^1+R_3^2\ dz^2\wedge d\ov z^2+{1\over\sqrt3}\,R_5^2\ dz^3\wedge d\ov z^3\cr
&+2\,R_1R_3\,[(e^{2\pi i/6}\cos\theta_{13}+i\cos\theta_{23})\ dz^2\wedge d\ov z^1+(e^{-2\pi i/6}\cos\theta_{13}-i\cos\theta_{23})\ dz^1\wedge d\ov z^2].
}}
To be able to read off the K\"ahler moduli, we must look at the real cohomology. The five untwisted $(1,1)$--forms that are invariant under this orbifold twist are
\eqn\invformsixi{\eqalign{
\om_1&=dx^1\wedge dx^2,\quad \om_2=dx^3\wedge dx^4,\quad \om_3=dx^5\wedge dx^6,\cr
\om_4&=dx^2\wedge dx^3-dx^1\wedge dx^4,\quad \om_5=3\ dx^1\wedge dx^3-3\ dx^1\wedge dx^4+dx^2\wedge dx^4.
}}
The $B$--field \bfieldsixi\ has the simple form
\eqn\bfieldsixico{
B=b_1\ \om_1+b_4\ \om_2+b_5\ \om_3+b_3\ \om_4+b_2\ \om_5.
}
The K\"ahler form expanded in the real cohomology is
\eqn\Kaehlersixireal{\eqalign{
J=&{1\over 2\sqrt3}\ \{\, R_1^2\ \om_1+R_3^2\ \om_2+3\,R_5^2\ \om_3\cr
&-2\,R_1R_3\,[\,\cos\theta_{13}\ \om_4-{2\over\sqrt3}(13\sqrt3\,\cos\theta_{13}+{29}\cos\theta_{23})\ \om_5]\}.
}}
Via $J+i\,B={\cal T}^i\,\om_i$ the K\"ahler moduli can now be easily read off:
\eqn\Kaehlermodsixi{\eqalign{
{\cal T}^1&={1\over2\sqrt3}\,R_1^2+i\,b_1,\quad {\cal T}^2={1\over2\sqrt3}\,R_3^2+i\,b_4,\quad {\cal T}^3={\sqrt3\over2}\,R_5^2+i\,b_5,\cr
{\cal T}^4&=-{1\over\sqrt3}\,R_1R_3\cos\theta_{13}+i\,b_3,\quad {\cal T}^5={1\over3}\,R_1R_3\,(13\sqrt3\,\cos\theta_{13}+{29}\,\cos\theta_{23})+b_2.
}}


\vskip0.8cm
\noindent{\it $G_2\times (SU(3))^2$--lattice with generalized Coxeter element}
\vskip0.5cm

On the root lattice of $G_2\times (SU(3))^2$, we act with the generalized Coxeter twist $Q=S_1S_2S_3S_4P_{36}P_{45}$, as explained in section 3. It has the following action:
\eqn\twistactionii{\eqalign{
Q\ e_1&=2\,e_1+3\,e_2\ \ \ ,\ \ \ Q\ e_2=-e_1 -e_2\ ,\cr
Q\ e_3&=e_6\ \ \ ,\ \ \ Q\ e_4=e_5\ ,\cr
Q\ e_5&=-e_3-e_4\ \ \ ,\ \ \ Q\ e_6=e_4\ .}}
As before, the twist $Q$ allows for five independent real deformations of the metric $g$
and five real deformations  of the
anti--symmetric tensor $b$:
\eqn\metricsixii{
g=\pmatrix{R_1^2&-{1\over 2}R_1^2&x&-x&-x&x\cr
-{1\over 2}R_1^2&{1\over 3}R_1^2&-y &x&y&z\cr
x&-y&R_5^2&-{1\over 2}R_5^2&-2\,R_5^2\cos\theta_{46}&R_5^2\cos\theta_{46}\cr
-x&x&-{1\over 2}R_5^2& R_5^2&R_5^2\cos\theta_{46}&R_5^2\cos\theta_{46}\cr
-x&y&-2\,R_5^2\cos\theta_{46}&R_5^2\cos\theta_{46}&R_5^2&-\half R_5^2\cr
x&z&R_5^2\cos\theta_{46}&R_5^2\cos\theta_{46}&-\half R_5^2&R_5^2},
}
with $x={1\over\sqrt3}(R_1R_5(\cos\theta_{25}+\cos\theta_{26}),\, y={1\over\sqrt3}R_1R_5\cos\theta_{25}, \,z={1\over\sqrt3}R_1R_5\cos\theta_{26}$ and the five real parameters $R_1^2,\ R_5^2,\ \theta_{25},\ \theta_{26}$ and $\theta_{46}$. 
For $b$ we find
\eqn\bfieldsixiI{
b=\pmatrix{0&b_1&2\,b_2+b_3&-b_2-2\,b_3&-2\,b_2-b_3&b_2-b_3\cr
-b_1&0&-b_2&b_2+b_3&b_2&b_3\cr
-2\,b_2-b_3&b_2&0&-b_5&0&-b_4\cr
b_2+2\,b_3&-b_2-b_3&b_5&0&-b_4&b_4\cr
2\,b_2+b_3&-b_2&0&b_4&0&b_5\cr
-b_2+b_3&-b_2&b_4&-b_4&-b_5&0}
}
with the five real parameters $b_1,\ b_2,\ b_3,\ b_4,\ b_5$. 

For the complex structure, we find
\eqn\cplxzsixi{\eqalign{
dz^1=&a\,(-(1+e^{2\pi i/6})\,dx^1+dx^2)+b\,(e^{-2\pi i/3}dx^3+dx^4+e^{2\pi i/6}dx^5+e^{-2\pi i/6}dx^6),\cr
dz^2=&c\,(-(1+e^{2\pi i/6})\,dx^1+dx^2)+d\,(e^{-2\pi i/3}dx^3+dx^4+e^{2\pi i/6}dx^5+e^{-2\pi i/6}dx^6),\cr
dz^3=&e\,(e^{-2\pi i/3}dx^3+dx^4+e^{-2\pi i/3}dx^5+e^{2\pi i/3}\,dx^6).
}}
$a,\,b,\,c,\,d$ and $e$ are constants left unfixed by the twist. We choose them such that we get
\eqn\cplxzsixiIsim{\eqalign{
dz^1=&3^{-1/3}\,(dx^1+{1\over\sqrt3}\,e^{5\pi i/6}\,dx^2),\cr
dz^2=&{1\over\sqrt2}\,(dx^3+e^{2\pi i/3}\,dx^4-dx^5+e^{2\pi i/6}\,dx^6),\cr
dz^3=&{1\over\sqrt2}\,(dx^3+e^{2\pi i/3}\,dx^4+dx^5-e^{2\pi i/6}\,dx^6).
}}
The five untwisted $(1,1)$--forms that are invariant under this orbifold twist are
\eqn\invformsixiI{\eqalign{
\om_1&=dx^1\wedge dx^2,\cr 
\om_2&=2\,dx^1\wedge dx^3-dx^1\wedge dx^4-2\,dx^1\wedge dx^5+dx^1\wedge dx^6-dx^2\wedge dx^3+dx^2\wedge dx^4+dx^2\wedge dx^5,\cr
\om_3&=dx^1\wedge dx^3-2\,dx^1\wedge dx^4-dx^1\wedge dx^5-dx^1\wedge dx^6+dx^2\wedge dx^4+dx^2\wedge dx^6,\cr
\om_4&=dx^3\wedge dx^6+dx^4\wedge dx^5-dx^4\wedge dx^6,\cr
\om_5&=dx^5\wedge dx^6.
}}
The $B$--field \bfieldsixiI\ has the simple form
$B=b_1\ \om_1+b_2\ \om_2+b_3\ \om_3+b_4\ \om_4+b_5\ \om_5$.
Looking at the K\"ahler form expanded in the real cohomology and with $J+i\,B={\cal T}^i\,\om_i$ the K\"ahler moduli are found to be
\eqn\Kaehlermodsixi{\eqalign{
{\cal T}^1&={1\over2\sqrt3}\,R_1^2+i\,b_1,\quad {\cal T}^2={1\over3}\,R_1R_5\,(\cos\theta_{25}+20\,\cos\theta_{26})+i\,b_2,\cr
{\cal T}^3&=-R_1R_5\,(4\,\cos\theta_{25}-\cos\theta_{26})+i\,b_3,\cr
{\cal T}^4&={3\sqrt3\over2}\,R_5^2\,\cos\theta_{46}+i\,b_4,\quad {\cal T}^5={-\sqrt3}\,R_5^2+i\,b_5.
}}


\subsec{$\IZ_{6-II}$--orbifold}

The action of the $\IZ_{6-II}$--twist on the complex coordinates is as follows:
\eqn\twistactionsixii{
z^i \lra e^{2\pi i v^i}\ z^i\ \ \ ,\ \ \ v^1=\fc{1}{6}\ ,\ v^2=\fc{2}{6}\ ,\ v^3=-\fc{3}{6}\ .}


\vskip0.8cm
\noindent{\it $\IZ_{6-II}$ on $(SU(2))^2\times SU(3)\times G_2$}
\vskip0.5cm

The lattices of $G_2$ and $SU(3)$ are each two--dimensional, while the lattice of $SU(2)$ is one-dimensional. We associate the lattice basis vectors  $e_1,\,e_2$ with $G_2$, $e_3,\, e_4$ with SU(3) and $e_5,\,e_6$ with the two $SU(2)$s.
The orbifold twist expressed in the complex basis reads
\eqn\twistactionsix{
z^i \lra e^{2\pi i v^i}\ z^i\ \ \ ,\ \ \ v^1=\fc{1}{6}\ ,\ v^2=\fc{2}{6}\ ,\ v^3=-\fc{3}{6}\ .}
When examining \twistactionsix, we see that the twist of $z^3$ is merely a $\IZ_2$--reflection. It must be associated with the one--dimensional lattices $SU(2)$. The twist action for this case is $e_5 \to -e_5,\quad e_6\to -e_6$.

The twist on $z^2$ corresponds to a $\IZ_3$--action. This twist must clearly live on the $SU(3)$--lattice, whose simple roots are $(1,0)$ and $(-1/2, \sqrt3/2)$. The action of the Coxeter element of $SU(3)$ results in $e_3\to e_4,\quad e_4\to -e_3-e_4$.

The twist on $z^1$ is the $\IZ_6$ twist, living on the $G_2$--root lattice. The Coxeter element acts as $e_1\to 2\,e_1+3\,e_2,\quad e_2\to -e_1-e_2$ which fulfills $q^6=1$.
So the full twist $Q$ acts on the six roots $e_i$ in the following way:
\eqn\twistactionsixr{\eqalign{
Q\, e_1&= 2\,e_1+3\,e_2,\quad Q\,e_2=-e_1-e_2,\cr
Q\,e_3&=e_4,\quad Q\,e_4=-e_3-e_4,\cr
Q\,e_i&=-e_i,\quad i=5,6.
}}
Now we solve for the metric and antisymmetric tensor, using $Q^tg\,Q=g$ and $Q^tb\,Q=b$. We find the following solution for $g$ (which is incidentally the same we get directly from the Cartan matrix):
\eqn\metric{
g=\ \pmatrix{R_1^2&-1/2R_1^2&0&0&0&0\cr
-1/2R_1^2&1/3 R_1^2&0&0&0&0\cr
0&0&R_3^2&-1/2R_3^3&0&0\cr
0&0&-1/2R_3^2&R_3^2&0&0\cr
0&0&0&0&R_5^2&R_5R_6\cos(\theta_{56})\cr
0&0&0&0&R_5R_6\cos(\theta_{56})&R_6^2}.}
As can be seen, $R_1,\,R_3,\, R_5,\, R_6$ and $\theta_{56}$ are (real) free parameters. The $\IZ_2$--twists leave their part of the metric completely undetermined. 
For the antisymmetric tensor $b$, we get
\eqn\bfield{
b=\pmatrix{0&b_1&0&0&0&\cr
-b_1&0&0&0&0&0\cr
0&0&0&b_2&0&0\cr
0&0&-b_2&0&0&0\cr
0&0&0&0&0&b_3\cr
0&0&0&0&-b_3&0},}
with the three real parameters  $b_1,b_2,b_3$. Three of the free parameters of $g$ can be combined with the three free parameters of $b$ into three complex K\"ahler moduli, the remaining two free parameters of $g$ form one complex structure modulus.

Using the ansatz \ansatz\ and normalizing the first term to one, we find the following complex structure:
\eqn\dzsix{\eqalign{
dz^1&=dx^1+{1\over \sqrt3}e^{5\pi i/6}dx^2,\cr
dz^2&=dx^3+e^{2\pi i/3}dx^4,\cr
dz^3&={1\over \sqrt{2{\rm Im}\, {\cal U}^3}}(\,dx^5+{\cal U}^3dx^6),\quad{\rm with}\quad {\cal U}^3={R_6\over R_5}e^{i\theta_{56}}
}}
the complex structure modulus.
The invariant 2-forms are in this relatively simple case $dx^1\wedge dx^2, \ dx^3\wedge dx^4$ and $dx^5\wedge dx^6$. 
The K\"ahler form in complex coordinates has the form
\eqn\kcplx{-i\,J=R_1^2\, dz^1\wedge d\ov z^1+R_3^2\, dz^1\wedge d\ov z^1+2\,R_5R_6\,\sin\theta_{56}\,dz^3\wedge d\ov z^3.}
Expressed in the real cohomology, it takes the form
\eqn\kreal{J={1\over2\sqrt3}\,R_1^2\, dx^1\wedge dx^2+ {\sqrt3\over2}\,R_3^2\, dx^3\wedge dx^4+R_5R_6\sin\theta_{56} \, dx^5\wedge dx^6,}
so via $J+i\,B={\cal T}^i\,\om_i$, we find the three K\"ahler moduli to take the following form:
\eqn\moduli{\eqalign{
\Tc^1&={1\over2\sqrt3}\,R_1^2+i\,b_1,\cr
\Tc^2&={\sqrt3\over2}\,R_3^2+i\,b_2,\cr
\Tc^3&=R_5R_6\sin\theta_{56}+i\,b_3.}}


\vskip0.5cm
\noindent{\it $SU(6)\times SU(2)$--lattice}
\br

On the root lattice of
$SU(6)\times SU(2)$, 
the twist $Q$ acts on the six roots $e_i$ in the following way:
\eqn\twistactioni{\eqalign{
Q\ e_i&=e_{i+1}\ \ \ ,\ \ \ i=1,\ldots 4\ ,\cr
Q\ e_5&=-e_1-e_2-e_3-e_4-e_5\ ,\cr
Q\ e_6&=-e_6\ .}}
The twist $Q$ allows for five independent real deformations of the metric $g$
and three real deformations  of the
anti--symmetric tensor $b$. As before, these results follow from solving the equations
$Q^tg\,Q=g$ and $Q^tb\,Q=b$ which leads to:
\eqn\metrici{\eqalign{
g=\pmatrix{R_1^2&R_1^2\ x& R_1^2\ \alpha_{15}&R_1^2\ \alpha_{14}&R_1^2\
\alpha_{15}&R_1R_6\ \alpha_{16}\cr
R_1^2\ x&R_1^2&R_1^2\ x&R_1^2\ \alpha_{15}&R_1^2\ \alpha_{14}&-R_1R_6\ \alpha_{16}\cr
R_1^2\ \alpha_{15}&R_1^2\ x&R_1^2&R_1^2\ x&R_1^2\ \alpha_{15}&R_1R_6\ \alpha_{16}\cr
R_1^2\ \alpha_{14}&R_1^2\ \alpha_{15}&R_1^2\ x&R_1^2&R_1^2\ x&-R_1R_6\ \alpha_{16}\cr
R_1^2\ \alpha_{15}&R_1^2\ \alpha_{14}&R_1^2\ \alpha_{15}&R_1^2\ x&R_1^2&R_1R_6\ \alpha_{16}\cr
R_1R_6\ \alpha_{16}&-R_1R_6\ \alpha_{16}&R_1R_6\ \alpha_{16}&-R_1R_6\ \alpha_{16}&R_1R_6\ 
\alpha_{16}&R_6^2}},}
$x=-\h(1+\alpha_{14}+2\alpha_{15}),$ 
with the arbitrary real parameters $R_1^2,R_6^2,\alpha_{14},\alpha_{15},\alpha_{16}$  and 
\eqn\bfieldi{
b=\pmatrix{0&b_1&b_2&0&-b_2&b_3\cr
-b_1&0&b_1&b_2&0&-b_3\cr
-b_2&-b_1&0&b_1&b_2&b_3\cr
0&-b_2&-b_1&0&b_1&-b_3\cr
b_2&0&-b_2&-b_1&0&b_3\cr
-b_3&b_3&-b_3&b_3&-b_3&0}\ ,}
with the arbitrary real parameters $b_1,b_2,b_3$.
The $\sigma$--model action for the closed string reads in this lattice basis:
\eqn\sigmaaction{
\Sc=\int d^2z\  \ov\p x^i(\ov z)\ (g+b)_{ij}\ \p x^j(z)\ .}
The aim is to transform the action \sigmaaction\ into a complex basis $\{z^i\}_{i=1,2,3}$, 
where the twist $Q$ acts diagonally on the complex coordinates.
There are several such bases, the second requirement is that the three K\"ahler moduli $\Tc^j$
and complex structure moduli have to decouple from each other.

In the complex basis, the $\sigma$--model has the following action\foot{In the case of 
rectangular complex coordinates, i.e. $z^i=x^i+iy^i$, the action takes the form 
${\cal S}={1\over 2}\int d^2z\,[\ {b_{(1,2)}}_{ij}\p z^i\ov\p z^j+{b_{(1,1)}}_{i\ov j}
\p z^i\ov\p\ov z^{\ov j}+{\rm h.c}\ ]$, where ${b_{(1,2)}}$ is the matrix containing the 
complex structure moduli. The form of the matrices is determined via the twist, and the number 
of untwisted moduli is manifest in the action (see \eg\CveticYW).}:
\eqn\complexsigma{
{\cal S}={1\over 2}\int d^2z\,[\,{b_{(1,1)}}_{i\ov j}\p z^i\ov\p\ov z^{\ov j}+{\rm h.c}\ ]\ ,}
where $b_{(1,1)}$ is the matrix containing the K\"ahler moduli, i.e. $(g+b)$ transformed to the 
complex basis. The action \complexsigma\  must be invariant under the twist and the form of the 
matrix $b_{(1,1)}$ changes depending on the twist.
Again, we transform the action \sigmaaction\ into a complex basis $\{z^i\}_{i=1,2,3}$, 
where the twist $Q$ acts diagonally on the complex coordinates.
We use the vielbein
\eqn\sechsbeini{\eqalign{
e_i&=\sum_{j=1,3,5} A_j\ \lf\{\cos[(i-1)\kappa_j\alpha+\phi_j]\ \tilde e_j+
\sin[(i-1)\kappa_j\alpha+\phi_j]\ \tilde e_{j+1}\ri\}\ \ \ ,\ \ \ i=1,\ldots,5\ ,\cr
e_6&=R_6\ \lf[\ \cos(\Delta+\phi_5)\ \tilde e_5+\sin(\Delta+\phi_5)\ \tilde e_6\ \ri]\ ,}}
with
\eqn\withi{\eqalign{
A_1^2&=\fc{1}{6}\ R_1^2\ (1-3\alpha_{14}-4\alpha_{15})\ ,\cr
A_3^2&=\fc{1}{2}\ R_1^2\ (1+\alpha_{14})\ ,\cr
A_5^2&=\fc{1}{3}\ R_1^2\ (1+2\alpha_{15})\ ,\cr
\cos(\Delta)&=\fc{\sqrt 3\ \alpha_{16}}{\sqrt{1+2\alpha_{15}}}\ ,}}
and $\alpha=\fc{\pi}{3}$, $\kappa_1=1,\ \kappa_3=2$ and $\kappa_5=3$.
The angles $\phi_i$ are arbitrary reflecting the freedom of how
to embed our six--dimensional lattice into the orthonormal system 
$\{\tilde e_i\}_{i=1,\ldots,6}$.
After having moved into the real orthonormal basis $\{\tilde e_i\}_{i=1,\ldots,6}$, we
have to find the unitary transformation, which brings us to the complex basis $\{z^i\}_{i=1,2,3}$.
Here, we find:
\eqn\worki{\eqalign{
dz^1&=dx^1+e^{2\pi i/6} dx^2+e^{2\pi i/3} dx^3-dx^4+e^{-2\pi i/3}dx^5\ ,\cr
dz^2&=dx^1+e^{2\pi i/3} dx^2+e^{-2\pi i/3}dx^3+dx^4+e^{2\pi i/3}dx^5\ ,\cr
dz^3&=\fc{1}{2\sqrt{3}}\ \lf[
\fc{1}{3}\,(dx^1-dx^2+dx^3-dx^4+dx^5)+\Uc^3\ dx^6\ \ri]\ .}}
We have set the three free angles $\phi_i$ to zero after having realized, that they act on each 
$z^i$ just as an overall phase. In the basis \worki\ the $\sigma$--model action \complexsigma\ 
takes the form 
\eqn\Sigmaaction{
\Sc=\h\int d^2z\  \sum_{j=1}^3\  \Tc^j\ \ov\p \ov z^j\, \p z^j+{\rm h.c.}\ ,}
where we defined ${\cal T}^i={b_{(1,1)}}_{i\ov i}$ and $
{\cal U}^3={b_{(1,2)}}_{33}$, with the three K\"ahler moduli $\Tc^i$:\foot{The three invariant 2-forms of the real cohomology are in this case
\eqn\invrealb{\eqalign{
\om_1&=dx^1\wedge dx^2+dx^2\wedge dx^3+dx^3\wedge dx^4+dx^4\wedge dx^5,\cr
\om_2&=dx^1\wedge dx^3-dx^1\wedge dx^5+dx^2\wedge dx^4+dx^3\wedge dx^5,\cr
\om_3&=dx^1\wedge dx^6-dx^2\wedge dx^6+dx^3\wedge dx^6-dx^4\wedge dx^6+dx^5\wedge dx^6.
}}
It is possible to re-parameterize this result obtained via the vielbein to the language of the other examples presented.}
\eqn\modulii{\eqalign{
\Tc^1&=R_1^2\ (1-3\ \alpha_{14}-4\ \alpha_{15})+2\ i\ \sqrt 3\ (b_1+b_2)\ ,\cr
\Tc^2&=\sqrt 3\ R_1^2\ (1+\alpha_{14})+2\ i\ (b_1-b_2)\ ,\cr
\Tc^3&=\sqrt 3\ R_1R_6\ \sqrt{1+2\ \alpha_{15}-3\ \alpha_{16}^2}+3\ i\ b_3\ ,}}
and the complex structure modulus $\Uc^3$:
\eqn\complexstr{
\Uc^3=\fc{R_6}{R_1}\ \fc{\alpha_{16}+\fc{i}{3}\sqrt 3
\sqrt{1+2\ \alpha_{15}-3\ \alpha_{16}^2}}{1+2\alpha_{15}}\ .}


\vskip0.8cm
\noindent{\it The $(SU(2))^2\times (SU(3))^2$--lattice with generalized Coxeter element}
\vskip0.5cm
 
This time, we associate $e_1$ and $e_2$ with $\IZ_2$. The generalized Coxeter element $Q=S_1S_2S_3P_{36}P_{45}$ contains transpositions of the roots of the $SU(3)$--factors. Using \Weyl\ and the Cartan matrix of $SU(3)$, we find the following for the total twist:
\eqn\twistactionsixrii{\eqalign{
Q\, e_1&= -e_1,\quad Q\,e_2=-e_2,\cr
Q\,e_3&=-e_5,\quad Q\,e_4=-e_5+e_6,\cr
Q\,e_5&=e_4,\quad Q\,e_6=e_3.}
}
{From} $Q^tg\,Q=g$ we find the following $g$:
\eqn\gsixii{ g=\pmatrix{R_1^2&R_1R_2\cos\theta_{12}&0&0&0&0\cr
R_1R_2\cos\theta_{12}& R_3^2&0&0&0&0\cr
0&0&R_2^2&-\half R_3^2&-2\,R_3^2\cos\theta_{46}&R_3^2\cos\theta_{46}\cr
0&0&-\half R_3^2&R_3^2&R_3^2\cos\theta_{46}&R_3^2\cos\theta_{46}\cr
0&0&-2\,R_3^2\cos\theta_{46}&R_3^2\cos\theta_{46}&R_3^2&-\half R_3^2\cr
0&0&R_3^2\cos\theta_{46}&R_3^2\cos\theta_{46}&-\half R_3^2&R_3^2},}
$R_1,\, R_2,\,R_3,\,\theta_{12}$ and $\theta_{46}$  being its five real deformation parameters.
For the antisymmetric tensor $b$, we get
\eqn\bfieldsixii{
b=\pmatrix{0&b_1&0&0&0&\cr
-b_1&0&0&0&0&0\cr
0&0&0&-b_3&0&b_2\cr
0&0&b_3&0&b_2&-b_2\cr
0&0&0&-b_2&0&b_3\cr
0&0&-b_2&b_2&-b_3&0}\ ,}
with the three real parameters  $b_1,\,b_2,\,b_3$. 
This leads to the complex structure
\eqn\cplxabc{
dz^1={1\over\sqrt3}(dx^3-e^{2\pi i/3}\,dx^4),\quad dz^2={1\over2}(dx^5-e^{2\pi i/3}\,dx^6),\quad dz^3={1\over\sqrt{2\,{\rm Im}\,{\cal U}^3}}(\,dx^1+{\cal U}^3\,dx^2\,),
}
with ${\cal U}^3=R_2/R_1\,e^{i\theta_{12}}$.
The invariant real 2--forms are
\eqn\invtwof{\eqalign{
\om_1&=dx^1\wedge dx^2,\quad \om_2=dx^3\wedge dx^6+dx^4\wedge dx^5-dx^4\wedge dx^6,\cr
\om_3&=-dx^3\wedge dx^4+dx^5\wedge dx^6.
}}
Via $B+i\,J={\cal T}^i\,\om_i$, we find
\eqn\kahlerm{
{\cal T}^1=b_1+R_1R_2\,\sin\theta_{12},\quad {\cal T}^2=b_2+i\,{3\over\sqrt3}{2}\,R_3^2,\quad {\cal T}^3=b_3-i\,2\sqrt3\,R_3^2\,\cos\theta_{46}.}


\vskip0.8cm
\noindent{\it The $SU(3)\times SO(8)$--lattice}
\vskip0.5cm

On the root lattice of
$SU(3)\times SO(8)$, 
the twist $Q$ acts on the six roots $e_i$ in the following way:
\eqn\twistactioni{\eqalign{
Q\ e_1&=e_2,\quad Q\ e_2=e_1+e_2+e_3+e_4 ,\cr
Q\ e_3&=-e_1-e_2-e_3,\quad Q\ e_4=-e_1-e_2-e_4,\cr
Q\ e_5&=e_6,\quad Q\ e_6=-e_5-e_6\ .}}
The twist $Q$ allows for five independent real deformations of the metric $g$
and three real deformations  of the
anti--symmetric tensor $b$:
\eqn\metrici{\eqalign{
g=\pmatrix{x&y&R_1^2\cos\theta_{13}&R_3^2+R_1^2(-1+\cos\theta_{13})&0&0\cr
y&x&z&z&0&0\cr
R_1^2\cos\theta_{13}&z&R_1^2&R_1R_3\cos\theta_{34}&0&0\cr
R_3^2+R_1^2(-1+\cos\theta_{13})&z&R_1R_3\cos\theta_{34}&R_3^2&0&0\cr
0&0&0&0&R_5^2&-\half R_5^2\cr
0&0&0&0&-\half R_5^2&R_5^2},}}
with $x=R_1^2\cos\theta_{13}+R_3(R_3+R_1\cos\theta_{34}),\ y=-\half (R_1^2-2R_3^2-R_1(3R_1\,\cos\theta_{13}+R_3\cos\theta_{34})),\ z=-\half R_1(R_1+R_1\cos\theta_{13}+R_3\cos\theta_{34})$ 
and the arbitrary real parameters $R_1^2,\, R_3^2\,R_5^2,\,\theta_{13 },\,\theta_{34}$  and 
\eqn\bfieldi{
b=\pmatrix{0&b_1+b_2&-b_2&b_2&0&0\cr
-b_1-b_2&0&b_1+2\,b_2&b_1&0&0\cr
b_2&-b_1-2\,b_2&0&b_2&0&0\cr
-b_2&-b_1&-b_2&0&0&0\cr
0&0&0&0&0&b_3\cr
0&0&0&0&-b_3&0}.}
This leads to the complex structure
\eqn\cplxabcd{\eqalign{
dz^1&=dx^1+e^{2\pi i/6}\,dx^2-dx^3-dx^4,\quad dz^2=dx^5+e^{2\pi i/3}\,dx^6,\cr
dz^3&={1\over\sqrt{{\rm Im}\,{\cal U}^3}}\,(dx^1-dx^2+dx^3+{\cal U}^3\,(dx^1-dx^2+dx^4)),
}}
with 
\eqn\U{\eqalign{
{\cal U}^3&=-2+{1\over R_1\,(2+\cos\theta_{13}-R_3\cos\theta_{34})}\left[\,3 \,R_1\,(1+\cos\theta_{13})-i\,\sqrt{3}\,(R_1)^{-1/2}\times\right.\cr
&\times(\,2\,R_1R_3^2-R_1^3+R_1\cos\theta_{13}\,(R_1^2+R_3^2-2R_2R_3\cos\theta_{34})\cr
&\left.-R_3\cos\theta_{34}(R_3^2-R_1^2+R_1R_3\cos\theta_{34}))^{1/2}\right].
}}
The invariant 2--forms of the real cohomology are
\eqn\invreal{\eqalign{
\om_1&=dx^1\wedge dx^2+dx^2\wedge dx^3+dx^2\wedge dx^4,\cr
\om_2&=dx^1\wedge dx^2-dx^1\wedge dx^3+dx^1\wedge dx^4+2\,dx^2\wedge dx^3+dx^3\wedge dx^4\cr
\om_3&=dx^5\wedge dx^6.
}}
With the invariant 2--forms above, we find the following K\"ahler moduli:
\eqn\km{\eqalign{
{\cal T}^1&={\sqrt3\over2}\,R_1\,(R_1\,(1-\cos\theta_{13})+R_3\,\cos\theta_{34})),\cr
{\cal T}^2&={1\over2\sqrt3 }\,(-3\,R_1^2\,(-1+\cos\theta_{13})+3\,R_1R_3\cos\theta_{34}+10\sqrt{R_1}\times\cr
&\times(2\,R_1R_3^2-R_1^3+R_1\cos\theta_{13}\,(R_1^2+R_3^2-2R_2R_3\cos\theta_{34})\cr
&-R_3\cos\theta_{34}(R_3^2-R_1^2+R_1R_3\cos\theta_{34}))^{1/2},\cr
{\cal T}^3&={\sqrt3\over2}\,R_5^2.}
}


\subsec{$\IZ_7$--orbifold}

The action of the twist on the complex coordinates has the form
\eqn\twistactiontw{
z^i \lra e^{2\pi i v^i}\ z^i\ \ \ ,\ \ \ v^1=\fc{1}{7}\ ,\ v^2=\fc{2}{7}\ ,\ v^3=-\fc{3}{7}\ .}

Here, the (only possible) torus lattice for the $\IZ_7$--orbifold is the root lattice of $SU(7)$, with
the twist $Q$ acting on the six roots $e_i$ in the following way:
\eqn\twistaction{\eqalign{
Q\ e_i&=e_{i+1}\ \ \ ,\ \ \ i=1,\ldots 5\ ,\cr
Q\ e_6&=-e_1-e_2-e_3-e_4-e_5-e_6\ .}}
The twist $Q$ allows for three independent real deformations of the metric $g$
and three  real deformations  of the
anti--symmetric tensor $b$:
\eqn\metric{
g=R^2\ \pmatrix{1&\alpha_{12}&\alpha_{13}&x&x&\alpha_{13}\cr
\alpha_{12}&1&\alpha_{12}&\alpha_{13}&x&x\cr
\alpha_{13}&\alpha_{12}&1&\alpha_{12}&\alpha_{13}&x\cr
x&\alpha_{13}&\alpha_{12}&1&\alpha_{12}&\alpha_{13}\cr
x&x&\alpha_{13}&\alpha_{12}&1&\alpha_{12}\cr
\alpha_{13}&x&x&\alpha^2_{13}&\alpha_{12}&1}\ \ \ ,\ \ \ x=-\h-\alpha_{12}-\alpha_{13}\ ,}
with the three real parameter $R^2,\alpha_{12},\alpha_{13}$ and  
\eqn\bfield{
b=\pmatrix{0&b_1&b_2&b_3&-b_3&-b_2\cr
-b_1&0&b_1&b_2&b_3&-b_3\cr
-b_2&-b_1&0&b_1&b_2&b_3\cr
-b_3&-b_2&-b_1&0&b_1&b_2\cr
b_3&-b_3&-b_2&-b_1&0&b_1\cr
b_2&b_3&-b_3&-b_2&-b_1&0}\ ,}
with the three real parameter  $b_1,b_2,b_3$.
The $\sigma$--model action for the closed string reads in this lattice basis:
\eqn\sigmaaction{
\Sc=\int d^2z\  \ov\p x^i(\ov z)\ (g+b)_{ij}\ \p x^j(z)\ .}
The aim is to transform the action \sigmaaction\ into a complex basis $\{z^i\}_{i=1,2,3}$, 
where the twist $Q$ acts diagonally on the complex coordinates:
\eqn\twistaction{
z^i \lra e^{2\pi i v^i}\ z^i\ \ \ ,\ \ \ v^1=\fc{1}{7}\ ,\ v^2=\fc{2}{7}\ ,\ v^3=-\fc{3}{7}\ .}
There are many such bases, but the second requirement is that the three K\"ahler moduli $\Tc^j$
and complex structure moduli have to decouple from each other.
As a first step, the metric \metric\ may be expressed through the Sechsbein $e$, \ie
$g=e^te$.
This may be obtained from {\it Ref.} \structure, where the lattice vectors $e_i$ are 
expressed as a linear combination of a set of six real orthonormal 
basis vectors $\tilde e_i$
\eqn\sechsbein{
e_i=\sum_{j=1,3,5} R_j\ \lf\{\cos[(i-1)\kappa_j\alpha+\phi_j]\ \tilde e_j+
\sin[(i-1)\kappa_j\alpha+\phi_j]\ \tilde e_{j+1}\ri\}\ ,}
with
\eqn\with{\eqalign{
R_1^2&=R^2\ 
[\alpha_{12}\ (\alpha_5^2-\alpha_1^2)+\alpha_{13}\ (\alpha_5^2-\alpha_3^2)+\h\ \alpha_5^2]\ ,\cr
R_3^2&=R^2\ 
[\alpha_{12}\ (\alpha_1^2-\alpha_3^2)+\alpha_{13}\ (\alpha_1^2-\alpha_5^2)+\h\ \alpha_1^2]\ ,\cr
R_5^2&=R^2\ 
[\alpha_{12}\ (\alpha_3^2-\alpha_5^2)+\alpha_{13}\ (\alpha_3^2-\alpha_1^2)+\h\ \alpha_3^2]\ ,\cr
\alpha_i^2&=\fc{4}{7}\ [1-\cos(b_i\alpha)]\ \ \ ,\ \ \ i=1,3,5\ ,}}
and $\alpha=\fc{2\pi}{7}$, $\kappa_1=1,\ \kappa_3=2$ and $\kappa_5=4$.
Furthermore, the angles $\phi_i$ are arbitrary reflecting the freedom of how
to embedd our six--dimensional lattice into the orthonormal system 
$\{\tilde e_i\}_{i=1,\ldots,6}$.
After having moved into the real orthonormal basis $\{\tilde e_i\}_{i=1,\ldots,6}$, we
have to find the unitary transformation, which brings us to the complex basis $\{z^i\}_{i=1,2,3}$.
After some work, we find:
\eqn\work{\eqalign{
dz^1&=dx^1+(-1)^{2/7}\ dx^2+(-1)^{4/7}\ dx^3+(-1)^{6/7}\ dx^4-(-1)^{1/7}\ dx^5-
(-1)^{3/7}\ dx^6\ ,\cr
dz^2&=dx^1+(-1)^{4/7}\ dx^2-(-1)^{1/7}\ dx^3-(-1)^{5/7}\ dx^4+(-1)^{2/7}\ dx^5+
(-1)^{6/7}\ dx^6\ ,\cr
dz^3&=dx^1-(-1)^{1/7}\ dx^2+(-1)^{2/7}\ dx^3-(-1)^{3/7}\ dx^4+(-1)^{4/7}\ dx^5-
(-1)^{5/7}\ dx^6\ .}}
We have set the three free angles $\phi_i$ to zero after having realized that they act on each 
$z^i$ just as an overall phase. Invariance under the $Z_7$--twist is fulfilled for a $\sigma$--model action of the form
\eqn\Sigmaactionseven{
\Sc=\h\int d^2z\  \sum_{j=1}^3\  \Tc^j\ \ov\p \ov z^j(\ov z)\ \p z^j(z)+{\rm hc.}\ ,}
with the three K\"ahler moduli:
\eqn\moduli{\eqalign{
\Tc^1&=R_1^2+\fc{4}{7}\ i\ \lf[b_3\ \sin\lf(\fc{\pi}{7}\ri)+b_1\ \sin\lf(\fc{2\pi}{7}\ri)+b_2\ 
\sin\lf(\fc{3\pi}{7}\ri)\ri]\ ,\cr
\Tc^2&=R_3^2-\fc{4}{7}\ i\ \lf[b_2\ \sin\lf(\fc{\pi}{7}\ri)+b_3\ \sin\lf(\fc{2\pi}{7}\ri)-b_1\ 
\sin\lf(\fc{3\pi}{7}\ri)\ri]\ ,\cr
\Tc^3&=R_5^2-\fc{4}{7}\ i\ \lf[b_1\ \sin\lf(\fc{\pi}{7}\ri)-b_2\ \sin\lf(\fc{2\pi}{7}\ri)+b_3\ 
\sin\lf(\fc{3\pi}{7}\ri)\ri]\ .}}
{From} the expression \work\ we may read off the complex numbers $\tau^j_i$ and $\rho^j_i$
appearing in \achieved\ (with the replacement $dx^{2i-1}\ra dx^i$ and  $dx^{2i}\ra dy^i$).
In particular, we see that all complex structures 
are already fixed through the orbifold twist.
Hence, no further restrictions follow from flux quantization rules.
Furthermore, the K\"ahler form is given by:
\eqn\kaehlerform{
J=\Tc^1\ dz^1\wedge d\ov z^1+\Tc^2\ dz^2\wedge d\ov z^2+\Tc^3\ dz^3\wedge d\ov z^3\ .}
Hence, the two flux components, which survive the orbifold twist (\cf Table 4),
fulfill the primitivity condition $G_3\wedge J=0$\foot{For completeness, we also give the invariant 2--forms of the real cohomology:
\eqn\realseven{\eqalign{
\om_1&=dx^1\wedge dx^2+dx^2\wedge dx^3+dx^3\wedge dx^4+dx^4\wedge dx^5+dx^5\wedge dx^6,\cr
\om_2&=dx^1\wedge dx^3-dx^1\wedge dx^6+dx^2\wedge dx^4+dx^3\wedge dx^5+dx^4\wedge dx^6,\cr
\om_3&=dx^1\wedge dx^4-dx^1\wedge dx^5+dx^2\wedge dx^5-dx^2\wedge dx^6+dx^3\wedge dx^6.
}}
}.


\subsec{$\IZ_{12-I}$--orbifold}

The action of the twist on the complex coordinates has the form
\eqn\twistactiontw{
z^i \lra e^{2\pi i v^i}\ z^i\ \ \ ,\ \ \ v^1=\fc{1}{12}\ ,\ v^2=\fc{4}{12}\ ,\ v^3=-\fc{5}{12}\ .}


\noindent{\it $SU(3)\times F_4$--lattice}
\br

Here, the torus lattice is the root lattice of $SU(3)\times F_4$. The action of the twist upon the roots is
\eqn\twelve{\eqalign{
Qe_1&=e_2,\quad Qe_2=-e_1-e_2,\quad Qe_3=e_4,\cr
Qe_4&=e_3+3_4+2e_5,\quad Qe_5=e_6,\quad Qe_6=-e_3-e_4-e_5-e_6.}}
The twist allows for 3 real deformations of the metric, which has the form
\eqn\mtwelve{
g=\pmatrix{R_1^2&-\half R_1^2&0&0&0&0\cr
\half R_1^2& R1^2&0&0&0&0\cr
0&0&R_3^2&R_3^2\,\cos\theta_{34}&x&x\cr
0&0&R_3^2\,\cos\theta_{34}&R_3^2& -\half R_3^2&x\cr
0&0&x&-\half R_3^2&\half R_3^2&\half R_3^2\,\cos\theta_{34}\cr
0&0&x&x&\half R_3^2\,\cos\theta_{34}&\half R_3^2
},}
with $x=R_3^2\,\cos\theta_{34}$ and $R_1,\,R_3$ and $\cos\theta_{34}$ arbitrary free parameters. Also for the $B$--field, the twist allows 3 real deformations:
\eqn\btwelve{
b=\pmatrix{0& b_1&0&0&0&0\cr
-b_1&0&0&0&0&0\cr
0&0&0&2b_3&b_2&-b_2\cr
0&0&-2b_3&0&2b_3&b_2\cr
0&0&-b_2&-2b_3&0&b_3\cr
0&0&b_2&-b_2&-b_3&0},}
with $b_1,\,b_2,\,b_3$ the arbitrary parameters, so we have three untwisted K\"ahler moduli and no complex structure moduli. 
The three invariant 2--forms in the real cohomology are
\eqn\realtwelvei{\eqalign{
\om_1&=dx^1\wedge dx^2,\cr
\om_2&=dx^3\wedge dx^5-dx^3\wedge dx^6+dx^4\wedge dx^6,\cr
\om_3&=2\,dx^3\wedge dx^4+2\,dx^4\wedge dx^5+dx^5\wedge dx^6
.}}
The complex structure turns out to be as follows:
\eqn\fcplx{\eqalign{
dz^1=&3^{-1/4}\,(dx^3+e^{2\pi i/6}dx^4+{1\over \sqrt2}[e^{11\pi i/12}dx^5+e^{\pi i/12}dx^6]),\cr
dz^2=&3^{-1/4}\,(dx^1+e^{2\pi i/3}dx^2),\cr
dz^3=&3^{-1/4}\,(dx^3-e^{2 \pi i/12}dx^4+{1\over\sqrt2}[\,e^{5\pi i/12}dx^5+e^{-5\pi i/12}dx^6]).}}
For the K\"ahler moduli, we find
\eqn\ffourk{\eqalign{
{\cal T}^1=&{{\sqrt3\over2}\,R_1^2}+{i\,b_1},\cr
{\cal T}^2=&{{3\over4}\,R_3^2}(-1+2\cos\theta_{34})+i\,b_2,\cr
{\cal T}^3=&{9\over4}\,R_3^2+i\,b_3.
}}


\vskip0.5cm
\noindent{\it The $E_6$--lattice}
\vskip0.5cm

When we choose the $E_6$--lattice for the torus lattice, the Coxeter twist acts as
\eqn\twelve{\eqalign{
Qe_1&=e_2,\quad Qe_2=e_3,\quad Qe_3=e_1+e_2+e_3+e_4+e_6,\cr
Qe_4&=e_5,\quad Qe_5=-e_1-e_2-e_3-e_4-e_5,\quad Qe_6=-e_1-e_2-e_3-e_6.}}
The twist allows for 3 real deformations of the metric, which has the form
\eqn\mtwelve{
g=\pmatrix{R_5^2&x&\half(R_6^2-R_5^2)&R_5^2-R_6^2&y&-z\cr
x&R_5^2&x&y&R_5^2-R_6^2&-z\cr
\half(R_6^2-R_5^2)&x&R_5^2&-\half R_5^2&y&-\half R_6^2\cr
R_5^2-R_6^2&y&-\half R_5^2&R_5^2&x&z\cr
y&R_5^2-R_6^2&y&x&R_5^2&z\cr
-R_5R_6\cos\theta_{56}&-z&-\half R_6^2&z&z&R_6^2
},}
with $x=-\half R_5^2+R_5R_6 \cos\theta_{56},\ y=\half(R_6^2-R_5^2)-R_5R_6\cos\theta_{56},\,z=R_5R_6\cos\theta_{56}$
and $R_5,\,R_6$ and $\cos\theta_{56}$ arbitrary free parameters. Also for the $B$--field, the twist allows 3 real deformations:
\eqn\btwelve{
b=\pmatrix{0& b_1&b_2-b_1-b_3&0&b_1-b_2&b_3\cr
-b_1&0&b_1&b_2-b_1&0&-b_3\cr
-b_2+b_1+b_3&-b_1&0&b_1-b_3&b_2-b_1&b_2\cr
0&-b_2+b_1&-b_1+b_3&0&b_1&-b_3\cr
b_2-b_1&0&-b_2+b_1&-b_1&0&b_3\cr
-b_3&b_3&-b_2&b_3&-b_3&0},}
with $b_1,\,b_2,\,b_3$ the arbitrary parameters, so we have as expected three untwisted K\"ahler moduli and no complex structure moduli. 
The invariant 2--form of the real cohomology are
\eqn\realtwelveii{\eqalign{
\om_1=&dx^1\wedge dx^2-dx^1\wedge dx^3+dx^1\wedge dx^5+dx^2\wedge dx^3-dx^2\wedge dx^4+dx^3\wedge dx^4\cr
&-dx^3\wedge dx^5+dx^4\wedge dx^5,\cr
\om_2=&dx^1\wedge dx^3-dx^1\wedge dx^5+dx^2\wedge dx^4+dx^3\wedge dx^5+dx^3\wedge dx^6,\cr
\om_3=&-dx^1\wedge dx^3+dx^1\wedge dx^6-dx^2\wedge dx^6-dx^3\wedge dx^4-dx^4\wedge dx^6+dx^5\wedge dx^6.
}}
The complex structure is
\eqn\cplxtwelveii{\eqalign{
dz^1=&{1\over\sqrt6}\,(dx^1+e^{2\pi i/12}\,dx^2+e^{2\pi i/6}\,dx^3-dx^4+e^{-10\pi i/12}dx^5-(1-i\,)\,e^{2\pi i/6}\,dx^6)\cr
dz^2=&{1\over2}\,(dx^1+e^{2\pi i/3}dx^2+e^{-2\pi i/3}\,dx^3+dx^4+e^{2\pi i/3}dx^5),\cr
dz^3=&{1\over3^{1/4}\sqrt6}\,(dx^1+e^{-2\pi i/12}dx^2+e^{2\pi i/6}\,dx^3-dx^4+e^{2\pi i/12}dx^5-(1+i\,)\,e^{2\pi i/6}\,dx^6).
}}
Again we pair $J+i\,B={\cal T}^i\,\om_i$ in the real cohomology and get the following K\"ahler moduli:
\eqn\kaehlertwelveii{\eqalign{
{\cal T}^1&=2\sqrt3\,(2\,R_5^2-R_6^2)+i\,b_1,\cr
{\cal T}^2&=-({\sqrt3}\,R_5^2+(1+\sqrt3)\,R_6^2+R_5R_6\,\cos\theta_{56})+i\,b_2,\cr
{\cal T}^3&={5\over6}\,R_6^2-6\,R_5R_6\,\cos\theta_{56}+i\,b_3
.}}


\subsec{$\IZ_3\times \IZ_3$--orbifold}

The two twists act on the complex coordinate as follows:
\eqn\twistactionthth{\eqalign{
Q_1:\ z^i& \lra e^{2\pi i v_1^i}\ z^i\ \ \ ,\ \ \ v_1^1=\fc{1}{3}\ ,\ v_1^2=0\ ,\ v_1^3=-\fc{1}{3}\ ,\cr
Q_2:\ z^i& \lra e^{2\pi i v_2^i}\ z^i\ \ \ ,\ \ \ v_2^1=0\ ,\ v_2^2=\fc{1}{3}\ ,\ v_2^3=-\fc{1}{3}\ .
}}
The combined twist $Q_3=Q_2Q_1$ is again the well-known $Z_{3}$--twist:
\eqn\twistcomb{
Q_3:\ z^i \lra e^{2\pi i v_3^i}\ z^i\ \ \ ,\ \ \ v_3^1=\fc{1}{3}\ ,\ v_3^2=\fc{1}{3}\ ,\ v^3=-\fc{2}{3}\ .
}
We expect the root lattice of $SU(3)\times SU(3)\times SU(3)$ to be the correct one.
Lead by past experience, we choose for the twists acting on the lattice basis:
\eqn\twistactionththl{\eqalign{
Q_1\, e_1&= e_2,\quad Q_1\,e_2=-e_1-e_2,\quad Q_1\,e_3=e_3,\quad Q_1\,e_4=e_4,\cr
Q_1\,e_5&=e_6,\quad Q_1\,e_6=-e_5-e_6,\cr
Q_2\, e_1&= e_1,\quad Q_2\,e_2=e_2,\quad Q_2\,e_3=e_4,\quad Q_2\,e_4=-e_3-e_4,\cr
Q_2\,e_5&=e_6,\quad Q_2\,e_6=-e_5-e_6.
}}
The twists are the usual Coxeter--twists on $SU(3)$ and reproduce the correct eigenvalues and the condition $Q^3=1$. The combined twist $Q_3$ has the form
\eqn\twico{\eqalign{
Q_3\, e_1&= e_2,\quad Q_3\,e_2=-e_1-e_2,\cr
Q_3\,e_3&=e_4,\quad Q_3\,e_4=-e_3-e_4,\cr
Q_3\,e_5&=-e_5-e_6,\quad Q_3\,e_6=e_5,
}}
and also reproduces the required eigenvalues. The twist on $e_5,\,e_6$ is the anti-twist of the usual Coxeter--twist. We require the metric to be invariant under all three twists, i.e. we impose the three conditions $Q_i^TgQ_i=g,\quad i=1,2,3$. This leads to the following solution:
\eqn\g{g=\pmatrix{R_1^2&-\half R_1^2&0&0&0&0\cr
-\half R_1^2&R_1^2&0&0&0&0\cr
0&0&R_3^2&-\half R_3^2&0&0\cr
0&0&-\half R_3^2&R_3^2&0&0\cr
0&0&0&0&R_5^2&-\half R_5^2\cr
0&0&0&0&-\half R_5^2&R_5^2.}}
This corresponds exactly to the metric of $SU(3)^3$ without any extra degrees of freedom.
The solution for $b$ matches the pattern:
\eqn\b{b=\pmatrix{0&b_1&0&0&0&0\cr
-b_1&0&0&0&0&0\cr
0&0&0&b_2&0&0\cr
0&0&-b_2&0&0&0\cr
0&0&0&0&0&-b_3\cr
0&0&0&0&-b_3&0},}
we therefore know to have three K\"ahler moduli whereas the complex structure is completely fixed (recall that in the simple $Z_3$--twist, we had nine K\"ahler moduli). For the complex structure we get
\eqn\dzs{\eqalign{
dz^1&=3^{1/4}\,(dx^1+e^{2\pi i/3}dx^2),\cr
dz^2&=3^{1/4}\,(dx^3+e^{2\pi i/3}dx^4),\cr
dz^3&=3^{1/4}\,(dx^5+e^{-2\pi i/3}dx^6).}}
Examination of the K\"ahler form yields
\eqn\ks{\eqalign{
\Tc^1&={\sqrt3\over2}\,R_1^2+{i}b_1,\cr
\Tc^2&={\sqrt3\over2}\,R_3^2+{i}b_2,\cr
\Tc^3&={\sqrt3\over2}\,R_5^2+{i}b_3.}}


\subsec{$\IZ_2\times \IZ_6$--orbifold}

The two twists act on the complex coordinate as follows:
\eqn\twistactionthth{\eqalign{
Q_1:\ z^i& \lra e^{2\pi i v_1^i}\ z^i\ \ \ ,\ \ \ v_1^1=\fc{1}{2}\ ,\ v_1^2=0\ ,\ v_1^3=-\fc{1}{2}\ ,\cr
Q_2:\ z^i& \lra e^{2\pi i v_2^i}\ z^i\ \ \ ,\ \ \ v_2^1=0\ ,\ v_2^2=\fc{1}{6}\ ,\ v_2^3=-\fc{2}{6}\ .
}}
The combined twist $Q_3=Q_2Q_1$ is a $Z_{6-II}$--twist:
\eqn\twistcomb{
Q_3:\ z^i \lra e^{2\pi i v_3^i}\ z^i\ \ \ ,\ \ \ v_3^1=-\fc{3}{6}\ ,\ v_3^2=\fc{1}{6}\ ,\ v^3=\fc{2}{6}\ .
}
We find the root lattice of $SU(2)^2\times SU(3)\times G_2$ to be the correct one.
We choose for the twists acting on the lattice basis:
\eqn\twistactionththl{\eqalign{
Q_1\, e_1&=- e_1,\quad Q_1\,e_2=-e_2,\quad Q_1\,e_3=e_3,\quad Q_1\,e_4=e_4,\cr
Q_1\,e_5&=-e_5,\quad Q_1\,e_6=-e_6,\cr
Q_2\, e_1&= e_1,\quad Q_2\,e_2=e_2,\quad Q_2\,e_3=2\,e_3+3\,e_4,\quad Q_2\,e_4=-e_3-e_4,\cr
Q_2\,e_5&=-e_6,\quad Q_2\,e_6=e_5+e_6.
}}
The twists reproduce the correct eigenvalues and the conditions $Q_1^2=1,\ Q_2^6=1$. While the other twists are the usual Coxeter-twists, the $Q_2$-twist on $e_5,\,e_6$ is a generalized Coxeter--twist on $SU(3)$, namely $S_1P_{12}$. The combined twist $Q_3$ has the form
\eqn\twico{\eqalign{
Q_3\, e_1&= -e_1,\quad Q_3\,e_2=-e_2,\cr
Q_3\,e_3&=2\,e_3+3\,e_4,\quad Q_3\,e_4=-e_3-e_4,\cr
Q_3\,e_5&=e_6,\quad Q_3\,e_6=-e_5-e_6,
}}
and also reproduces the required eigenvalues. We require the metric to be invariant under all three twists, i.e. we impose the three conditions $Q_i^Tg\,Q_i=g,\quad i=1,2,3$. This leads to the following solution:
\eqn\g{g=\pmatrix{R_1^2&R_1R_2\,\cos\theta_{12}&0&0&0&0\cr
R_1R_2\,\cos\theta_{12}&R_1^2&0&0&0&0\cr
0&0&R_3^2&-\half R_3^2&0&0\cr
0&0&-\half R_3^2&{1\over3} R_3^2&0&0\cr
0&0&0&0&R_5^2&-\half R_5^2\cr
0&0&0&0&-\half R_5^2&R_5^2.}}
The solution for $b$ matches the pattern:
\eqn\b{b=\pmatrix{0&b_1&0&0&0&0\cr
-b_1&0&0&0&0&0\cr
0&0&0&b_2&0&0\cr
0&0&-b_2&0&0&0\cr
0&0&0&0&0&-b_3\cr
0&0&0&0&-b_3&0},}
we therefore know to have three K\"ahler moduli whereas the complex structure is completely fixed. For the complex structure we get
\eqn\dzs{\eqalign{
dz^1&={1\over\sqrt{2\,{\rm Im}\,{\cal U}^3}}\,(dx^1+{\cal U}^3\, dx^2),\cr
dz^2&=dx^3+{1\over\sqrt3}e^{10\pi i/12}dx^4,\cr
dz^3&=dx^5+e^{2\pi i/3}dx^6,}}
with ${\cal U}^3=R_2/R_1 e^{i\theta_{12}}$.
Examination of the K\"ahler form yields
\eqn\ks{\eqalign{
\Tc^1&=R_1R_2\,\sin\theta_{12}+{i}b_1,\cr
\Tc^2&={1\over2\sqrt3}\,R_3^2+{i}b_2,\cr
\Tc^3&={\sqrt3\over2}\,R_5^2+{i}b_3.}}


\subsec{$\IZ_2\times \IZ_{6'}$--orbifold}

The two twists act on the complex coordinate as follows:
\eqn\twistactionthth{\eqalign{
Q_1:\ z^i& \lra e^{2\pi i v_1^i}\ z^i\ \ \ ,\ \ \ v_1^1=\fc{1}{2}\ ,\ v_1^2=0\ ,\ v_1^3=-\fc{1}{2}\ ,\cr
Q_2:\ z^i& \lra e^{2\pi i v_2^i}\ z^i\ \ \ ,\ \ \ v_2^1=\fc{1}{6}\ ,\ v_2^2=\fc{1}{6}\ ,\ v_2^3=-\fc{2}{6}\ .
}}
The combined twist $Q_3=Q_2Q_1$ is a $Z_{6-I}$--twist:
\eqn\twistcomb{
Q_3:\ z^i \lra e^{2\pi i v_3^i}\ z^i\ \ \ ,\ \ \ v_3^1=-\fc{2}{6}\ ,\ v_3^2=\fc{1}{6}\ ,\ v^3=\fc{1}{6}\ .
}
We find the root lattice of $SU(3)\times G_2^2$ to be the correct one.
We choose for the twists acting on the lattice basis:
\eqn\twistactionththl{\eqalign{
Q_1\, e_1&=- e_1,\quad Q_1\,e_2=-e_2,\quad Q_1\,e_3=e_3,\quad Q_1\,e_4=e_4,\cr
Q_1\,e_5&=-e_5,\quad Q_1\,e_6=-e_6,\cr
Q_2\, e_1&= -e_2,\quad Q_2\,e_2=e_1+e_2,\quad Q_2\,e_3=2\,e_3+3\,e_4,\quad Q_2\,e_4=-e_3-e_4,\cr
Q_2\,e_5&=-2\,e_5-3\,e_6,\quad Q_2\,e_6=e_5+e_6.
}}
Here, the $Q_2$--twist on $e_5,\,e_6$ is minus the usual Coxeter--twist on $SU(3)$. The twists reproduce the correct eigenvalues and the conditions $Q_1^2=1,\ Q_2^6=1$. The combined twist $Q_3$ has the form
\eqn\twico{\eqalign{
Q_3\, e_1&= e_2,\quad Q_3\,e_2=-e_1-e_2,\cr
Q_3\,e_3&=2\,e_3+3\,e_4,\quad Q_3\,e_4=-e_3-e_4,\cr
Q_3\,e_5&=2\,e_5+3\,e_6,\quad Q_3\,e_6=-e_5-e_6,
}}
and also reproduces the required eigenvalues. We require the metric to be invariant under all three twists, i.e. we impose the three conditions $Q_i^Tg\,Q_i=g,\quad i=1,2,3$. This leads to the following solution:
\eqn\g{g=\pmatrix{R_1^2&-\half R_1^2&0&0&0&0\cr
-\half R_1^2&R_1^2&0&0&0&0\cr
0&0&R_3^2&-\half R_3^2&0&0\cr
0&0&-\half R_3^2&{1\over3} R_3^2&0&0\cr
0&0&0&0&R_5^2&-\half R_5^2\cr
0&0&0&0&-\half R_5^2&{1\over3} R_5^2.}}
The solution for $b$ matches the pattern:
\eqn\b{b=\pmatrix{0&b_1&0&0&0&0\cr
-b_1&0&0&0&0&0\cr
0&0&0&b_2&0&0\cr
0&0&-b_2&0&0&0\cr
0&0&0&0&0&-b_3\cr
0&0&0&0&-b_3&0},}
we therefore know to have three K\"ahler moduli whereas the complex structure is completely fixed. For the complex structure we get
\eqn\dzs{\eqalign{
dz^1&=3^{1/4}\,(dx^1+e^{-2\pi i/3}\, dx^2),\cr
dz^2&=dx^3+{1\over\sqrt3}e^{10\pi i/12}\,dx^4,\cr
dz^3&=dx^5+{1\over\sqrt3}e^{10\pi i/12}\,dx^6.}}
Examination of the K\"ahler form yields
\eqn\ks{\eqalign{
\Tc^1&={\sqrt3\over2}\, R_1^2+{i}b_1,\cr
\Tc^2&={1\over2\sqrt3}\,R_3^2+{i}b_2,\cr
\Tc^3&={1\over2\sqrt3}\,R_5^2+{i}b_3.}}


\subsec{$\IZ_6\times \IZ_{6}$--orbifold}

The two twists act on the complex coordinate as follows:
\eqn\twistactionthth{\eqalign{
Q_1:\ z^i& \lra e^{2\pi i v_1^i}\ z^i\ \ \ ,\ \ \ v_1^1=\fc{1}{6}\ ,\ v_1^2=0\ ,\ v_1^3=-\fc{1}{6}\ ,\cr
Q_2:\ z^i& \lra e^{2\pi i v_2^i}\ z^i\ \ \ ,\ \ \ v_2^1=0\ ,\ v_2^2=\fc{1}{6}\ ,\ v_2^3=-\fc{1}{6}\ .
}}
The combined twist $Q_3=Q_2Q_1$ is a $Z_{6-I}$--twist:
\eqn\twistcomb{
Q_3:\ z^i \lra e^{2\pi i v_3^i}\ z^i\ \ \ ,\ \ \ v_3^1=\fc{1}{6}\ ,\ v_3^2=\fc{1}{6}\ ,\ v^3=-\fc{2}{6}\ .
}
We find the root lattice of $G_2\times G_2\times G_2$ to be the correct one.
We choose for the twists acting on the lattice basis:
\eqn\twistactionththl{\eqalign{
Q_1\, e_1&=2\,e_1+3\,e_2,\quad Q_1\,e_2=-e_1-e_2,\quad Q_1\,e_3=e_3,\quad Q_1\,e_4=e_4,\cr
Q_1\,e_5&=2\,e_5+3\,e_6,\quad Q_1\,e_6=-e_5-e_6,\cr
Q_2\, e_1&= e_1,\quad Q_2\,e_2=e_2,\quad Q_2\,e_3=2\,e_3+3\,e_4,\quad Q_2\,e_4=-e_3-e_4,\cr
Q_2\,e_5&=2\,e_5+3\,e_6,\quad Q_2\,e_6=-e_5-e_6.
}}
The twists reproduce the correct eigenvalues and the conditions $Q_1^6=1,\ Q_2^6=1$. The combined twist $Q_3$ has the form
\eqn\twico{\eqalign{
Q_3\, e_1&=2\, e_1+3\,e_3,\quad Q_3\,e_2=-e_1-e_2,\cr
Q_3\,e_3&=2\,e_3+3\,e_4,\quad Q_3\,e_4=-e_3-e_4,\cr
Q_3\,e_5&=e_5+3\,e_6,\quad Q_3\,e_6=-e_5-2\,e_6,
}}
 where the twist on $e_5,\,e_6$ is twice the Coxeter--twist on $G_2$. $Q_3$ also reproduces the required eigenvalues. We require the metric to be invariant under all three twists, i.e. we impose the three conditions $Q_i^Tg\,Q_i=g,\quad i=1,2,3$. This leads to the following solution:
\eqn\g{g=\pmatrix{R_1^2&-\half R_1^2&0&0&0&0\cr
-\half R_1^2&{1\over3}R_1^2&0&0&0&0\cr
0&0&R_3^2&-\half R_3^2&0&0\cr
0&0&-\half R_3^2&{1\over3} R_3^2&0&0\cr
0&0&0&0&R_5^2&-\half R_5^2\cr
0&0&0&0&-\half R_5^2&{1\over3} R_5^2.}}
The solution for $b$ matches the pattern:
\eqn\b{b=\pmatrix{0&b_1&0&0&0&0\cr
-b_1&0&0&0&0&0\cr
0&0&0&b_2&0&0\cr
0&0&-b_2&0&0&0\cr
0&0&0&0&0&-b_3\cr
0&0&0&0&-b_3&0},}
we therefore know to have three K\"ahler moduli whereas the complex structure is completely fixed. For the complex structure we get
\eqn\dzs{\eqalign{
dz^1&=3^{1/4}\,(dx^1+{1\over\sqrt3}e^{10\pi i/12}\,dx^2),\cr
dz^2&=3^{1/4}\,(dx^3+{1\over\sqrt3}e^{10\pi i/12}\,dx^4),\cr
dz^3&=3^{1/4}\,(dx^5+{1\over\sqrt3}e^{-10\pi i/12}\,dx^6).}}
Examination of the K\"ahler form yields
\eqn\ks{\eqalign{
\Tc^1&={1\over2\sqrt3}\, R_1^2+{i}b_1,\cr
\Tc^2&={1\over2\sqrt3}\,R_3^2+{i}b_2,\cr
\Tc^3&={1\over2\sqrt3}\,R_5^2+{i}b_3.}}


\appendix\appB{Flux solutions for $\IZ_N$ and $\IZ_N\times \IZ_M$--orbifolds}

In this appendix, we present the flux solutions of the remaining orbifolds we have considered in this paper.

\subsec{Invariant Fluxes}

\vskip0.5cm
{\vbox{\ninepoint{$$
\vbox{\offinterlineskip\tabskip=0pt
\halign{\strut\vrule#
&~$#$~\hfil
&\vrule$#$
&~$#$~\hfil
&~$#$~\hfil
&~$#$~\hfil
&~$#$~\hfil
&~$#$~\hfil
&~$#$~\hfil
&~$#$~\hfil
&~$#$~\hfil
&~$#$~\hfil
&\vrule$#$\cr
\noalign{\hrule}
&\ \  G_3&&\ \IZ_3
&\IZ_4&\IZ_{6-I}&\IZ_{6-II}&\IZ_7&\IZ_{8-I}&\IZ_{8-II}&\IZ_{12-I}&\IZ_{12-II}&\cr
\noalign{\hrule}\noalign{\hrule}
&dz^1\wedge dz^2\wedge dz^3&&\ +&+&+&+&+&+&+&+&+&\cr
&d\ov z^1\wedge dz^2\wedge dz^3&&\ -&-&-&-&-&-&-&-&-&\cr
&dz^1\wedge d\ov z^2\wedge dz^3&&\ -&-&-&-&-&-&-&-&-&\cr
&dz^1\wedge dz^2\wedge d\ov z^3&&\ -&+&-&+&-&-&+&-&+&\cr
&d z^1\wedge d\ov z^2\wedge d\ov z^3&&\ -&-&-&-&-&-&-&-&-&\cr
&d\ov z^1\wedge d z^2\wedge d\ov z^3&&\ -&-&-&-&-&-&-&-&-&\cr
&d\ov z^1\wedge d\ov z^2\wedge d z^3&&\ -&+&-&+&-&-&+&-&+&\cr
&d\ov z^1\wedge d\ov z^2\wedge d\ov z^3&&\ +&+&+&+&+&+&+&+&+&\cr
\noalign{\hrule}}}$$
\vskip-6pt
\centerline{\noindent{\bf Table 4:}
{\sl Allowed $3$--form fluxes for point group $\IZ_N$}}
\vskip10pt}}}
\vskip-0.5cm 
\vskip0.5cm
{\vbox{\ninepoint{$$
\vbox{\offinterlineskip\tabskip=0pt
\halign{\strut\vrule#
&~$#$~\hfil
&\vrule$#$
&~$#$~\hfil
&~$#$~\hfil
&~$#$~\hfil
&~$#$~\hfil
&~$#$~\hfil
&~$#$~\hfil
&~$#$~\hfil
&~$#$~\hfil
&\vrule$#$\cr
\noalign{\hrule}
&\ \ G_3&&\IZ_2\times\IZ_2&\IZ_3\times \IZ_3&\IZ_2\times \IZ_4
&\IZ_4\times \IZ_4&\IZ_2\times \IZ_{6-I}&\IZ_2 \times \IZ_{6-II}&\IZ_3\times \IZ_6
&\IZ_6\times \IZ_6&\cr
\noalign{\hrule}\noalign{\hrule}
&dz^1\wedge dz^2\wedge dz^3&&\ +&+&+&+&+&+&+&+&\cr
&d\ov z^1\wedge dz^2\wedge dz^3&&\ +&-&+&-&+&-&-&-&\cr
&dz^1\wedge d\ov z^2\wedge dz^3&&\ +&-&-&-&-&-&-&-&\cr
&dz^1\wedge dz^2\wedge d\ov z^3&&\ +&-&-&-&-&-&-&-&\cr
&d z^1\wedge d\ov z^2\wedge d\ov z^3&&\ +&-&+&-&+&-&-&-&\cr
&d\ov z^1\wedge d z^2\wedge d\ov z^3&&\ +&-&-&-&-&-&-&-&\cr
&d\ov z^1\wedge d\ov z^2\wedge d z^3&&\ +&-&-&-&-&-&-&-&\cr
&d\ov z^1\wedge d\ov z^2\wedge d\ov z^3&&\ +&+&+&+&+&+&+&+&\cr
\noalign{\hrule}}}$$
\vskip-8pt
\centerline{\noindent{\bf Table 5:}
{\sl Allowed $3$--form fluxes for point group $\IZ_M\times \IZ_N$}}
\vskip10pt}}}
\vskip-0.5cm \ \br
The remaining 12 fluxes of the form $dz^a\wedge d\ov z^a\wedge dz^b$ and 
$dz^a\wedge d\ov z^a\wedge d \ov z^b$, respectively are always projected out
and therefore do not appear in the tables. Note, 
that we have for completeness also listed the orbifold groups
$\Gamma\in \{\IZ_4,\,\IZ_{8-I},\,\IZ_{8-II},\,\IZ_{12-II}\}$, 
whose tadpoles may only be cancelled in the more general orbifold setups with discrete
torsion or vector structure.


\subsec{$\IZ_{3}$--orbifold}

On this orbifold, only the $(3,0)$-- and $(3,0)$--components survive, so the flux takes in complex notation the form
$${1\over (2\pi)^2\alpha'}\,G_3=A_0\,\om_{A_0}+B_0\,\om_{B_0}.$$
We choose the free constants such that $dz^1=dx^{1}+e^{2\pi i/3}\,dx^2$, $dz^2=dx^3+e^{2\pi i/3}\,dx^4$ and $dz^3=dx^5+e^{2\pi i/3}\,dx^6$, as suggested in appendix \appA.
The $(3,0)$--form on this orbifold takes the form
\eqn\threezeroiii{\eqalign{
\om_{A_0}=\alpha_0+\beta_0+e^{2\pi i/3}\,(\alpha_1+\alpha_2+\alpha_3)+e^{2\pi i/6}\,(\beta_1+\beta_2+\beta_3\,).
}}
$\om_{B_0}$ is the complex conjugate of the above.
For the complex coefficients we find
\eqn\cocoeffi{\eqalign{
A_0=&{1\over \sqrt3}\left\{-e^{2\pi i/12}\,a^1+e^{-2\pi i/12}\,b_1-iS\,[e^{2\pi i/12}\,c^1-e^{-2\pi i/12}\,d_1]\right\},\cr
B_0=&{1\over \sqrt3}\left\{-e^{-2\pi i/12}\,a^1+e^{2\pi i/12}\,b_1+iS\,[e^{-2\pi i/12}\,c^1-e^{2\pi i/12}\,d_1]\right\}.
}}
Expressed in the real 3--forms, the flux takes the form
\eqn\realtwelveii{\eqalign{
{1\over (2\pi)^2\alpha'}\,G_3=&(-a^1+b_1+iS\,(-c^1+d_1))\,(\,\alpha_0+\beta^0)\cr
&+(a^1+iS\,c^1)\,(\alpha_1+\alpha_2+\alpha_3)+(b_1+iS\,d_1)\,(\beta^1+\beta^2+\beta^3).
}}


\subsec{$\IZ_{6-I}$--orbifold}

Again, only the $(3,0)$-- and $(3,0)$--components survive, so the flux takes in complex notation the form
$${1\over (2\pi)^2\alpha'}\,G_3=A_0\,\om_{A_0}+B_0\,\om_{B_0}.$$


\vskip0.5cm
\noindent{\it The $G_2^2\times SU(3)$--lattice}
\vskip0.5cm

The $(3,0)$--form on this orbifold takes the form
\eqn\threezeroiii{\eqalign{
\om_{A_0}=&\alpha_0+{1\over\sqrt 3}\,e^{5\pi i/6}\,(\alpha_1+\alpha_2)+e^{2\pi i/3}\,\alpha_3+{1\over 3}\,e^{2\pi i/6}\,\beta_0\cr
&+{i\over\sqrt3}\,(\beta_1+\beta_2)+{1\over3}\,e^{2\pi i/3}\,\beta_3.
}}
$\om_{B_0}$ is the complex conjugate of the above.
For the complex coefficients we find
\eqn\cocoeffi{\eqalign{
A_0=&{1\over\sqrt3}\,(e^{2\pi i/12}\,a^0-i\sqrt3\,b_0+iS\,[e^{2\pi i/12}\,c^0-{i\sqrt3}\,d_0]),\cr
B_0=&{1\over\sqrt3}\,(e^{-2\pi i/12}\,a^0+i\sqrt3\,b_0-iS\,[e^{-2\pi i/12}\,c^0+{i\sqrt3}\,d_0]).
}}
Expressed in the real 3--forms, the flux takes the form
\eqn\realtwelveii{\eqalign{
{1\over (2\pi)^2\alpha'}\,G_3=&(a^0+iS\,c^0)\,\alpha_0+{1\over 3}(-2\,a^0+3\,b_0-iS\,[2\,c^0-3\,d_0])\,(\alpha_1+\alpha_2)\cr
&+(-a^0+3\,b_0-iS\,(c^0-3\,d_0))\,\alpha_3+(b_0+iS\,d_0)\,\beta^0\cr
&+{1\over 3}(-a^0+6\,b_0-iS\,[c^0-6\,d_0])\,(\beta^1+\beta^2)+(-a^0+3\,b_0-iS\,[c^0-3\,d_0])\,\beta^3.
}}
In order for all the real coefficients to be integer, $a^0,b_0,c^0$ and $d_0$ should be chosen to be multiples of 3.


\vskip0.5cm
\noindent{\it The $G_2\times (SU(3))^2$--lattice with generalized Coxeter element}
\vskip0.5cm

The $(3,0)$--form on this orbifold takes the form
\eqn\threezeroiii{\eqalign{
\om_{A_0}=&2\,(\alpha_0+{1\over\sqrt 3}\,e^{5\pi i/6}\,\alpha_1+e^{2\pi i/3}\,\alpha_2-e^{2\pi i/6}\,\alpha_3+{1\over \sqrt3}\,e^{5\pi i/6}\,\beta_0\cr
&-\beta_1-{1\over\sqrt3}e^{2\pi i/6}\,\beta_2+{i\over\sqrt3}\,\beta_3).
}}
$\om_{B_0}$ is the complex conjugate of the above.
For the complex coefficients we find
\eqn\cocoeffi{\eqalign{
A_0=&-{1\over2}\,(e^{2\pi i/3}\,a^0+\sqrt3\,b_0-iS\,[e^{5\pi i/6}\,c^0+{\sqrt3}\,d_0]),\cr
B_0=&-{1\over2}\,(e^{-2\pi i/3}\,a^0+\sqrt3\,b_0+iS\,[e^{-5\pi i/6}\,c^0+{\sqrt3}\,d_0]).
}}
Expressed in the real 3--forms, the flux takes the form
\eqn\realtwelveii{\eqalign{
{1\over (2\pi)^2\alpha'}\,G_3=&(a^0+iS\,c^0)\,\alpha_0+(a^1+iS\,C^1)\,\alpha_1+(a^0+3\,a^1+iS\,[\,c^0+3\,c^1])\alpha_2\cr
&+(-2\,a^0-3\,a^1-iS\,(2\,c^0+3\,c^1))\,\alpha_3+(a^1+iS\,c^1)\,\beta^0\cr
&+(-a^0-iS\,c^0)\,\beta^1+(-a^0-a^1-iS\,[\,c^0+c^1])\,\beta^2\cr
&+(a^0+2\,a^1+iS\,[\,c^0+2\,c^1])\,\beta^3.
}}


\subsec{$\IZ_{6-II}$--orbifold}

This is a case with one complex structure modulus left unfixed. Therefore one $(2,1)$-- and one $(1,2)$--component survive and the flux takes the form
$${1\over (2\pi)^2\alpha'}\,G_3=A_0\,\om_{A_0}+A_3\,\om_{A_3}+B_0\,\om_{B_0}+B_3\,\om_{B_3}.$$


\vskip0.5cm
\noindent{\it The $(SU(6))\times SU(2)$--lattice}
\vskip0.5cm

The $(3,0)$--form on this orbifold takes the form
\eqn\threezerosixi{\eqalign{
\omega_{A_0}=&\fc{1}{6}\left\{(-3i\alpha_0+(i+\sqrt 3)(\alpha_1+\alpha_2-\delta_6)+(-i+\sqrt 3)(\beta_3-\gamma_2-\gamma_3)\right.\cr
&\left. -i(\gamma_1+\gamma_4-\delta_5) - U\left[3i\alpha_3-3i\beta_0+2\sqrt 3 \beta_1+\sqrt 3 \beta_2\right.\right.\cr
&\left.\left.-3i\gamma_5+3i\gamma_6+\sqrt 3(\delta_1-\delta_2-2\delta_3-\delta_4)\right]\right\}.
}}
The one $(2,1)$--form surviving the twist takes the form
\eqn\twoone{\eqalign{
\omega_{A_3}=&\fc{1}{6}\left\{(3i\alpha_0+(-i+\sqrt 3)(\alpha_1+\alpha_2-\delta_6)+(i+\sqrt 3)(\beta_3-\gamma_2-\gamma_3)\right.\cr
&\left. +i(\gamma_1+\gamma_4-\delta_5) - U\left[-3i\alpha_3+3i\beta_0 +2\sqrt 3 \beta_1+\sqrt 3 \beta_2\right.\right.\cr
&\left.\left.+3i\gamma_5-3i\gamma_6+\sqrt 3(\delta_1-\delta_2-2\delta_3-\delta_4)\right]\right\}.
}}
$\om_{B_0}$ and $\om_{B_3}$ are the complex conjugate of the above.
For the complex coefficients we find
\eqn\cocoeffsixii{\eqalign{
A_0=&{1\over 2 {\rm Im}{{\cal U}^3}}\left\{-b_0+{i\sqrt 3\over 2}b_1-S\left({\sqrt 3\over 2}d_1+i d_0\right)\right.\cr
     &\left.+U\left[-a_0+i\left({1\over \sqrt3}a_0+\sqrt 3 a_1\right) -S\left({1\over \sqrt 3}c_0+\sqrt 3 c_1+ic_0\right)\right]\right\} \ ,\cr       
B_0=&{1\over 2 {\rm Im}{{\cal U}^3}}\left\{-b_0-{i\sqrt 3\over 2}b_1+S\left({\sqrt 3\over 2}d_1-i d_0\right)\right.\cr
     &\left.+\bar U\left[-a_0-i\left({1\over \sqrt3}a_0+\sqrt 3 a_1\right) +S\left({1\over \sqrt 3}c_0+\sqrt 3 c_1-ic_0\right)\right]\right\} \ , \cr  
A_3=&{1\over 2 {\rm Im}{{\cal U}^3}}\left\{b_0+{i\sqrt 3\over 2}b_1-S\left({\sqrt 3\over 2}d_1-i d_0\right)\right.\cr
     &\left.+U\left[a_0+i\left({1\over \sqrt3}a_0+\sqrt 3 a_1\right) -S\left({1\over \sqrt 3}c_0+\sqrt 3 c_1-ic_0\right)\right]\right\}  ,\cr
B_3=&{1\over 2 {\rm Im}{{\cal U}^3}}\left\{b_0-{i\sqrt 3\over 2}b_1+S\left({\sqrt 3\over 2}d_1+i d_0\right)\right.\cr
     &\left.+\bar U\left[a_0-i\left({1\over \sqrt3}a_0+\sqrt 3 a_1\right) +S\left({1\over \sqrt 3}c_0+\sqrt 3 c_1+ic_0\right)\right]\right\}    \ , 
}}
Expressed in the real 3--forms, the flux takes the form
\eqn\realsixii{\eqalign{
\fc{1}{(2\pi)^2\alpha'}G_3=&\fc{1}{3}(a_0+iS c_0)(3\alpha_0+2\beta_3+\gamma_1-2\gamma_2-2\gamma_3+\gamma_4-\delta_5)\cr
&+(b_0+iSd_0)(-\alpha_3+\beta_0+\gamma_5-\gamma_6) \cr
&+\fc{1}{2}(b_1+iSd_1)(2\beta_1+\beta_2+\delta_1-\delta_2-2\delta_3-\delta_4)\cr
&+(a_1+iSc_1)(\alpha_1+\alpha_2+\beta_3-\gamma_2-\gamma_3-\delta_6) \ .
}}


\vskip0.5cm
\noindent{\it The $SU(3)\times SO(8)$--lattice}
\vskip0.5cm

The $(3,0)$--form on this orbifold takes the form
\eqn\threezerosixi{\eqalign{
\om_{A_0}=&-( 2+{\cal U}^3)\,\alpha_0+(e^{-2\pi i/6}+{\cal U}^3)\,\alpha_1-(1+2\,{\cal U}^3)\,\alpha_2+e^{2\pi i/3}\,(2+{\cal U}^3)\,\alpha_3\cr
&+({\cal U}^3+e^{2\pi i/3}\,(1+{\cal U}^3))\,\beta^0+e^{2\pi i/3}\,(1+2\,{\cal U}^3)\,\beta_1-e^{2\pi i/}\,(1+{\cal U}^3)\beta_2\cr
&+(-1+e^{2\pi i/3}\,{\cal U}^3)\,\beta^3+(1+e^{2\pi i/6})\,(1+{\cal U}^3)\,\gamma_2+(-1+{\cal U}^3)\,\gamma_4\cr
&+(-1+e^{2\pi i/3})(1+{\cal U}^3)\,\delta^2+e^{2\pi i/3}\,(-1-{\cal U}^3)\,\delta^4.
}}
The one $(2,1)$--form surviving the twist takes the form
\eqn\twoone{\eqalign{
\om_{A_3}=&-( 2+\ov{\cal U}^3)\,\alpha_0+(e^{-2\pi i/6}+\ov{\cal U}^3)\,\alpha_1-(1+2\,\ov{\cal U}^3)\,\alpha_2+e^{2\pi i/3}\,(2+\ov{\cal U}^3)\,\alpha_3\cr
&+({\cal U}^3+e^{2\pi i/3}\,(1+\ov{\cal U}^3))\,\beta^0+e^{2\pi i/3}\,(1+2\,\ov{\cal U}^3)\,\beta_1-e^{2\pi i/}\,(1+\ov{\cal U}^3)\,\beta_2\cr
&+(-1+e^{2\pi i/3}\,\ov{\cal U}^3)\,\beta^3+(1+e^{2\pi i/6})\,(1+\ov{\cal U}^3)\,\gamma_2+(-1+\ov{\cal U}^3)\,\gamma_4\cr
&+(-1+e^{2\pi i/3})(1+\ov{\cal U}^3)\,\delta^2+e^{2\pi i/3}\,(-1-\ov{\cal U}^3)\,\delta^4.
}}
$\om_{B_0}$ and $\om_{B_3}$ are the complex conjugate of the above.
For the complex coefficients we find
\eqn\cocoeffsixii{\eqalign{
A_0=&{
i\over6\,{\rm Im}\,{\cal U}^3}\left\{e^{2\pi i/6}\,(1+2\,\ov{\cal U}^3)\,a^1+(1+e^{2\pi i/6}\,\ov{\cal U}^3)\,a_2-(e^{-2\pi i/6}+\ov{\cal U}^3)\,b_1-(1+2\,\ov{\cal U}^3)\,b_2\right.\cr
&\left.-iS\,[-e^{2\pi i/6}(1+2\,\ov{\cal U}^3)\,c_1+(1+e^{2\pi i/6}\ov{\cal U}^3)\,c_2-(e^{-2\pi i/6}+\ov{\cal U}^3)\,d_1+(1+2\,\ov{\cal U}^3)\,d_2)]\right\},\cr
B_0=&{
-i\over6\,{\rm Im}\,{\cal U}^3}\left\{e^{-2\pi i/6}\,(1+2\,{\cal U}^3)\,a^1+(1+e^{-2\pi i/6}\,{\cal U}^3)\,a_2-(e^{2\pi i/6}+{\cal U}^3)\,b_1-(1+2\,{\cal U}^3)\,b_2\right.\cr
&\left.-iS\,[-e^{-2\pi i/6}(1+2\,{\cal U}^3)\,c_1-(1+e^{-2\pi i/6}\,{\cal U}^3)\,c_2+(e^{2\pi i/6}+{\cal U}^3)\,d_1+(1+2\,{\cal U}^3)\,d_2)]\right\},\cr
A_3=&{
i\over6\,{\rm Im}\,{\cal U}^3}\left\{-e^{2\pi i/6}\,(1+2\,{\cal U}^3)\,a^1-(1+e^{2\pi i/6}\,{\cal U}^3)\,a_2+(e^{-2\pi i/6}+{\cal U}^3)\,b_1+(1+2\,{\cal U}^3)\,b_2\right.\cr
&\left.-iS\,[e^{2\pi i/6}(1+2\,{\cal U}^3)\,c_1+(1+e^{2\pi i/6}\,{\cal U}^3)\,c_2-(e^{-2\pi i/6}+{\cal U}^3)\,d_1-(1+2\,{\cal U}^3)\,d_2)]\right\},\cr
B_3=&{
-i\over6\,{\rm Im}\,{\cal U}^3}\left\{-e^{-2\pi i/6}\,(1+2\,\ov{\cal U}^3)\,a^1-(1+e^{-2\pi i/6}\,\ov{\cal U}^3)\,a_2+(e^{2\pi i/6}+\ov{\cal U}^3)\,b_1+(1+2\,\ov{\cal U}^3)\,b_2\right.\cr
&\left.-iS\,[-e^{-2\pi i/6}(1+2\,\ov{\cal U}^3)\,c_1+(1+e^{-2\pi i/6}\,\ov{\cal U}^3)\,c_2+(e^{2\pi i/6}+\ov{\cal U}^3)\,d_1-(1+2\,\ov{\cal U}^3)\,d_2)]\right\}.
}}
Expressed in the real 3--forms, the flux takes the form
\eqn\realsixii{\eqalign{
{1\over (2\pi)^2\alpha'}\,G_3=&\,(-a^1+b_1+2\,b_2+iS\,(-c^1+d_1+2\,d_2)
)\,\alpha_0+(a^1+iS\,c^1)\,\alpha_1\cr
&+(a^2+iS\,c^2)\,\alpha_2+(2\,a^1+a^2-b_1-b_2+iS\,(2\,c^1+c^2-d_1-d_2))\,\alpha_3\cr
&+(-a^1-a^2+b_1+b_2+iS\,(-c^1-c^2+b_1+b_2))\beta^0+(b_1+iS\,d_1)\,\beta^1\cr 
&+(b_2+iS\,d_2)\,\beta^2+(b_1+b_2+iS\,(d_1+d_2))\,\beta^3-(a^2+b^2+iS\,(c^2+d^2))\,\gamma_2 \cr
&+(-a^1-a^2+b_1+2\,b_2+iS\,(-c^1-c^2+d_1+2\,d^2))\,\gamma_4\cr
&+(-a^1+b_1+b_2+iS\,(-c^1+d_1+d_2))\,\delta_2\cr
&+(2\,a^1+\,a^2-b_2+iS\,(2\,c^1+c^2-d_2))\,\delta_4.
}}


\subsec{$\IZ_7$--orbifold}

In section 2 we learned that only the $(3,0)$-- and the $(0,3)$--flux component survive the $Z_7$--twist, so
$${1\over (2\pi)^2\alpha'}\,G_3=A_0\,\om_{A_0}+B_0\,\om_{B_0}.$$
In appendix A, we calculated the complex structure. With this, we find for the $(3,0)$--form:
\eqn\threezero{\eqalign{
\om_{A_0}=&-i\sqrt7\, \alpha_0+\half(7+i\sqrt7)\,\alpha_1+i\sqrt7\,\alpha_2-i\sqrt7\, \alpha_3-i\sqrt7 \,\beta_0+i\sqrt7\, \beta_1-i\sqrt7\, \beta_2\cr
&+\half(7-i\sqrt7 )\,\beta_3-i\sqrt7\, \gamma_1+i\sqrt7\, \gamma_2+\half(-7+i\sqrt7 )\,\gamma_3-i\sqrt7\,\gamma_4+(7+i\sqrt7 )\,\gamma_5 \cr
&-(7-i\sqrt7 )\,\gamma_6-i\sqrt7\, \delta_1-(7+i\sqrt7 )\,\delta_2-i\sqrt7 \,\delta_3+(1-i\sqrt7)\,\delta_4+i\sqrt7 \,\delta_5-(7+i\sqrt7)\,\delta_6.
}}
$\om_{B_0}$ is simply the complex conjugate of the above.
It is possible to express the two complex coefficients of the 3--form flux through four of the real coefficients which we may choose freely. The other real coefficients are constrained by the form of the flux.
For the complex coefficients we find
\eqn\cocoeff{\eqalign{
A_0&={1\over 14}\,[\,(1+i\sqrt7 )\,a^0+2\,a^1-iS\,((1+i\sqrt7 )\,c^0+2\,c^1)\,],\cr
B_0&={1\over 14}\,[\,(1-i\sqrt7 )\,a^0+2\,a^1+iS\,((1-i\sqrt7 )\,c^0+2\,c^1)\,].
}}
Expressed in real coordinates, the flux takes the form
\eqn\recoeff{\eqalign{
{1\over (2\pi)^2\alpha'}\,G_3=&(a^0+iS\,c^0)\,\alpha_0+(a^1+iS\,c^1)\,\alpha_1+(-a^0-iS\,c^0)\,\alpha_2+(a^0+iS\,c^0)\,\alpha_3\cr
&+(a^0+iS\,c^0)\,\beta^0+(-a^0-iS\,c^0)\,\beta^1+(a^0+iS\,c^0)\,\beta^2\cr
&+(a^0+a^1+iS\,(c^0+c^1)))\,\beta^3+(a^0+iS\,c^0)\,\gamma_1+(-a^0+iS\,c^0)\,\gamma_2
\cr
&+(-a^0-a^1-iS\,(c^0+c^1))\,\gamma_3+(a^0+iS\,c^0)\,\gamma_4\cr
&+(a^1+iS\,c^1)\,\gamma_5+(-a^0-a^1-iS\,(c^0+c^1))\,\gamma_6-(a^1+iS\,c^1)\,\delta^6\cr
&-(-a^0-a^1-iS\,(c^1+c^1))\,\delta^4-(a^0+iS\,c^0)\,\delta^5-(a^1+iS\,c^1)\,\delta^2\cr
&-(-a^0-iS\,c^0)\,\delta^3-(a^0+iS\,c^0)\,\delta^1.
}}


\subsec{$\IZ_{12-I}$--orbifold}

This is another case where only the $(3,0)$-- and the $(0,3)$--flux component survive the twist, so
$${1\over (2\pi)^2\alpha'}\,G_3=A_0\,\om_{A_0}+B_0\,\om_{B_0}.$$


\vskip0.5cm
\noindent{\it  The $SU(3)\times F_4$--lattice}
\vskip0.5cm

Using the complex structure found in appendix \appA, we find for the $(3,0)$--form:
\eqn\threezeroiv{\eqalign{
\om_{A_0}=&-e^{2\pi i/12}\, \alpha_0+e^{-2\pi i/12}\,\alpha_1+{1\over\sqrt2}\,e^{2\pi i/8}\,(\sqrt3\,e^{-2\pi i/6}\,\alpha_2+e^{-2\pi i/12}\, \alpha_3-i\sqrt3\,\beta_0\cr
&+\sqrt3\,e^{-2\pi i/12}\,\beta_1+(1-i)\,\beta_2-\sqrt3\,e^{2\pi i/6}\,\beta_3+(1+\sqrt3)\,\gamma_3\cr
&-{1\over\sqrt2}e^{2\pi i/8}\,(1+i (2+\sqrt3))\,\delta^5).
}}
$\om_{B_0}$ is the complex conjugate of the above.
For the complex coefficients we find
\eqn\cocoeffiv{\eqalign{
A_0=&{1\over\sqrt2}\,e^{5\pi i/12}\,a^1+{1\over\sqrt3}\,e^{2\pi i/6}\,b_1+iS\,[\,{1\over\sqrt2}\,e^{5\pi i/12}\,c^1+{1\over\sqrt3}e^{2\pi i/6}\,d_1],\cr
B_0=&{1\over\sqrt2}\,e^{-5\pi i/12}\,a^1+{1\over\sqrt3}\,e^{-2\pi i/6}\,b_1-iS\,[\,{1\over\sqrt2}\,e^{-5\pi i/12}\,c^1+{1\over\sqrt3}e^{-2\pi i/6}\,d_1].
}}
Expressed in real coordinates, the flux takes the form
\eqn\recoeffi{\eqalign{
{1\over (2\pi)^2\alpha'}\,G_3=&-3\,(b_1+iS\,d_1)\,\alpha_0+(a^1+iS\,c^1)\,\alpha_1+{1\over3}\,(a^1+3\,b_1+iS\,(c^1+3\,d_1))\,\alpha_2\cr
&+(-a^1+{2\over\sqrt3}\,b_1+iS\,(c^1+{2\over\sqrt3}\,d_1))\,\alpha_3+{1\over3}\,(a^1+iS\,c^1)\,\beta^0+(b_1+iS\,d_1)\,\beta^1\cr
&+{1\over3}\,(a^1+3\,b_1+iS\,(c^1+3\,d_1))\,\beta^2+(\sqrt3\,a^1-b_1+iS\,(\sqrt3\,c^1-d_1))\,\beta^3.
}}
Here we have again some factors of $\sqrt 3$, which in this case can only be got rid of by further constraining the real coefficients and fixing $S$. $S=i$ and $a^1=c^1,\ b_1=d_1$ does the job.


\subsec{$\IZ_3\times \IZ_3$--orbifold}

Only the $(3,0)$-- and the $(0,3)$--flux component survive the twist, so
$${1\over (2\pi)^2\alpha'}\,G_3=A_0\,\om_{A_0}+B_0\,\om_{B_0}.$$
Using the complex structure found in chapter 3, we find for the $(3,0)$--form:
\eqn\threezeroi{\eqalign{
\om_{A_0}=&\, \alpha_0+e^{2\pi i/3}\,\alpha_1+e^{2\pi i/3}\,\alpha_2+e^{-2\pi i/3}\, \alpha_3+e^{2\pi i/3}\,\beta_0-\beta_1-\beta_2+e^{2\pi i/3}\beta_3.
}}
$\om_{B_0}$ is the complex conjugate of the above.
For the complex coefficients we find
\eqn\cocoeffi{\eqalign{
A_0&={1\over\sqrt3}\{ -{i}\,a^1+e^{5\pi i/6}\,b_1+iS\,[-{i}\,c^1+e^{5\pi i/6}\,d_1\,]\},\cr
B_0&={1\over\sqrt3}\{ {i}\,a^1+e^{-5\pi i/6}\,b_1-iS\,[{i}\,c^1+e^{-5\pi i/6}\,d_1\,]\}.
}}
Expressed in real coordinates, the flux takes the form
\eqn\recoeffi{\eqalign{
{1\over (2\pi)^2\alpha'}\,G_3=&(-b_1-iS\,d_1)\,\alpha_0+(a^1+iS\,c^1)\,(\alpha_1+\alpha_2)+(-a^1+b_1+iS\,(-c^1+d_1))\,\alpha_3\cr
&+(a^1+iS\,c^1)\,\beta^0+(b_1+iS\,d_1)\,(\beta^1+\beta^2)+(a^1-b_1+iS\,(c^1-d_1)))\,\beta^3.
}}


\subsec{$\IZ_2\times \IZ_6$--orbifold}

This is a case with one complex structure modulus left unfixed. Therefore one $(2,1)$-- and one $(1,2)$--component survive and the flux takes the form
$${1\over (2\pi)^2\alpha'}\,G_3=A_0\,\om_{A_0}+A_3\,\om_{A_3}+B_0\,\om_{B_0}+B_3\,\om_{B_3}.$$

The $(3,0)$--form on this orbifold takes the form
\eqn\threezerosixi{\eqalign{
\om_{A_0}=&\alpha_0+{1\over\sqrt3}\,e^{5\pi i/6}\,\alpha_2+e^{2 \pi i/3}\,\alpha_3+{i\over\sqrt3}\,\beta^1\cr
&+{\cal U}^3\,[\, \alpha_1-{i\over\sqrt3}\,\beta_0+e^{-2\pi i/6}\,\beta_2+{1\over\sqrt3}\,e^{-\pi i/6}\,\beta_3].
}}
The one $(2,1)$--form surviving the twist takes the form
\eqn\twoone{\eqalign{
\om_{A_3}=&\alpha_0+{1\over\sqrt3}\,e^{-5\pi i/6}\,\alpha_2+e^{-2 \pi i/3}\,\alpha_3+{1\over\sqrt3}e^{2 \pi i/3}\,\beta^1\cr
&+{\cal U}^3\,[\, \alpha_1+{1\over\sqrt3}e^{\pi i/6}\,\beta_0+e^{2\pi i/6}\,\beta_2+{1\over\sqrt3}\,e^{-\pi i/6}\,\beta_3].
}}
$\om_{B_0}$ and $\om_{B_3}$ are the complex conjugate of the above.
For the complex coefficients we find
\eqn\cocoeffsixii{\eqalign{
A_0=&{1\over {\cal U}^3}\left\{-b_0+{1\over\sqrt3} e^{\pi i/6}\,b_2+{\cal U}^3\,\left[\,-{i\over\sqrt3}\,a_0-e^{2\pi i/6}\,a_2+iS\,({-i\over\sqrt3}\,c_0-e^{2\pi i/6}\,c_2\right.\right.\cr
&\left.\left.-{1\over\sqrt3}e^{\pi i/6}\,c^3+d_1)\right]\right\},\cr
B_0=&{1\over\ov{\cal U}^3}\left\{-b_0+{1\over\sqrt3} e^{-\pi i/6}\,b_2+\ov{\cal U}^3\,\left[\,{i\over\sqrt3}\,a_0-e^{-2\pi i/6}\,a_2-iS\,({i\over\sqrt3}\,c_0+e^{-2\pi i/6}\,c_2\right.\right.\cr
&\left.\left.+{1\over\sqrt3}e^{-\pi i/6}\,c^3-d_1)\right]\right\},\cr
A_3=&{1\over {\cal U}^3}\left\{b_0-{1\over\sqrt3} e^{\pi i/6}\,b_2+{\cal U}^3\,\left[\,{1\over\sqrt3}e^{-\pi i/6}\,a_0+e^{-2\pi i/6}\,a_2+iS\,(-{1\over\sqrt3}e^{-\pi i/6}\,c_0\right.\right.\cr
&\left.\left.+e^{2\pi i/3}\,c_2
-{1\over\sqrt3}e^{\pi i/6}\,c^3+d_1)\right]\right\},\cr
B_3=&{1\over \ov{\cal U}^3}\left\{b_0-{1\over\sqrt3} e^{-\pi i/6}\,b_2+\ov{\cal U}^3\,\left[\,{1\over\sqrt3}e^{\pi i/6}\,a_0+e^{2\pi i/6}\,a_2+iS\,(-{1\over\sqrt3}e^{\pi i/6}\,c_0\right.\right.\cr
&\left.\left.+e^{-2\pi i/3}\,c_2
-{1\over\sqrt3}e^{-\pi i/6}\,c^3+d_1)\right]\right\},\cr
}}
Expressed in the real 3--forms, the flux takes the form
\eqn\realsixii{\eqalign{
{1\over (2\pi)^2\alpha'}\,G_3=&\,(a^0+iS\,c^0)\,\alpha_0+[{\rm Re}\,{\cal U}^3\,(a^0+iS\,c^0)\cr
&+\sqrt3\,{\rm Im}\,{\cal U}^3\,(a^0+2\,a^2+iS\,(c^0+2\,c^2))]\,\alpha_1\cr
&+(a^2+iS\,c^2)\,\alpha_2+{1\over|\,{\cal U}^3|^2}[\sqrt3\,{\rm Im}\,{\cal U}^3\,(-2\,b_0+b_2)+i({\rm Im}\,{\cal U}^3)^2\,S\,c^3)\cr
&-{\rm Re}\,{\cal U}^3\,(b_2-i{\rm Re}\,{\cal U}^3\,S\,c^3)]\,\alpha_3+[b_0-i{1\over 3}\,S\,(\sqrt3\,{\rm Im}\,{\cal U}^3\,(2\,c^3-3\,d_1)\cr
&+3\,{\rm Re}\,{\cal U}^3\,d_1]\,\beta^0-{1\over|\,{\cal U}^3|^2}[{\rm Re}\,{\cal U}^3\,b_0+{\rm Im}\,{\cal U}^3(\sqrt3\,\,b_0-{2\over\sqrt3}\,b_2)\cr
&-i\,|\,{\cal U}^3|^2\,S\,d_1]\,\beta^1+[b_2+iS\,({\rm Re}\,{\cal U}^3\,c^3+\sqrt3\,{\rm Im}\,{\cal U}^3(c^3-2\,d_1))]\,\beta^2\cr
&+[{\rm Re}\,{\cal U}^3\,(-a^2-iS\,c^2)+{1\over\sqrt3}\,{\rm Im}\,{\cal U}^3\,(2\,a^0+3\,a^2+iS(2\,c^0+3\,c_2))]\,\beta^3
}}
This again presents only a valid flux for certain choices of ${\cal U}^3$ like for example $e^{2\pi i/6}$ or ${1\over\sqrt3}e^{\pi i/6}$.


\subsec{$\IZ_2\times \IZ_{6'}$--orbifold}

Again, the flux is of the form
$${1\over (2\pi)^2\alpha'}\,G_3=A_0\,\om_{A_0}+B_0\,\om_{B_0}.$$
Using the complex structure found in appendix \appA, we find for the $(3,0)$--form:
\eqn\threezeroii{\eqalign{
\om_{A_0}=&\, \alpha_0+e^{-2\pi i/3}\,\alpha_1+{1\over\sqrt3}\,e^{5\pi i/6}\,(\alpha_2+\alpha_3)-{1\over3}\,\beta_0+{1\over3}e^{2\pi i/3}\,\beta_1+{1\over\sqrt3}\,e^{-5\pi i/6}\,(\beta_2+\beta_3).
}}
$\om_{B_0}$ is the complex conjugate of the above.
For the complex coefficients we find
\eqn\cocoeffii{\eqalign{
A_0&={1\over \sqrt 3}e^{5\pi i/6}\,a^1+\sqrt3e^{-5\pi i/6}\,b_1-iS\,[{1\over \sqrt 3}e^{5\pi i/6}\,c^1+\sqrt3\,e^{-5\pi i/6}\,d_1\,]\},\cr
B_0&={1\over \sqrt 3}e^{-5\pi i/6}\,a^1+\sqrt3\,e^{5\pi i/6}\,b_1+iS\,[{1\over \sqrt 3}e^{-5\pi i/6}\,c^1+\sqrt3\,e^{5\pi i/6}\,d_1\,]\}.
}}
Expressed in real coordinates, the flux takes the form
\eqn\recoeffii{\eqalign{
{1\over (2\pi)^2\alpha'}\,G_3=&-(a^1+3\,b_1+iS\,(c^1+3\,d_1))\,\alpha_0+(a^1+iS\,c^1)\,\alpha_1\cr
&+{1\over3}\,(a^1+6\,b_1+iS\,(c^1+6\,d_1))\,\alpha_2\cr
&+{1\over3}\,(a^1+6\,b_1+iS\,(c^1+6\,d_1))\,\alpha_3+{1\over3}\,(a^1+3\,b_1+iS\,(c^1+3\,d_1))\,\beta^0\cr
&+(b_1+iS\,d_1)\,\beta^1+{1\over3}\,(2\,a^1+3\,b_1+iS\,(2\,c^1+3\,d_1)\,(\beta^2+\beta^3).
}}


\subsec{$\IZ_3\times \IZ_6$--orbifold}

Again, the flux is of the form
$${1\over (2\pi)^2\alpha'}\,G_3=A_0\,\om_{A_0}+B_0\,\om_{B_0}.$$
Using the complex structure found in chapter 2, we find for the $(3,0)$--form:
\eqn\threezeroii{\eqalign{
\om_{A_0}=&\, \alpha_0+e^{2\pi i/3}\,\alpha_1+{1\over\sqrt3}\,e^{5\pi i/6}\,\alpha_2+{1\over\sqrt3}\,e^{-5\pi i/6}\, \alpha_3\cr
&+{1\over3}e^{2\pi i/3}\,\beta_0-2\,\beta_1+{1\over\sqrt3})\,e^{5\pi i/6}\,\beta_2+{i\over\sqrt3}\,\beta_3.
}}
$\om_{B_0}$ is the complex conjugate of the above.
For the complex coefficients we find
\eqn\cocoeffii{\eqalign{
A_0&=-{i\over \sqrt 3}\,a^1+\sqrt3\,e^{5\pi i/6}\,b_1+iS\,[-{i\over \sqrt3}\,c^1+\sqrt3\,e^{5\pi i/6}\,d_1\,]\},\cr
B_0&={i\over \sqrt 3}\,a^1+\sqrt3\,e^{-5\pi i/6}\,b_1-iS\,[{i\over \sqrt3}\,c^1+\sqrt3\,e^{-5\pi i/6}\,d_1\,]\}.
}}
Expressed in real coordinates, the flux takes the form
\eqn\recoeffii{\eqalign{
{1\over (2\pi)^2\alpha'}\,G_3=&3\,(-\,b_1-iS\,d_1)\,\alpha_0+(a^1+iS\,c^1)\,\alpha_1+{1\over3}\,(a^1+3\,b_1+iS\,(c^1+3\,d_1))\,\alpha_2\cr
&+{1\over3}\,(-a^1+6\,b_1-iS\,(c^1-6\,d_1))\,\alpha_3+{1\over3}\,(a^1+iS\,c^1)\,\beta^0+(b_1+iS\,d_1)\,\beta^1\cr
&+{1\over3}\,(a^1+3\,b_1+iS\,(c^1+3\,d_1)\,\beta^2+{1\over3}\,(2\,a^1-3\,b_1+iS\,(2\,c^1-3\,d_1)))\,\beta^3.
}}


\subsec{$\IZ_6\times \IZ_{6}$--orbifold}

Again, the flux is of the form
$${1\over (2\pi)^2\alpha'}\,G_3=A_0\,\om_{A_0}+B_0\,\om_{B_0}.$$
Using the complex structure found in appendix \appA, we find for the $(3,0)$--form:
\eqn\threezeroii{\eqalign{
\om_{A_0}=&\, \alpha_0+{1\over\sqrt3}\,e^{5\pi i/6}\,(\alpha_1+\alpha_2)+{1\over\sqrt3}\,e^{-5\pi i/6}\alpha_3-{1\over3\sqrt3}\,e^{5\pi i/6}\,\beta_0\cr
&-{1\over3}\,(\beta_1+\beta_2)+{1\over\sqrt3}\,e^{5\pi i/6}\beta_3.
}}
$\om_{B_0}$ is the complex conjugate of the above.
For the complex coefficients we find
\eqn\cocoeffii{\eqalign{
A_0&={-i\sqrt 3}\,a^1+3\,e^{2\pi i/3}\,b_1+iS\,[{-i\sqrt 3}\,c^1+3\,e^{2\pi i/3}\,d_1\,]\},\cr
B_0&={i\sqrt 3}\,a^1+3\,e^{-2\pi i/3}\,b_1-iS\,[i\sqrt 3\,c^1+3\,e^{-2\pi i/3}\,d_1\,]\}.
}}
Expressed in real coordinates, the flux takes the form
\eqn\recoeffii{\eqalign{
{1\over (2\pi)^2\alpha'}\,G_3=&-3\,(b_1+iS\,d_1)\,\alpha_0+(a^1+iS\,c^1)\,(\alpha_1+\alpha_2)+(-a^1+3\,b_1+iS\,(-c^1+3\,d_1))\,\alpha_3\cr
&+{1\over3}\,(a^1+iS\,c^1)\,\beta^0+(b_1+iS\,d_1)\,(\beta^1+\beta^2)+(a^1-b_1+iS\,(c^1-d_1))\,\beta^3.
}}


\appendix\appC{Cartan matrices of the relevant Lie groups}

$SU(n+1)$
\eqn\sun{A=\pmatrix{2&-1&0&.&.&.&0&0\cr
-1&2&-1&.&.&.&0&0\cr
0&-1&2&-1&.&.&0&0\cr
.&.&.&.&.&.&.&.\cr
0&0&0&.&.&-1&2&-1\cr
0&0&0&.&.&0&-1&2}}

$SO(2n)$
\eqn\sun{A=\pmatrix{2&-1&0&.&.&.&0&0\cr
-1&2&-1&.&.&.&0&0\cr
0&-1&2&-1&.&.&0&0\cr
.&.&.&.&.&.&.&.\cr
0&0&0&.&.&-1&2&-2\cr
0&0&0&.&.&0&-1&2}}

$G_2$
\eqn\gtwo{A=\pmatrix{2&-3\cr-1&2}}

$F_4$
\eqn\ff{A=\pmatrix{2&-1&0&0\cr
-1&2&-2&0\cr
0&-1&2&-1\cr
0&0&-1&2}}

$E_6$
\eqn\esix{A=\pmatrix{2&-1&0&0&0&0\cr
-1&2&-1&0&0&0\cr
0&-1&2&-1&0&-1\cr
0&0&-1&2&-1&0\cr
0&0&0&-1&2&0\cr
0&0&-1&0&0&2}}

\appendix\appD{Anti--symmetric tensor and K\"ahler moduli for heterotic orbifolds}

For completeness we shall present for the two cosets $\fc{SU(2,2)}{SU(2)\times SU(2)\times U(1)}
\times \fc{SU(1,1)}{U(1)}$ 
and $\fc{SU(3,3)}{SU(3)\times SU(3)}$ the parameterization of the K\"ahler moduli
on the heterotic side.

\subsec{Coset space $\fc{SU(2,2)}{SU(2)\times SU(2)\times U(1)}\times \fc{SU(1,1)}{U(1)}$}

On the heterotic side, the twist--invariant  $2$--form $B_2$
\eqn\twoform{
B_2=b_1\ dx^1\wedge dy^1+b_2\ dx^2\wedge dy^2+b_5\ dx^3\wedge  dy^3+b_3\ (dx^1\wedge dx^2+
dy^1\wedge dy^2)+b_4\ (dx^1\wedge dy^2-dy^1\wedge dx^2)} 
is relevant. The twist invariant $2$--forms are the Hodge--dual of the invariant $4$--forms, given in 
\fourform.
Hence, there is a direct correspondence between the expansion of $C_4$ and $B_2$.
W.r.t. the complex coordinates \introducecomplex\ the $2$--form $B_2$ becomes:
\eqn\Twoform{
B_2=ib_1\ dz^1\wedge d\ov z^1+ib_2\ dz^2\wedge d\ov z^2+ib_5\ dz^3\wedge d\ov z^3+
(b_3-ib_4)\ d\ov z^1\wedge d z^2+(b_3+ib_4)\ dz^1\wedge d\ov z^2\ .} 
We read off the 
definition of the (complexified) K\"ahler moduli $\Tc^i$ from expanding
the form $J+iB$ w.r.t. a basis $\omega_j\ ,\ j=1,\ldots,h_{(1,1)}$ for $H^2(X_6,\IZ)$, \ie
$J+iB=\Tc^j\ \omega_j$. Hence we would conclude:
\eqn\conress{\eqalign{
\Tc^{1}&=g_{11}+i\ b_1\ \ \ ,\ \ \ \Tc^{2}=g_{22}+i\ b_2\ \ \ ,\ \ \ \Tc^5=\re(\Tc^5)+i\ b_5\ ,\cr
\Tc^{3}&=g_{13}+i\ b_4\ \ \ ,\ \ \ \Tc^{4}=g_{14}+i\ b_3\ .}}
The K\"ahler potential $K_\K$ takes the same form as in \Kaehlerpot.
In fact, it may be checked, that from \Kaehlerpot\ the correct metric for the background $g,b$ 
(given in \eqqs \allowme\ and \twoform), \ie 
$$\fc{1}{8}\ \Tr(d g^{-1}\ dg)+
\fc{1}{8}\ \Tr(g^{-1} d b\ g^{-1} d b)=\sum_{j,k} \fc{\p^2 K_\K}{\p \Tc^j\p\ov \Tc^k} d\Tc^j
d\ov \Tc^k$$ follows.

\subsec{Coset space $\fc{SU(3,3)}{SU(3)\times SU(3)}$}

On the heterotic side we consider the twist--invariant $2$--form $B_2$
\eqn\bbfield{\eqalign{
B_2&=b_1\ dx^1\wedge dy^1+b_2\ dx^2\wedge dy^2+b_3\ dx^3\wedge dy^3\cr
&+b_4\ (dx^1\wedge dx^2-dy^1\wedge dx^2+dy^1\wedge dy^2)+b_5\ (dx^1\wedge dy^2-dy^1\wedge dx^2)\cr
&+b_6\ (dx^1\wedge dx^3-dy^1\wedge dx^3+dy^1\wedge dy^3)+b_7\ (dx^1\wedge dy^3-dy^1\wedge dx^3)\cr
&+b_8\ (dx^2\wedge dx^3-dy^2\wedge dx^3+dy^2\wedge dy^3)+b_9\ (dx^2\wedge dy^3-dy^2\wedge dx^3)\ ,}}
which becomes
\eqn\bbfieldcompl{\eqalign{
B_2&=i\sqrt 3\ b_1\ dz^1\wedge d\ov z^1+i\sqrt 3\ b_2\ dz^2\wedge d\ov z^2+
i\sqrt 3\ b_3\ dz^3 \wedge d\ov z^3 \cr
&-i\sqrt 3\ (\rho\ b_4-b_5)\ dz^1\wedge d\ov z^2
-i\sqrt 3\ (\rho\ b_6-b_7) \ dz^1\wedge d\ov z^3
-i\sqrt3\ (\rho\ b_8-b_9) \ dz^2\wedge d\ov z^3\cr
&-i\sqrt 3\ (\ov \rho\ b_4-b_5)\ d z^2\wedge d\ov z^1 
-i\sqrt 3\ (\ov\rho\ b_6-b_7)\ d z^3\wedge d\ov z^1 
-i\sqrt3\ (\ov\rho\ b_8-b_9) \ d z^3\wedge d\ov z^2}}
in complex coordinates \Introduce.
On the heterotic side, we decompose the form $B+iJ$ w.r.t. the integral 
(twist--invariant) basis of $H^2(X_6,\IZ)$, \ie $J+iB=\Tc^j \omega_j$ to 
read of the nine complexified K\"ahler moduli 
$\Tc^i$. This gives:
\eqn\ninemoduli{\eqalign{
\Tc^1&=\fc{\sqrt 3}{2}\ g_{11}+i\ b_1\ \ \ ,\ \ \ \Tc^2=\fc{\sqrt 3}{2}\ g_{33}+i\ b_2\ \ \ ,\ \ \ 
\Tc^3=\fc{\sqrt 3}{2}\ g_{55}+i\ b_3\ ,\cr
\Tc^4&=-\fc{1}{\sqrt 3}\ (g_{13}+2\ g_{14})+i\ b_4\ \ \ ,\ \ \ 
\Tc^5=\fc{1}{\sqrt 3}\ (2\ g_{13}+g_{14})+i\ b_5\ ,\cr
\Tc^6&=-\fc{1}{\sqrt 3}\ (g_{15}+2\ g_{16})+i\ b_6\ \ \ \ ,\ \ \ 
\Tc^7=\fc{1}{\sqrt 3}\ (2\ g_{15}+g_{16})+i\ b_7\ ,\cr
\Tc^8&=-\fc{1}{\sqrt 3}\ (g_{35}+2\ g_{36})+i\ b_8\ \ \ ,\ \ \ 
\Tc^9=\fc{1}{\sqrt 3}\ (2\ g_{35}+g_{36})+i\ b_9\ .}}
Again, the K\"ahler potential $K_\K$ assumes the same form as in \Kaehlerpoti.


\listrefs

\end